\documentclass[prd,showpacs,floatfix,amsmath,amssymb,floatfix]{revtex4}
\usepackage{graphicx,dcolumn,booktabs,bm}
\usepackage{longtable,lscape}
\usepackage{hyperref}
\usepackage{txfonts}
\usepackage{overpic}
\usepackage{multirow}
\usepackage{amssymb}
\usepackage{indentfirst}
\usepackage{feynmf}   
\usepackage{slashed}  
\usepackage{cases}
\usepackage{color,ulem}
\usepackage{graphicx}
\usepackage{epsfig}
\usepackage{subfigure}
\usepackage{calligra}
\usepackage{mathtools}
\usepackage{dcolumn}

\graphicspath{{Figures/}}

\newcommand{\udots}{\mathinner{\mskip1mu\raise1pt\vbox{\kern7pt\hbox{.}}  
        \mskip2mu\raise4pt\hbox{.}\mskip2mu\raise7pt\hbox{.}\mskip1mu}} 

\begin{document}


\title{Tritonlike molecules of three identical baryons}
\author{Li Ma$^{1}$}\email{ma.li@bjtu.edu.cn}

\affiliation{
$^1$Department of physics, Beijing Jiaotong University,\\
Beijing 100044, China}

\date{\today}

\begin{abstract}
In nuclear physics, triton and helium-3 nucleus can be understood as three-body hadronic molecules. 
Analogous to the loosely bound structures for the triton and helium-3 nucleus, whether there is a bound state formed by three hadrons leaves us an open issue. Based on the one-boson exchange model as well as the adoption of the variational approach, we make a comprehensive investigation on the tritonlike systems of three identical baryons $NNN$, $\Lambda\Lambda\Lambda$, $\Xi\Xi\Xi$ and $\Sigma\Sigma\Sigma$. We predict that the three-body molecular states for the systems of three identical hadrons of baryon octet are probably existent as long as their two-body subsystems have bound states. The numerical results of this work may be helpful for the theoretical and experimental researches on the tri-hadron molecules in future. 
\end{abstract}

\pacs{14.40.Rt, 36.10.Gv} \maketitle

\section{Introduction}\label{sec1}

Numerous exotic states call "XYZ" reported in the last few decades pose great challenges to the traditional quark model \cite{Chen:2016qju,Chen:2016spr,Oset:2016lyh,Esposito:2016noz,Lebed:2016hpi,Hosaka:2016pey,Olsen:2017bmm,Francis:2016hui,Karliner:2017qjm,Eichten:2017ffp,Olsen:2014qna}. 
Some of them sit in the vicinity of the open charm threshold can be viewed as good candidates of di-hadron molecules \cite{Guo:2017jvc}. 
The scenario of di-hadron molecule was first motivated by the deuteron, in which a proton and a neutron form a loosely bound state by colorless strong interactions.  
The analog of deuteron as well as the largely accumulated experimental data on the exotic states arouse interest in the explorations on two-body hadronic systems. 
Currently, lots of effort has been spent on searching for the bound states of two-hadron systems \cite{Dong:2017gaw,Li:2012cs,Wang:2013cya,Chen:2015loa,Chen:2015moa,Meissner:2015mza,Xiao:2019mst,Du:2019pij,Zhao:2014gqa,Zhao:2015mga,He:2013nwa,Sun:2011uh,Wang:2019aoc,Chen:2019uvv}. 
In nuclear physics, triton and helium-3 nucleus can be understood as tri-hadron molecules, where the three nucleons bind together via colorless strong interactions with the binding energy 8.40 MeV for triton and 7.80 MeV for helium-3 nucleus. 
Along the same line for deuteron, the existence of triton and helium-3 nucleus leaves us an open questions that whether three-hadron systems have loosely bound states. 

Generally, the Faddeev Equation provide us a rigorous tool to explore the bound solution of a three-body system \cite{Malfliet:1968tj,Eichmann:2009qa,Ishii:1995bu,Eichmann:2011vu,Huang:1993yd,Ishii:1993np,Ishikawa:2002ti,SanchisAlepuz:2011jn,Elster:2008hn,Eichmann:2008ef,Popovici:2010ph,Fujiwara:2003wr,Stadler:1991zz,Valderrama:2018knt,Revai:2014twa}. 
One of the common methods to simplify the Faddeev Equation of specific systems is the Fixed Center Approximation (FCA). For instance, the study of the $X(2175)$ as a resonance of the $\phi K\bar{K}$ has been performed through the FCA method \cite{MartinezTorres:2008gy}. 
The $\pi K\bar{K}$ and $\pi\pi\eta$ via the FCA on the unitary chiral dynamics were also performed \cite{MartinezTorres:2011vh}. 
Theoretical studies on the $KK\bar{K}$ \cite{Torres:2011jt}, the $NK\bar{K}$ \cite{Xie:2010ig}, the $J/\psi K\bar{K}$ \cite{MartinezTorres:2009xb}, the $NDK$, the $\bar{K}DN$ and $ND\bar{D}$ \cite{Xiao:2011rc}, the $BDD$ and $BD\bar{D}$ \cite{Dias:2017miz}, $D\bar{D}^*K$ and $\bar{D}D^*K$ \cite{Ren:2018pcd} were published in recent years. 
Discussions on the $BB^*B^*$ \cite{Garcilazo:2018rwu}, the $\Xi NN$ \cite{Garcilazo:2016ylj}, the $\Omega NN$ and $\Omega\Omega N$ \cite{Garcilazo:2019igo} with similar method also can be found. Other calculations by using FCA method list in Refs \cite{Jido:2008kp,MartinezTorres:2008kh,Bayar:2012rk,Oset:2012gi,MartinezTorres:2010ax,Liang:2013yta,Bayar:2015oea,YamagataSekihara:2010qk,Roca:2010tf,Xie:2011uw,Debastiani:2017vhv}. 
Another method for three-body systems is isobar formalism, which has been adopted to discuss the three-nucleon system with $\Delta(1236)$ isobar \cite{Hajduk:1979yn}. Some other applications via isobar formalism can be found in Refs \cite{Ikeda:2007nz,Ikeda:2008ub,Gal:2013dca,Dreissigacker:1981az}. 
The third useful tool to describe a three-body system called dimer formalism were present in a series of studies, where a composite field is introduced to describe the two-body subsystem when rescattering with a third particle \cite{Konig:2015aka,Konig:2016yka,Wilbring:2017fwy,Schmidt:2018vvl,Hammer:2017uqm,Hammer:2017kms,Meng:2017jgx}.   
The Gauss Expansion Method is another effective tool for few-body system \cite{Hiyama:2018ivm}, which has been applied to discuss few-nucleon systems \cite{Kameyama:1989zz}, and $DDDK$ system \cite{Wu:2019vsy}.  

The One-Boson Exchange (OBE) model works well in describing nuclear force \cite{Nagels:1977ze,Machleidt:1987hj}. It contains long-range force from $\pi$ and $\eta$ exchange, medium-range force from $\sigma$ exchange and short-range force from $\rho/\omega/\phi$. 
There are many theoretical studies on di-hadron systems in the framework of the OBE model \cite{Li:2012cs,Chen:2015loa,Zhao:2014gqa,Zhao:2015mga,Sun:2011uh}. The OBE interaction with the exchange of $\pi$, $\eta$, $\sigma$, $\rho$, $\omega$ and $\phi$ plays an important role in the formation of di-hadron molecules.  
One may wonder how the OBE interaction works in a tri-hadron system and whether the tri-hadron system has a loosely bound state. 
As we know, a triton contains one proton and two neutron. 
Under SU(3) chiral symmetry, proton and neutron belong to the octet of the $1/2^+$ baryon. The existence of triton leave us an interesting question that whether an identical three-body system composed of the other members in the octet of the $1/2^+$ baryon has a bound state. 
Since the success of the OBE mechanism in the description of deuteron, It is quite natural to extend the mechanism to a tri-hadron system consists of three baryons. 
In the present work we shall perform investigations on the three-body systems composed of identical hadrons from the octet of the $1/2^+$ baryon. For simplicity, we denote the three-body systems as "$\mho\mho\mho$", i.e. $NNN$, $\Lambda\Lambda\Lambda$, $\Xi\Xi\Xi$ and $\Sigma\Sigma\Sigma$. Other configurations will be studied in a future work.  

This work is organized as follows. After the introduction, we present the formalism for an identical tri-fermion system within the framework of potential interactions in Sec.~\ref{sec2}. We apply our formalism to the three-nucleon system to verify its feasibility in Sec.~\ref{subsec32}. Then we extend the formalism to the tri-hyperon systems composed of three identical hadrons from the octet of the $1/2^+$ baryon for searching their possible molecular states in Sec.~\ref{subsec33}-\ref{subsec35}. The last section is brief summary. Some technicalities are relegated to the appendix. 

\section{Formalism}\label{sec2}

\begin{figure}[t!]
  \begin{center}
  \rotatebox{0}{\includegraphics*[width=0.35\textwidth]{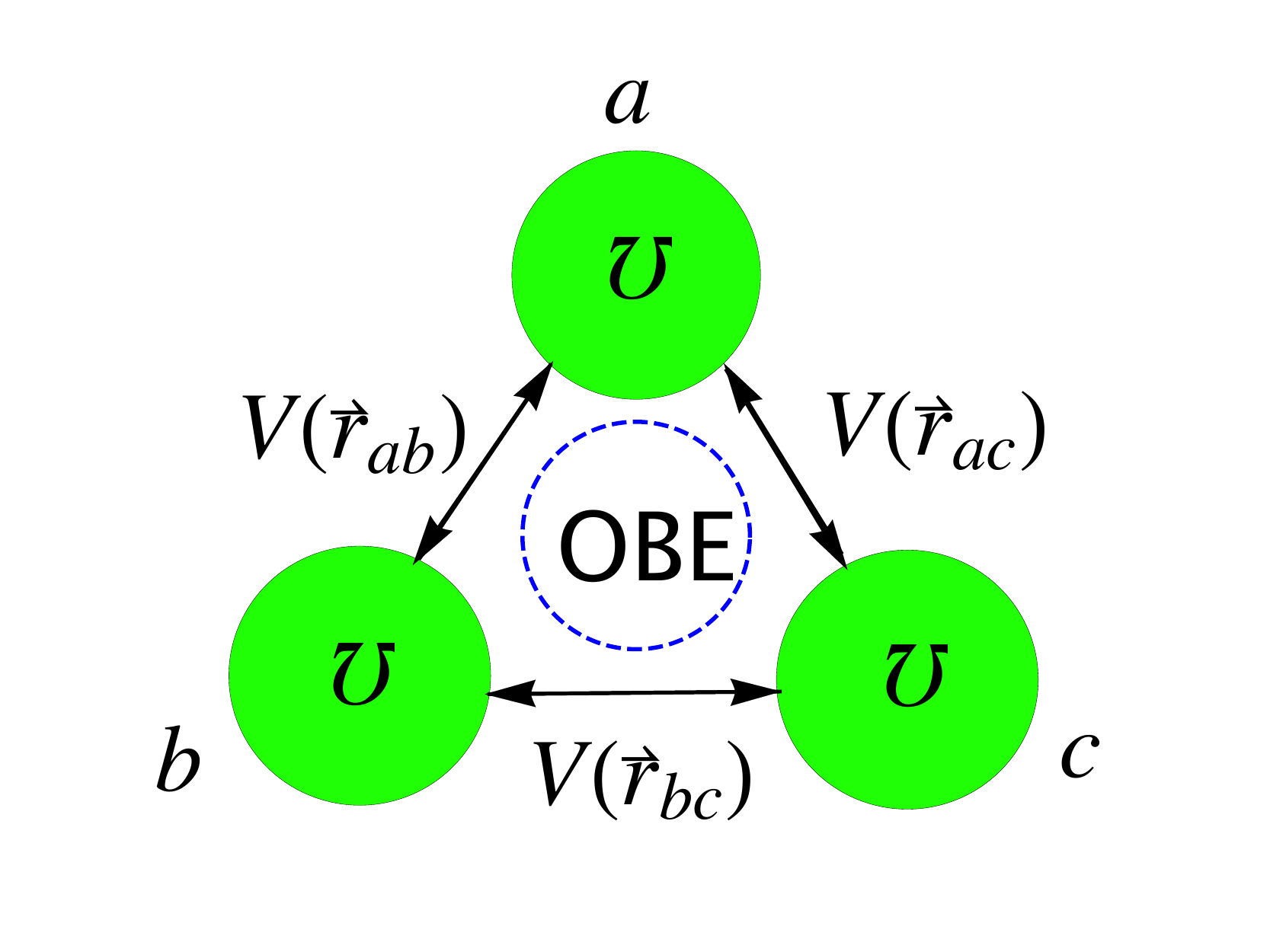}}
  \caption{Dynamical illustration of the identical tri-fermion system $\mho\mho\mho$ with a circle describing the delocalized OBE interaction inside.}
    \label{Fig01}
  \end{center}
\end{figure}

In this section, we construct the general formalism for an identical tri-fermion system. As illustrated in Fig.~\ref{Fig01}, we can use $a$, $b$ and $c$ to label the three fermions, i.e. $\mho_a\mho_b\mho_c$. 
Here the labels are artificial, as the system is invariant under the change of the order of $a$, $b$ and $c$. 
Since the system contains three identical fermions, its total wave function should be antisymmetry under exchange of its two constituents. We use $V(\vec{r}_{ab})$, $V(\vec{r}_{bc})$ and $V(\vec{r}_{ac})$ to denote the effective potential between $\mho_a$ and $\mho_b$, $\mho_b$ and $\mho_c$, $\mho_c$ and $\mho_a$, respectively. $\vec{r}_{ij}$ is the relative displacement between the $i$-th and $j$-th fermions. $T_a$, $T_b$ and $T_c$ are the kinetic energy for the fermions $\mho_a$, $\mho_b$ and $\mho_c$ in their center-of-mass frame, respectively.    

For a better description of the interactions between any two constituents in the system, we divide the channel of $\mho_a\mho_b\mho_c$ into $\{\hat{\mho_{a}}\mho_{b}\mho_{c},\mho_{a}\hat{\mho}_{b}\mho_{c},\mho_{a}\mho_{b}\hat{\mho}_{c}\}$. We make a convention that a fermion $\mho$ absorbs a virtual boson will acquire a "hat" over it. Conversely, a fermion $\hat{\mho}$ emits a virtual boson will lose its "hat". Within this convention, the divided three-channel space is equivalent to the original channel $\{ \mho\mho\mho \}$. The virtual boson is exchanged between any two constituents of the system, and it is not localized between any two constituents but rather shared by the whole system. It is very similar with the delocalized $\Pi$ bond in benzene molecule where a pair of electrons shared by the six carbon atoms. 

The effective Hamiltonian of the three-fermion system in the channel
space $|\mho_{B}\mho_{B}\mho_{B}\rangle\vcentcolon=\{\hat{\mho_{a}}\mho_{b}\mho_{c},\mho_{a}\hat{\mho}_{b}\mho_{c},\mho_{a}\mho_{b}\hat{\mho}_{c}\}$ takes
the following form, 
\begin{eqnarray}
H_{T}=\begin{bmatrix}T_{a}+T_{b}+T_{c} & V(\vec{r}_{ab}) & V(\vec{r}_{ac})\\
V(\vec{r}_{ab}) & T_{a}+T_{b}+T_{c} & V(\vec{r}_{bc})\\
V(\vec{r}_{ac}) & V(\vec{r}_{bc}) & T_{a}+T_{b}+T_{c}
\end{bmatrix},\label{HT}
\end{eqnarray}
where $V=V_{\pi}+V_{\eta}+V_{\rho}+V_{\omega}+V_{\phi}+V_{\sigma}$. 
The $V_{\pi}$, $V_{\eta}$, $V_{\rho}$, $V_{\omega}$, $V_{\phi}$ and $V_{\sigma}$ are the effective potentials from $\pi$, $\eta$, $\rho$, $\omega$, $\phi$ and $\sigma$ exchange, respectively. 

One may wonder whether we can obtain the binding solutions by diagonalizing the Hamiltonian In Eq. (\ref{HT}) directly. In principle, we can do it theoretically, but the procedure contains lots of complicated integrations when the potentials are some combinations of Yukawa potentials. There are two independent degrees of freedom for the motion of a three-body system which can be expressed as a set of Jacobi coordinates generally. Thus, the Hamiltonian elements will contain integrations in several directions. It is quite hard if the potential is complicate.   
Therefore, we should resort approximation method to simplify the problem. The core question to solve the three-body system is how to handle the effect of one of the particle on the scattering of the other two. 

The Born-Oppenheimer Approximation works well on the system composed of light particles and heavy particles \cite{Ma:2017ery,Moroz:2014eba,Braaten:2014qka,Zhao:2014gqa}. Through introducing the Born-Oppenheimer (BO) potential to describe the contribution of light particle on the remaining two heavy particles, we can simplify the three-body system into a two-body system. 
If the BO potential is strong enough, the three-body system may have a bound state. Within this scenario, the light particle works like "glue" to bind the two heavy particles. In fact, the Born-Oppenheimer Approximation is a kind of adiabatic approximation that we divided the degrees of freedom of the three-body system into a heavy one and light one. It inspires us that the BO potential can reflect the influence of one of the particles on the dynamics of the remaining two.     

Even though the application of the BO approximation is not straightforward for the system of three identical particles, one can still adopt the BO potential to describe the influence one of the particles on the scattering of the other two. Since the system $\mho_a\mho_b\mho_c$ is invariant under the interchange of the $a$, $b$ and $c$, one should count the contribution of each fermion on the dynamics of the other two fermions one by one. 
Within this scenario, one can simplify the system into three two-body subsystems $\mho_a\mho_b$, $\mho_b\mho_c$ and $\mho_c\mho_a$ and add the corresponding BO potential from the remaining one in each subsystem instead. If the three subsystems have a negative common eigenvalue, the whole  system may have a three-body bound state. We can call this method as the Born-Oppenheimer potential method (BOP method) for simplicity \cite{Ma:2018vhp}.

\subsection{Born-Oppenheimer Potential}\label{subsec21}

As the discussion above, we can use the BO potential to describe the influence of one of the fermions on the dynamics of the other two. 
That is to say, one can separate the $\mho_a$ from the three-fermion system and add the BO potential from $\mho_a$ when explore the dynamics of the subsystem $\mho_b\mho_c$. 
The derivation procedure of the BO potential is very similar with the BO approximation. 
First we investigate the dynamics of the $\mho_a$ with the assumption that the distance of the $\mho_b$ and $\mho_c$ is a fixed parameter. Here the $\mho_a$ havs one-boson interactions with $\mho_b$ and $\mho_c$ which can be regarded as two static sources. 
We can derive the binding energy of the $\mho_a$ which should be the function of $r_{bc}$. Then we substract the binding energy of the $\mho_a$ in the limit $r_{bc}\rightarrow\infty$ which is trivial for the tri-fermion system. 
One can seperate the effective Hamiltonian from fermion $\mho_a$, 
\begin{eqnarray*}
H_{a}=\begin{bmatrix}T_{a} & V_{ab} & V_{ac}\\
V_{ab} & T_{a} & 0\\
V_{ac} & 0 & T_{a}
\end{bmatrix},
\end{eqnarray*}
then the remaining part of the Hamiltonian in Eq. (\ref{HT}) is
\begin{eqnarray*}
H_{bc}=\begin{bmatrix}T_{b}+T_{c} & 0 & 0\\
0 & T_{b}+T_{c} & V_{bc}\\
0 & V_{bc} & T_{b}+T_{c}
\end{bmatrix},
\end{eqnarray*}
where we have used the abbreviations $V_{ab}$, $V_{bc}$, $V_{ac}$
for $V(\vec{r}_{ab})$, $V(\vec{r}_{bc})$, $V(\vec{r}_{ac})$, respectively.
Assuming the wave function between $\mho_{a}$ and $\mho_{b}$ is
$\Psi_{2}$ , one can write the two-body wave function in the channel
space $|\mho_{B}\mho_{B}\rangle\vcentcolon=\{\hat{\mho_{a}}\mho_{b},\mho_{a}\hat{\mho}_{b}\}$,
which is
\begin{eqnarray*}
\Psi_{2}=\psi(\vec{r}_{ab})|\mho_{B}\mho_{B}\rangle=\frac{1}{\sqrt{2}}\psi(\vec{r}_{ab})|\hat{\mho_{a}}\mho_{b}\rangle+\frac{1}{\sqrt{2}}\psi(\vec{r}_{ab})|\mho_{a}\hat{\mho}_{b}\rangle. 
\end{eqnarray*}
The $\psi(\vec{r}_{ab})$ satisfies the Schrödinger Equation
\begin{eqnarray}
(T_{ab}+V_{ab})\psi_{ab}=E_{2}\psi_{ab},\label{SE}
\end{eqnarray}
where the $E_{2}$ is the two-body eigenenergy of the $\mho_{a}\mho_{b}$. The abbreviations $\psi_{ab}$ and $\psi_{ac}$ are used for $\psi(\vec{r}_{ab})$ and $\psi(\vec{r}_{ac})$, respectively. The Eq. (\ref{SE}) in the channel space $|\mho_{B}\mho_{B}\rangle\vcentcolon=\{\hat{\mho_{a}}\mho_{b},\mho_{a}\hat{\mho}_{b}\}$
reads 
\begin{eqnarray*}
\begin{pmatrix}T & V_{ab}\\
V_{ab} & T
\end{pmatrix}\begin{pmatrix}\frac{1}{\sqrt{2}}\psi_{ab}\\
\frac{1}{\sqrt{2}}\psi_{ab}
\end{pmatrix}=E_{2}\begin{pmatrix}\frac{1}{\sqrt{2}}\psi_{ab}\\
\frac{1}{\sqrt{2}}\psi_{ab}
\end{pmatrix}.
\end{eqnarray*}

Given the wave function $\psi_{ab}$, we can obtain the two-body energy
$E_{2}$ , which is $E_{2}=\psi_{ab}T_{ab}\psi_{ab}+\psi_{ab}V_{ab}\psi_{ab}$
. The final wave function of the $\mho_{a}$ should be the superposition
of the two components 
\begin{eqnarray}
\psi_{a}(r_{ab},r_{ac}) & = & N\Big\{[\frac{1}{\sqrt{2}}\psi(r_{ab})+\frac{1}{\sqrt{2}}\psi(r_{ac})]|\hat{\mho_{a}}\mho_{b}\mho_{c}\rangle\nonumber\\
 & + & \frac{1}{\sqrt{2}}\psi(r_{ab})|\mho_{a}\hat{\mho}_{b}\mho_{c}\rangle+\frac{1}{\sqrt{2}}\psi(r_{ac})|\mho_{a}\mho_{b}\hat{\mho}_{c}\rangle\Big\},\label{Wa}
\end{eqnarray}
where $N$ is normalization coefficient, and we have $|N|^{2}=\left[2+\langle\psi_{ab}|\psi_{ac}\rangle\right]^{-1}$. 
Accordingly, one can obtain the energy value of the $\mho_{a}$ 
\begin{eqnarray}
E_{a}(\Lambda,\vec{r}_{bc}) & = & \langle\psi_{a}(\vec{r}_{ab},\vec{r}_{ac})|H_{a}|\psi_{a}(\vec{r}_{ab},\vec{r}_{ac})\rangle\nonumber\\
 & = & \frac{1}{1+\frac{1}{2}\langle\psi_{ab}|\psi_{ac}\rangle}\left\{ E_{2}+\frac{1}{4}\langle\psi_{ab}|T_{a}|\psi_{ac}\rangle+\frac{1}{4}\langle\psi_{ac}|T_{a}|\psi_{ab}\rangle+\frac{1}{2}\langle\psi_{ab}|V_{ab}|\psi_{ac}\rangle+\frac{1}{2}\langle\psi_{ac}|V_{ab}|\psi_{ab}\rangle\right\},\label{Ea}
\end{eqnarray}
where in the second step we have used Eq. (\ref{SE}) and the symmetry between
$b$ and $c$. The $\lambda$ is a scaling parameter which can be determined by the experimental data for the two-body binding energy $E_2$. Since the interactions should be relevant to the scaling, both the wave functions and binding energy should depend on the scaling parameter $\lambda$. Thus,  $E_{a}$ is also relevant to the $\lambda$. 

As shown in Fig.~\ref{Fig01}, when the $r_{bc}=0$, the energy value of fermion $\mho_a$ reach to its minimum. It corresponds to the limit that the $\mho_b$ and $\mho_c$ are on top of each other, then the system is reduced to the $\mho_b\mho_c-\mho_a$ quasi-two-body system. If we put the fermion $\mho_b$ infinitely far away from the fermion $\mho_c$, which corresponds the limit $r_{bc}\rightarrow\infty$, then the energy $E_a$ will tend to the two-body energy eigenvalue $E_2$. It can be easily seen from the Eq. (\ref{Ea}), where the overlap integration $\langle\psi_{ab}|\psi_{ac}\rangle$, $\langle\psi_{ab}|T_{a}|\psi_{ac}\rangle$, $\langle\psi_{ac}|T_{a}|\psi_{ab}\rangle$, $\langle\psi_{ab}|V_{ab}|\psi_{ac}\rangle$ and $\langle\psi_{ac}|V_{ab}|\psi_{ab}\rangle$ tends to $0$. It corresponds to the case that a two-body system plus a free fermion. If the attraction provided by the one-boson exchange is strong enough, the fermion $\mho_a$ can form a two-body bound state with either $\mho_b$ or $\mho_c$. In fact, it is the break-up state for the tri-fermion system. In general, the $E_2$ is a number quantity irrelevant to the $r_{bc}$, which is trivial for the tri-fermion system. Thus, we define the BO potential as
\begin{eqnarray*}
V_{BO}(\lambda,\vec{r}_{bc})=E_{a}(\lambda,\vec{r}_{bc})-E_{a}(\lambda,\infty).
\end{eqnarray*}
Therefore, the physical meaning of the BO potential between the $\mho_b$ and $\mho_c$ is the energy eigenvalue of fermion $\mho_a$ relative to that of the break-up state.

\subsection{The configurations of tritonlike systems}\label{subsec22}

Assuming the fermions $\mho_b$ and $\mho_c$ are much heavier than the $\mho_a$, one can use the BO approximation to separate the motion of fermion $\mho_a$ from the three-body system. The BO approximation is a kind of adiabatic approximation that we divide the degrees of freedom into a heavy one and light one. 
The heavy degree of freedom is the relative motion between $\mho_b$ and $\mho_c$. The light degree of freedom is the motion of fermion $\mho_a$ relative to the three-body center of mass. 
The BO potential can reflect the influence of one fermion on the dynamics of the other two. 
Then one can simplify the three-body system into a two-body system $\mho_b\mho_c$ with the BO potential created by the fermion $\mho_a$. 
The light degree of freedom can be described by the wave function of the $\mho_a$ we have derived in Eq. (\ref{Wa}). 
The heavy degree of freedom can be described by a wave function $\Phi(\vec{r}_{bc})$ which should be determined by three-body Schrödinger Equation. 
Then, in this case, the total wave function of the system has the form $\Phi(\vec{r}_{bc})\psi(\vec{r}_{ab},\vec{r}_{ac})$.        

For the system $\mho_a\mho_b\mho_c$, the application of the BO approximation is not straightforward. However, we can use the BOP method where the BO potential is still employed to describe the contribution of one of the particles on the dynamics of the other two. 
Based on the BOP method, one should count the influence of each fermion on the dynamics of the other two fermions one by one for the system $\mho_a\mho_b\mho_c$. 
The system has three basic simplification schemes. One is that we divide the system into the two-body subsystem $\mho_b\mho_c$ with the BO potential created by the fermion $\mho_a$. One is $\mho_c\mho_a$ with the BO potential created by the fermion $\mho_b$. The other one is $\mho_a\mho_b$ with the BO potential created by the fermion $\mho_c$. The three different simplification schemes lead to three different configurations. For simplicity, we use $\psi_{\slashed{a}}
=\Phi(\vec{r}_{bc})\psi(\vec{r}_{ab},~\vec{r}_{ac})$, $\psi_{\slashed{b}}=\Phi(\vec{r}_{ac})\psi(\vec{r}_{ab},~\vec{r}_{bc})$ and $\psi_{\slashed{c}}=
\Phi(\vec{r}_{ab})\psi(\vec{r}_{bc},~\vec{r}_{ac})$ represent the three basic configuration wave functions, where $\Phi(\vec{r}_{bc})$, $\Phi(\vec{r}_{ac})$ and $\Phi(\vec{r}_{ab})$ are undetermined functions that need to be solved. The $\psi(\vec{r}_{ab},~\vec{r}_{bc})$ and $\psi(\vec{r}_{bc},~\vec{r}_{ac})$ can be obtained by the similar procedures for the $\psi(\vec{r}_{ab},~\vec{r}_{ac})$. 
The configuration wave function $\psi_{\slashed{a}}$ denote the scheme that we omit the fermion $\mho_a$ and add the BO potential provided by the fermion $\mho_a$ instead. The $\psi_{\slashed{b}}$ and $\psi_{\slashed{c}}$ denote the BO potential provided by the fermion $\mho_b$ and $\mho_c$, respectively. 

The three configuration functions can be regarded as a set of basis states and constitute a configuration space $\{ \psi_{\slashed{a}},~\psi_{\slashed{b}},~\psi_{\slashed{c}} \}$. We expect that 
a common approximation for the three-body eigenstate can be expressed as a superposition of the three kinds of basic configurations. Thus, the interpolating wave function of the three-body wave function can be written as
\begin{eqnarray}
\Psi_T &=& 
 \alpha\psi_{\slashed{a}}+\beta\psi_{\slashed{b}}+\gamma\psi_{\slashed{c}}  
 = \left(
\begin{array}{c}
   \alpha  \\
   \beta  \\
   \gamma 
\end{array}
\right), \label{Basis0}
\end{eqnarray}
where the $\alpha$, $\beta$ and $\gamma$ are the expansion coefficients. The three basic configuration functions in the channel space $\{\hat{\mho_{a}}\mho_{b}\mho_{c},\mho_{a}\hat{\mho}_{b}\mho_{c},\mho_{a}\mho_{b}\hat{\mho}_{c}\}$ according to Eq.~(\ref{Wa}) read 
\begin{eqnarray}
\psi_{\slashed{a}} = N \Phi(\vec{r}_{bc}) \left(
\begin{array}{c}
   \frac{1}{\sqrt{2}}[\psi(\vec{r}_{ab})+\psi(\vec{r}_{ac})] \\
   \frac{1}{\sqrt{2}}\psi(\vec{r}_{ab})  \\
   \frac{1}{\sqrt{2}}\psi(\vec{r}_{ac})  
  \end{array}
\right), 
~
\psi_{\slashed{b}} = N \Phi(\vec{r}_{ac}) \left(
\begin{array}{c}
   \frac{1}{\sqrt{2}}\psi(\vec{r}_{ab})  \\
   \frac{1}{\sqrt{2}}[\psi(\vec{r}_{ab})+\psi(\vec{r}_{bc})] \\
   \frac{1}{\sqrt{2}}\psi(\vec{r}_{bc})  
\end{array}
\right), 
~
\psi_{\slashed{c}} = N \Phi(\vec{r}_{ab}) \left(
\begin{array}{c}
   \frac{1}{\sqrt{2}}\psi(\vec{r}_{ac})  \\
   \frac{1}{\sqrt{2}}\psi(\vec{r}_{bc})  \\
   \frac{1}{\sqrt{2}}[\psi(\vec{r}_{bc})+\psi(\vec{r}_{ac})] 
\end{array}
\right), \label{Phi}
\end{eqnarray}
where the $\Phi(\vec{r}_{bc})$ can be expanded as a set of Laguerre Polynomials $\Phi(\vec{r}_{bc})= \sum_{i} a_i\phi_i(\vec{r}_{bc})$. $N$ is a normalization constant. We define the $i\rm{th}$ order of the
configuration functions as $\psi_{\slashed{a}}^i = \phi_i(\vec{r}_{bc})\psi(\vec{r}_{ab},~\vec{r}_{ac})$,
$\psi_{\slashed{b}}^i = \phi_i(\vec{r}_{ac})\psi(\vec{r}_{ab},~\vec{r}_{bc})$ and $\psi_{\slashed{c}}^i
= \phi_i(\vec{r}_{ab})\psi(\vec{r}_{bc},~\vec{r}_{ac})$. 

Then the configuration space $\{ \psi_{\slashed{a}},~\psi_{\slashed{b}},~\psi_{\slashed{c}} \}$ can be expanded as $\{ \psi_{\slashed{a}}^i,\cdots,~\psi_{\slashed{b}}^i,\cdots,~\psi_{\slashed{c}}^i \}$. 
In order to get an orthogonal basis, we should orthonormalize the $\{ \psi_{\slashed{a}}^i,\cdots,~\psi_{\slashed{b}}^i,\cdots,~\psi_{\slashed{c}}^i \}$
into a new basis $\{ \tilde{\psi}_{\slashed{a}}^i,\cdots,~\tilde{\psi}_{\slashed{b}}^i,\cdots,~\tilde{\psi}_{\slashed{c}}^i \}$.
The $\tilde{\psi}_{\slashed{a}}^i$, $\tilde{\psi}_{\slashed{b}}^i$ and $\tilde{\psi}_{\slashed{c}}^i$ are used to denote the $i\rm{th}$ order of the new configuration functions $\tilde{\psi}_{\slashed{a}}$,
$\tilde{\psi}_{\slashed{b}}$ and $\tilde{\psi}_{\slashed{c}}$, respectively. Then the interpolating wave function of the three-body system can be written as 
\begin{eqnarray}
\left(
\begin{array}{c}
\tilde{\psi}_{\slashed{a}}^i \\
\vdots \\
\tilde{\psi}_{\slashed{b}}^i \\
\vdots \\
\tilde{\psi}_{\slashed{c}}^i 
\end{array}
\right) = 
\left(
\begin{array}{c}
\frac{1}{N_i}\big[ (\psi_{\slashed{a}}^i+\psi_{\slashed{b}}^i+\psi_{\slashed{c}}^i)-\sum_{i} x_{ij}\psi_{\slashed{a}}^j \big] \\
\vdots \\
\frac{1}{N_i}\big[ (\psi_{\slashed{a}}^i+\psi_{\slashed{b}}^i+\psi_{\slashed{c}}^i)-\sum_{i} x_{ij}\psi_{\slashed{b}}^j \big] \\
\vdots \\
\frac{1}{N_i}\big[ (\psi_{\slashed{a}}^i+\psi_{\slashed{b}}^i+\psi_{\slashed{c}}^i)-\sum_{i} x_{ij}\psi_{\slashed{c}}^j \big]
\end{array}
\right),  \label{PsiTT}
\end{eqnarray}
where the $x_{ij}$ and $N_i$ are the parameter matrix and normalization coefficients, respectively, which need to be determined later. Because of the interchange symmetry for the $\mho\mho\mho$ system, the parameter matrix $x_{ij}$ in the three configuration functions are the same.

In Eq. (\ref{PsiTT}), the $i\rm{th}$ order of configuration function $\tilde{\psi}_{\slashed{a}}^i$ should be orthogonal with the any order of the other configuration function $\tilde{\psi}_{\slashed{a}}^j$,
We get the orthomoraliztion condition
\begin{eqnarray*}
\langle \tilde{\psi}_{\slashed{a}}^i | \tilde{\psi}_{\slashed{b}}^j \rangle= \Big{\langle} \frac{1}{N_i}\big[ (\psi_{\slashed{a}}^i+\psi_{\slashed{b}}^i+\psi_{\slashed{c}}^i)-\sum_{i} x_{ik}\psi_{\slashed{a}}^k \big] \Big{|} \frac{1}{N_j}\big[ (\psi_{\slashed{a}}^j+\psi_{\slashed{b}}^j+\psi_{\slashed{c}}^j)-\sum_{i} x_{jl}\psi_{\slashed{b}}^l \big] \Big{\rangle} = \delta_{ij},
\end{eqnarray*}
which yields
\begin{eqnarray}
x_{ik}\langle \psi_{\slashed{a}}^k | \psi_{\slashed{b}}^l \rangle x_{lj} - x_{ik}(\delta_{kj}+2\langle \psi_{\slashed{a}}^k | \psi_{\slashed{b}}^j \rangle) - x_{jl}(\delta_{il}+2\langle \psi_{\slashed{a}}^i | \psi_{\slashed{b}}^l \rangle) + 3\delta_{ij}+6\langle \psi_{\slashed{a}}^i | \psi_{\slashed{b}}^j \rangle = 0~, \label{orth}
\end{eqnarray}
It will determine the parameter matrix $x_{ij}$. The orthomoraliztion condition also gives
\begin{eqnarray}
\frac{1}{N^{\ast}_i N_j} \big[  3\delta_{ij}+ 6 \langle \psi_{\slashed{a}}^i | \psi_{\slashed{b}}^j \rangle -2x_{ij}-4 \sum_{m} x_{im} \langle \psi_{\slashed{a}}^m | \psi_{\slashed{b}}^j \rangle + \sum_{n} x_{in}x_{nj}  \big]=\delta_{ij}~. \label{norm}
\end{eqnarray}
It will determine the normalization coefficients $N_i$. We get an orthonormalized configuration basis after solving the equations for $x_{ij}$ and $N_i$. Then the eigenvector for the three-body system $\mho\mho\mho$ can be written as a vector in the new configuration space $\{ \tilde{\psi}_{\slashed{a}},~\tilde{\psi}_{\slashed{b}},
 ~\tilde{\psi}_{\slashed{c}} \}$. Thus, the eigenvector for the three-body system has the from
\begin{eqnarray*}
\tilde{\Psi}_T= \sum_{i}\tilde{\alpha}_{i}\tilde{\psi}_{\slashed{a}}^i+\sum_{j}\tilde{\beta}_{j}\tilde{\psi}_{\slashed{b}}^j+\sum_{k}\tilde{\gamma}_{k}\tilde{\psi}_{\slashed{c}}^k
 = \left(
\begin{array}{c}
   \tilde{\alpha}_{i}  \\
   \vdots \\
   \tilde{\beta}_{i} \\
   \vdots \\ 
   \tilde{\gamma}_{i}
\end{array}
\right),
\end{eqnarray*}
where the $\tilde{\alpha}_{i}$, $\tilde{\beta}_{i}$ and $\tilde{\gamma}_{i}$ are the $i\rm{th}$ order
expansion coefficients of the new configuration basis.

\subsection{Three-body Schrödinger Equation and its corrections}\label{subsec23}
We define a reduced Hamiltonian as
\begin{equation*}
\mathcal{H}=H-E_2. 
\end{equation*}
Without the kinetic energy of the center of mass for the three-body system, the explicit Hamiltonian has the form 
\begin{eqnarray*}
H_{T}=\begin{bmatrix}T_{\ast}+T'_{\ast} & V(\vec{r}_{ab}) & V(\vec{r}_{ac})\\
V(\vec{r}_{ab}) & T_{\ast}+T'_{\ast} & V(\vec{r}_{bc})\\
V(\vec{r}_{ac}) & V(\vec{r}_{bc}) & T_{\ast}+T'_{\ast}
\end{bmatrix},
\end{eqnarray*}
where $V=V_{\pi}+V_{\eta}+V_{\rho}+V_{\omega}+V_{\phi}+V_{\sigma}$.  $T_{\ast}=-({1}/{2\mu})\nabla_{ab}^2$, $T'_{\ast}=-({1}/{2\mu'})\nabla_{\xi}^2$ are the kinetic energy operators, and the corresponding reduced  masses are $\mu=M/2$, $\mu'=\frac{2}{3}M$. Here $\nabla_{ab}^2=({1}/{r_{ab}})({d^2}/{dr_{ab}^2})r_{ab}-({\overrightarrow{L}_{ab}^2}/{r_{ab}^2})$ and
$\nabla_{\xi}^2=({1}/{\xi})({d^2}/{d\xi^2})\xi-({\overrightarrow{L}_{\xi}^2}/{\xi^2})$
with $\vec{\xi}={\vec{r}_{ab}}/{2}-\vec{r}_{bc}$, where $\vec{r}_{bc}$ is the direction of the fermion $\mho_b$ relative to the fermion $\mho_c$. $\overrightarrow{L}_{\xi}$ is the relative angular momentum operator between two-body centre of mass for the fermion $a$ and $b$ and the fermion $c$.
$\overrightarrow{L}_{ab}$ is the angular momentum operator between fermions $\mho_a$ and $\mho_b$. 

We can rewrite the total Hamiltonian for the three-body system in the configuration space $\{ \tilde{\psi}_{\slashed{a}}^i,\cdots,~\tilde{\psi}_{\slashed{b}}^i,\cdots,~\tilde{\psi}_{\slashed{c}}^i \}$ as
\begin{eqnarray}
H_T = \left(
        \begin{array}{ccccc}
 H_{\slashed{a}\slashed{a}}^{ij}&\cdots & H_{\slashed{a}\slashed{b}}^{ij} &\cdots  & H_{\slashed{a}\slashed{c}}^{ij} \\
 \vdots &\ddots &\vdots&\udots & \vdots  \\
 H_{\slashed{b}\slashed{a}}^{ij}&\cdots & H_{\slashed{b}\slashed{b}}^{ij} &\cdots & H_{\slashed{b}\slashed{c}}^{ij}  \\
 \vdots&\udots &\vdots&\ddots & \vdots  \\
 H_{\slashed{c}\slashed{a}}^{ij} &\cdots & H_{\slashed{c}\slashed{b}}^{ij}  &\cdots & H_{\slashed{c}\slashed{c}}^{ij}  
       \end{array}
       \right)=
       \left(
        \begin{array}{ccccc}
 \mathcal{H}_{\slashed{a}\slashed{a}}^{ij}&\cdots & \mathcal{H}_{\slashed{a}\slashed{b}}^{ij}  &\cdots& \mathcal{H}_{\slashed{a}\slashed{c}}^{ij} \\
  \vdots &\ddots &\vdots&\udots & \vdots  \\
 \mathcal{H}_{\slashed{b}\slashed{a}}^{ij} &\cdots& \mathcal{H}_{\slashed{b}\slashed{b}}^{ij}  &\cdots& \mathcal{H}_{\slashed{b}\slashed{c}}^{ij}  \\
  \vdots&\udots &\vdots&\ddots & \vdots  \\
 \mathcal{H}_{\slashed{c}\slashed{a}}^{ij} &\cdots& \mathcal{H}_{\slashed{c}\slashed{b}}^{ij}  &\cdots& \mathcal{H}_{\slashed{c}\slashed{c}}^{ij}  
       \end{array}
       \right)+E_2
       \left(
        \begin{array}{ccccc}
        1 &\cdots& 0 &\cdots& 0 \\
          \vdots &\ddots &\vdots&\udots & \vdots  \\
        0 &\cdots& 1 &\cdots& 0 \\
           \vdots&\udots &\vdots&\ddots & \vdots  \\
        0 &\cdots& 0 &\cdots& 1
        \end{array}
       \right),
\end{eqnarray}
with $H_{\slashed{m}\slashed{n}}=\langle \tilde{\psi}_{\slashed{m}}| H | \tilde{\psi}_{\slashed{m}}\rangle$
$(m,~n=a,~b,~c)$.

The total reduced Hamiltonian for the three-body system $\mho\mho\mho$ in
the configuration space $\{ \tilde{\psi}_{\slashed{a}}^i,\cdots,~\tilde{\psi}_{\slashed{b}}^i,\cdots,~\tilde{\psi}_{\slashed{c}}^i \}$ reads
\begin{eqnarray}
\mathcal{H}_T = \left(
        \begin{array}{ccccc}
 \mathcal{H}_{\slashed{a}\slashed{a}} &\cdots& \mathcal{H}_{\slashed{a}\slashed{b}}  &\cdots& \mathcal{H}_{\slashed{a}\slashed{c}} \\
   \vdots &\ddots &\vdots&\udots & \vdots  \\
 \mathcal{H}_{\slashed{b}\slashed{a}} &\cdots& \mathcal{H}_{\slashed{b}\slashed{b}}  &\cdots& \mathcal{H}_{\slashed{b}\slashed{c}}  \\
 \vdots&\udots &\vdots&\ddots & \vdots  \\
 \mathcal{H}_{\slashed{c}\slashed{a}} &\cdots& \mathcal{H}_{\slashed{c}\slashed{b}}  &\cdots& \mathcal{H}_{\slashed{c}\slashed{c}}  
       \end{array}
       \right),
\end{eqnarray}
with $\mathcal{H}_{\slashed{m}\slashed{n}}=\langle \tilde{\psi}_{\slashed{m}}| \mathcal{H} |
\tilde{\psi}_{\slashed{m}}\rangle$ $(m,~n=a,~b,~c)$. Thus we have $H_T=\mathcal{H}_{T}+E_2$, The explicit form of the matrix element $\mathcal{H}_{\slashed{a}\slashed{a}}$ will be listed in Appendix. 
Given the above definitions, the three-body Schr\"{o}dinger equation $\mathcal{H}_{T}\Psi_T=E_3\Psi_T$ reads
\begin{eqnarray}
 \left(
        \begin{array}{ccccc}
 \mathcal{H}_{\slashed{a}\slashed{a}} &\cdots& \mathcal{H}_{\slashed{a}\slashed{b}}  &\cdots& \mathcal{H}_{\slashed{a}\slashed{c}} \\
  \vdots &\ddots &\vdots&\udots & \vdots  \\
 \mathcal{H}_{\slashed{b}\slashed{a}} &\cdots& \mathcal{H}_{\slashed{b}\slashed{b}}  &\cdots& \mathcal{H}_{\slashed{b}\slashed{c}}  \\
  \vdots&\udots &\vdots&\ddots & \vdots  \\
 \mathcal{H}_{\slashed{c}\slashed{a}} &\cdots& \mathcal{H}_{\slashed{c}\slashed{b}}  &\cdots& \mathcal{H}_{\slashed{c}\slashed{c}}  
       \end{array}
       \right)
       \left(
\begin{array}{c}
   \tilde{\alpha}_i  \\
   \vdots \\
   \tilde{\beta}_i  \\
   \vdots \\
   \tilde{\gamma}_i 
\end{array}
\right)
= E_3
\left(
\begin{array}{c}
     \tilde{\alpha}_i  \\
   \vdots \\
   \tilde{\beta}_i  \\
   \vdots \\
   \tilde{\gamma}_i 
\end{array}
\right),  \label{SDE}
\end{eqnarray}
where the energy eigenvalue $E_3$ is the reduced three-body energy eigenvalue relative to the break-up state. The total energy eigenvalue relative to the $\mho\mho\mho$ mass threshold is $E_T=E_3+E_2$. 

One may wonder the wave function $\psi_a(\vec{r}_{ab},\vec{r}_{ac})$ we used for calculating the BO potential is too rough. Through simply superpositions of the two components as Eq. (\ref{Wa}), we get the interpolating wave functions of the fermion $\mho_a$. The $\psi(\vec{r}_{ab})$ and $\psi(\vec{r}_{ac})$ are simply obtained by solving the two-body Schrödinger Equations for the $\mho_a\mho_b$ and $\mho_a\mho_c$. In fact, the existence of the $\mho_c$ will distort the shape of the wave function for the $\mho_a\mho_b$. We should consider the distortion created by the $\mho_c$ while calculating the BO potential between the $\mho_a$ and $\mho_b$. After solving the three-body Schrödinger Equation as Eq. (\ref{SDE}), we obtain the distorted wave functions $\Phi(\vec{r}_{bc})$, $\Phi(\vec{r}_{ab})$ and $\Phi(\vec{r}_{ac})$ in Eq. (\ref{Phi}). Therefore, we should use $\Phi(\vec{r}_{ab})$ and $\Phi(\vec{r}_{ac})$ to calculate the BO potential created by the $\mho_a$, which can be regarded as the first order corrections. With the distortion effect, the wave function for the $\mho_a$ reads
\begin{eqnarray*}
\psi'_{a}(r_{ab},r_{ac}) & = & N'\Big\{[\frac{1}{\sqrt{2}}\Phi(r_{ab})+\frac{1}{\sqrt{2}}\Phi(r_{ac})]|\hat{\mho_{a}}\mho_{b}\mho_{c}\rangle\\
 & + & \frac{1}{\sqrt{2}}\Phi(r_{ab})|\mho_{a}\hat{\mho}_{b}\mho_{c}\rangle+\frac{1}{\sqrt{2}}\Phi(r_{ac})|\mho_{a}\mho_{b}\hat{\mho}_{c}\rangle\Big\},
\end{eqnarray*}
where $N'$ is normalization coefficient, and we have $|N'|^{2}=\left[2+\langle\Phi_{ab}|\Phi_{ac}\rangle\right]^{-1}$. 
Accordingly, Considering the distortion effect, one can obtain the energy value of the $\mho_{a}$ 
\begin{eqnarray*}
E'_{a}(\Lambda,\vec{r}_{bc}) & = & \langle\psi'_{a}(\vec{r}_{ab},\vec{r}_{ac})|H_{a}|\psi'_{a}(\vec{r}_{ab},\vec{r}_{ac})\rangle\\
 & = & \frac{1}{1+\frac{1}{2}\langle\Phi_{ab}|\Phi_{ac}\rangle}\left\{ E_{2}+\frac{1}{4}\langle\Phi_{ab}|T_{a}|\Phi_{ac}\rangle+\frac{1}{4}\langle\Phi_{ac}|T_{a}|\Phi_{ab}\rangle+\frac{1}{2}\langle\Phi_{ab}|V_{ab}|\Phi_{ac}\rangle+\frac{1}{2}\langle\Phi_{ac}|V_{ab}|\Phi_{ab}\rangle\right\} .
\end{eqnarray*}
 
We define the BO potential with the distortion correction as
\begin{eqnarray*}
V'_{BO}(\Lambda,\vec{r}_{bc})=E'_{a}(\Lambda,\vec{r}_{bc})-E'_{a}(\Lambda,\infty).
\end{eqnarray*}
Then with the similar procedures discussed in Subsec.~\ref{subsec22} and Subsec.~\ref{subsec23}, we can arrive the final three-body Schrödinger Equation. The solutions of this three-body Schrödinger Equation may unveil the binding information of the three-body system $\mho_a\mho_b\mho_c$.

\section{Applications to the baryon octet}\label{sec3}

We have construct the formalism for the system of three identical fermions above. 
Now we return to the systems mainly concerned in this work, which are the tritonlike systems composed of identical hadrons from the octet of the $1/2^+$ baryon, i.e., $NNN$, $\Lambda\Lambda\Lambda$, $\Xi\Xi\Xi$ and $\Sigma\Sigma\Sigma$. We shall perform investigations on the bound states of these tritonlike systems, and study the dependence of their three-body binding energies on the two-body binding energies of their two-body subsystems.  

\subsection{The dynamics of the baryon octet}\label{subsec31}

\begin{figure}[t!]
  \begin{center}
  \rotatebox{0}{\includegraphics*[width=0.45\textwidth]{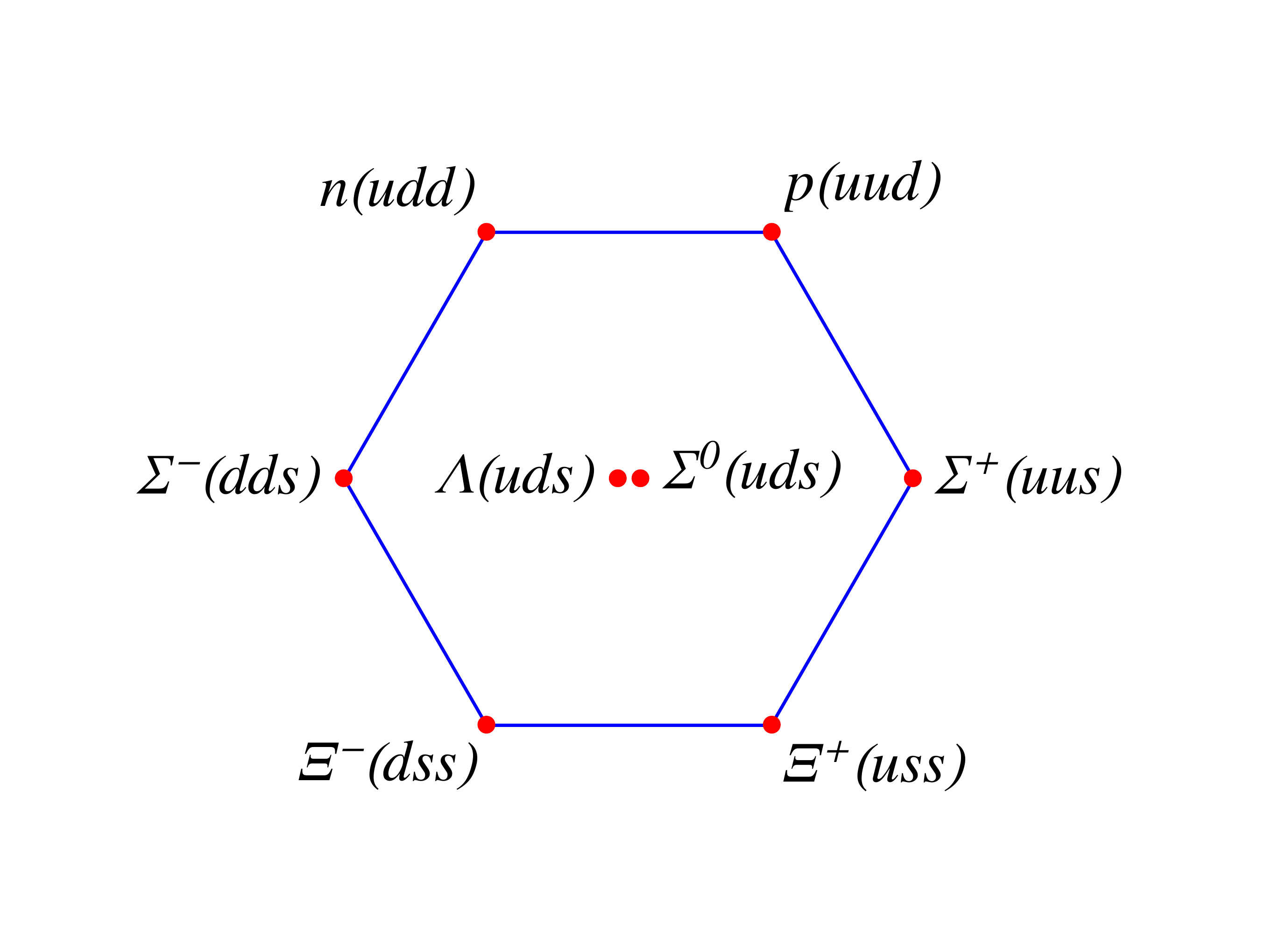}}
  \caption{Diagram of baryon octet. We have included the quark content.}
    \label{Fig02}
  \end{center}
\end{figure}

Under the SU(3) flavor symmetry, the states $p$, $n$, $\Xi^0$, $\Xi^-$, $\Sigma^+$, $\Sigma^0$, $\Sigma^-$ and $\Lambda$ are grouped into the ground state baryon octet. We draw the diagram of the baryon octet in Fig.~\ref{Fig02}, where we have included the quark content. Among them, the $\Lambda$ is an isoscalar; $\{\Xi^0,\Xi^-\}$ and $\{p,n\}$ are isospinors; and the $\{\Sigma^+,\Sigma^0,\Sigma^-\}$ is an isovector. We denote these state using $\Lambda$, $\Xi$, $N$ and $\Sigma$. The wave function of a tritonlike system consists of its isospin, spin and spatial wave functions. For simplicity, we define the isospin wave function of a tritonlike system as $|I_2,I_3,I_{3z}\rangle$, where the $I_2$ is the isospin of its two-body subsystem, the $I_3$ and $I_{3z}$ denote the total isospin of the whole system and its $z$ direction, respectively. We collect the isospin wave functions of them in Appendix B. Since the strong interactions conserve isospin symmetry, the effective potentials and eigenvalues for a specific system do not depend on the third components of its isospin. It is adequate to take the isospin wave function of a tritonlike system as $|I_2,I_3\rangle$ in our calculations. 

Since we focus on the systems composed of three identical baryon, the generalized identity principle  constrict their isospin and spin wave functions to be antisymmetry. Thus, only some specific combinations of the wave functions can survive. 
Generally speaking, the two-body force in these tritonlike systems depends on their isospin and spin. However, it cannot be arbitrary due to the constrain from the generalized identity principle. 
To illuminate it specifically, we use the $(I_2,S_2)$ to denote the isospin and spin of the interacting baryons. For the two-body force in the $NNN$ system, only the cases $(1,0)$ and $(0,1)$ survive. For the $\Lambda\Lambda\Lambda$ system, there is only one case $(0,0)$ need to be considered. For the $\Xi\Xi\Xi$ system, the cases $(1,0)$ and $(0,1)$ should be taken into account. The $\Sigma\Sigma\Sigma$ system only have three cases $(0,0)$, $(1,1)$ and $(2,0)$ after considering the generalized identity principle.  

Since our computation is based on the OBE model, we should build the Lagrangians for $N/\Lambda/\Xi/\Sigma$ with interactions from $\pi/\eta/\rho/\omega/\phi/\sigma$. 
The Lagrangians under SU(3)-flavor symmetry read
\begin{eqnarray}
\mathcal{L}_{NN} & = & g_{\pi NN}\bar{N}i\gamma_{5}\mathcal{\boldsymbol{\tau}}N\cdot\boldsymbol{\pi}+g_{\eta NN}\bar{N}i\gamma_{5}N\eta +  g_{\rho NN}\bar{N}\gamma_{\mu}\boldsymbol{\tau}N\cdot\boldsymbol{\rho}^{\mu}+\frac{f_{\rho NN}}{2M_{N}}\bar{N}\sigma_{\mu\nu}\boldsymbol{\tau}N\cdot\partial^{\mu}\boldsymbol{\rho}^{\nu}\nonumber\\
 & + & g_{\omega NN}\bar{N}\gamma_{\mu}N\omega^{\mu}+\frac{f_{\omega NN}}{2M_{N}}\bar{N}\sigma_{\mu\nu}N\partial^{\mu}\omega^{\nu} +  g_{\sigma NN}\bar{N}N\sigma,\label{LNN}
\end{eqnarray}
\begin{eqnarray}
\mathcal{L}_{\Lambda\Lambda} & = & g_{\pi\Lambda\Lambda}\bar{\Lambda}i\gamma_{5}\mathcal{\boldsymbol{\tau}}\Lambda\cdot\boldsymbol{\pi}+g_{\eta\Lambda\Lambda}\bar{\Lambda}i\gamma_{5}\Lambda\eta +  g_{\rho\Lambda\Lambda}\bar{\Lambda}\gamma_{\mu}\boldsymbol{\tau}\Lambda\cdot\boldsymbol{\rho}^{\mu}+\frac{f_{\rho\Lambda\Lambda}}{2M_{\Lambda}}\bar{\Lambda}\sigma_{\mu\nu}\boldsymbol{\tau}\Lambda\cdot\partial^{\mu}\boldsymbol{\rho}^{\nu}\nonumber\\
 & + & g_{\omega NN}\bar{\Lambda}\gamma_{\mu}\Lambda\omega^{\mu}+\frac{f_{\omega\Lambda\Lambda}}{2M_{\Lambda}}\bar{\Lambda}\sigma_{\mu\nu}\Lambda\partial^{\mu}\omega^{\nu}  +  g_{\phi NN}\bar{\Lambda}\gamma_{\mu}\Lambda\phi^{\mu}+\frac{f_{\phi\Lambda\Lambda}}{2M_{\Lambda}}\bar{\Lambda}\sigma_{\mu\nu}\Lambda\partial^{\mu}\phi^{\nu}  +  g_{\sigma NN}\bar{\Lambda}\Lambda\sigma,\label{LLL}
\end{eqnarray}
\begin{eqnarray}
\mathcal{L}_{\Xi\Xi} & = & g_{\pi\Xi\Xi}\bar{\Xi}i\gamma_{5}\mathcal{\boldsymbol{\tau}}\Xi\cdot\boldsymbol{\pi}+g_{\eta\Xi\Xi}\bar{\Xi}i\gamma_{5}\Xi\eta  +  g_{\rho\Xi\Xi}\bar{\Xi}\gamma_{\mu}\boldsymbol{\tau}\Xi\cdot\boldsymbol{\rho}^{\mu}+\frac{f_{\rho\Xi\Xi}}{2M_{\Xi}}\bar{\Xi}\sigma_{\mu\nu}\boldsymbol{\tau}\Xi\cdot\partial^{\mu}\boldsymbol{\rho}^{\nu}\nonumber\\
 & + & g_{\omega\Xi\Xi}\bar{\Xi}\gamma_{\mu}\Xi\omega^{\mu}+\frac{f_{\omega\Xi\Xi}}{2M_{\Xi}}\bar{\Xi}\sigma_{\mu\nu}\Xi\partial^{\mu}\omega^{\nu} +  g_{\phi\Xi\Xi}\bar{\Xi}\gamma_{\mu}\Xi\phi^{\mu}+\frac{f_{\phi\Xi\Xi}}{2M_{\Xi}}\bar{\Xi}\sigma_{\mu\nu}\Xi\partial^{\mu}\phi^{\nu} +  g_{\sigma NN}\bar{\Xi}\Xi\sigma,\label{LXX}
\end{eqnarray}
\begin{eqnarray}
\mathcal{L}_{\Sigma\Sigma} & = & g_{\pi\Sigma\Sigma}(-i)\bar{\boldsymbol{\Sigma}}i\gamma_{5}\times\boldsymbol{\Sigma}\cdot\boldsymbol{\pi}+g_{\eta\Sigma\Sigma}\bar{\boldsymbol{\Sigma}}\cdot i\gamma_{5}\boldsymbol{\Sigma}\eta +  g_{\rho\Sigma\Sigma}\bar{(-i)\boldsymbol{\Sigma}}\gamma_{\mu}\times\boldsymbol{\Sigma}\cdot\boldsymbol{\rho}^{\mu}+\frac{f_{\rho\Sigma\Sigma}}{2M_{\Sigma}}\bar{\boldsymbol{\Sigma}}\sigma_{\mu\nu}\times\boldsymbol{\Sigma}\cdot\partial^{\mu}\boldsymbol{\rho}^{\nu}\nonumber\\
 & + & g_{\omega\Sigma\Sigma}\bar{\boldsymbol{\Sigma}}\gamma_{\mu}\cdot\boldsymbol{\Sigma}\omega^{\mu}+\frac{f_{\omega\Sigma\Sigma}}{2M_{\Sigma}}\bar{\boldsymbol{\Sigma}}\sigma_{\mu\nu}\cdot\boldsymbol{\Sigma}\partial^{\mu}\omega^{\nu} +  g_{\phi\Sigma\Sigma}\bar{\boldsymbol{\Sigma}}\gamma_{\mu}\cdot\boldsymbol{\Sigma}\phi^{\mu}+\frac{f_{\phi\Sigma\Sigma}}{2M_{\Sigma}}\bar{\boldsymbol{\Sigma}}\sigma_{\mu\nu}\cdot\boldsymbol{\Sigma}\partial^{\mu}\phi^{\nu} +  g_{\sigma\Sigma\Sigma}\bar{\boldsymbol{\Sigma}}\cdot\boldsymbol{\Sigma}\sigma,\label{LSS}
\end{eqnarray}
where we have introduced the notations
\begin{eqnarray*}
 & N=\begin{pmatrix}p\\
n
\end{pmatrix}, & \Xi=\begin{pmatrix}\Xi^{0}\\
\Xi^{-}
\end{pmatrix},\\
 & \boldsymbol{\Sigma}=\left\{ \frac{1}{\sqrt{2}}(-\Sigma^{+}+\Sigma^{-}),\frac{i}{\sqrt{2}}(-\Sigma^{+}-\Sigma^{-}),\Sigma^{0}\right\} , & \Lambda=\Lambda^{0}, 
\end{eqnarray*}
to represent the corresponding baryon fields. 
The coefficients $g_{\pi NN}$,
$g_{\pi\Xi\Xi}$, $g_{\pi\Sigma\Sigma}$, $f_{\rho NN}$, etc. are the coupling constants.
$\boldsymbol{\tau}=\{\tau_{1},\tau_{2},\tau_{3}\}$ are the Pauli
matrices, and $\boldsymbol{\pi}=\{\frac{1}{\sqrt{2}}(\pi^{+}+\pi^{-}),\frac{i}{\sqrt{2}}(\pi^{+}-\pi^{-}),\pi^{0}\}$
are the $\pi$ fields. $\boldsymbol{\rho}=\{\frac{1}{\sqrt{2}}(\rho^{+}+\rho^{-}),\frac{i}{\sqrt{2}}(\rho^{+}-\rho^{-}),\rho^{0}\}$ are the $\rho$ fields. 

The well-known coupling constants for the nucleon in Eq. (\ref{LNN}) can be extracted from experiment data. For the coupling constants in Eqs. (\ref{LLL})-(\ref{LSS}), we will estimate them using the nucleon-meson coupling constant as inputs with the help of quark model. The details of the procedure and the specific expression of the nucleon-meson coupling constants at the quark level can be found in Ref.~\cite{Li:2012cs}. We list the formulas which relate the coupling constants for the hyperons to coupling constants for the nucleon below.   

\begin{eqnarray*}
&&\Lambda:~~~~~~~~~~~g_{\eta\Lambda\Lambda}=-2g_{\eta NN}\frac{M_{\Lambda}}{M_N},~~~~~g_{\sigma\Lambda\Lambda}=g_{\sigma NN},~~~~~g_{\omega\Lambda\Lambda}=\frac{2}{3}g_{\omega NN},~~~~~f_{\omega\Lambda\Lambda}=-\frac{2}{3}g_{\omega NN},~~~~~g_{\phi\Lambda\Lambda}=\sqrt{2}g_{\rho NN},\\
&&~~~~~~~~~~~~~~~~~f_{\phi\Lambda\Lambda}=2\sqrt{2}g_{\rho NN}, \\
&&\Xi:~~~~~~~~~~~g_{\pi\Xi\Xi}=-\frac{1}{5}g_{\pi NN}\frac{M_{\Xi}}{M_N},~~~~~g_{\eta\Xi\Xi}=-3g_{\eta N}\frac{M_{\Xi}}{M_N},~~~~~g_{\sigma\Xi\Xi}=g_{\sigma NN},~~~~~g_{\rho\Xi\Xi}=g_{\rho NN},~~~~~f_{\rho\Xi\Xi}=-g_{\rho NN}(\frac{M_{\Xi}}{M_N}+1), \\
&&~~~~~~~~~~~~~~~~~g_{\omega\Xi\Xi}=\frac{1}{3}g_{\omega NN},~~~~~f_{\omega\Xi\Xi}=-\frac{1}{3}g_{\omega NN}(\frac{M_{\Xi}}{M_N}+1),~~~~~g_{\phi\Xi\Xi}=2\sqrt{2}g_{\rho NN},~~~~~f_{\phi\Xi\Xi}=2\sqrt{2}g_{\rho NN}(2\frac{M_{\Xi}}{M_N}-1), \\
&&\Sigma:~~~~~~~~~~~g_{\pi\Sigma\Sigma}=\frac{4}{5}g_{\pi NN}\frac{M_{\Sigma}}{M_N},~~~~~g_{\eta\Sigma\Sigma}=2g_{\eta NN}\frac{M_{\Sigma}}{M_N},~~~~~g_{\sigma\Sigma\Sigma}=g_{\sigma NN},~~~~~g_{\rho\Sigma\Sigma}=2g_{\rho NN},~~~~~f_{\rho\Sigma\Sigma}=g_{\rho NN}(4\frac{M_{\Sigma}}{M_N}-1), \\
&&~~~~~~~~~~~~~~~~~g_{\omega\Sigma\Sigma}=\frac{2}{3}g_{\omega NN},~~~~~f_{\omega\Sigma\Sigma}=\frac{2}{3}g_{\omega NN}(2\frac{M_{\Sigma}}{M_N}-1),~~~~~g_{\phi\Sigma\Sigma}=\sqrt{2}g_{\rho NN},~~~~~f_{\phi\Sigma\Sigma}=-\sqrt{2}g_{\rho NN}(\frac{M_{\Sigma}}{M_N}+1). 
\end{eqnarray*}

By fitting to experimental data, we adopt the values $g_{\pi NN}=13.07$, $g_{\eta NN}=2.24$, $g_{\sigma NN}=8.46$, $g_{\rho NN}=3.25$, $f_{\rho NN}=6.1 g_{\rho NN}$, $g_{\omega NN}=15.85$ and $f_{\omega NN}=0$ from Refs. \cite{Li:2012cs,Machleidt:1987hj,Machleidt:2000ge,Cao:2010km}. We collect the numerical values of them in Table I. 

\begin{table}[htbp]
\caption{The coupling constants and masses in our calculation. The masses are taken
    from the PDG \cite{Tanabashi:2018oca}.}
\begin{center}
\begin{tabular}{c || c || c | c | c | c | c | c   }
\hline \hline   
 & {mass(MeV)} & $\pi$ & $\eta$ & $\sigma$ & $\rho$ & $\omega$ & $\phi$  \\
\hline
 {mass(MeV)} & & $m_{\pi}=139.00$ & $m_{\eta}=547.85$ & $m_{\sigma}=600.00$ & $m_{\rho}=775.49$ & $m_{\omega}=782.65$ & $m_{\phi}=1019.50$ \\
\hline\hline
\multirow{2}{*}{$N$}  & \multirow{2}{*}{$M_N=939.00$} & \multirow{2}{*}{$g_{\pi NN}=13.07$} & \multirow{2}{*}{$g_{\eta NN}=2.24$} & \multirow{2}{*}{$g_{\sigma NN}=8.46$} & $g_{\rho NN}=3.25$ & $g_{\omega NN}=15.85$ & \\
& & & & & $f_{\rho NN}=19.83$ & $f_{\omega NN}=0$ & \\
\hline
\multirow{2}{*}{$\Lambda$} & \multirow{2}{*}{$M_{\Lambda}=1115.68$} & \multirow{2}{*}{$g_{\pi \Lambda\Lambda}=0$} & \multirow{2}{*}{$g_{\eta \Lambda\Lambda}=-5.33$} & \multirow{2}{*}{$g_{\sigma \Lambda\Lambda}=8.46$} & $g_{\rho \Lambda\Lambda}=0$ & $g_{\omega \Lambda\Lambda}=10.57$ & $g_{\phi \Lambda\Lambda}=4.60$ \\
& & & & & $f_{\rho \Lambda\Lambda}=0$ & $f_{\omega \Lambda\Lambda}=-10.57$ & $f_{\phi \Lambda\Lambda}=9.19$ \\
\hline
\multirow{2}{*}{$\Xi$} & \multirow{2}{*}{$M_{\Lambda}=1318.28$} & \multirow{2}{*}{$g_{\pi \Lambda\Lambda}=-3.67$} & \multirow{2}{*}{$g_{\eta \Lambda\Lambda}=-9.44$} & \multirow{2}{*}{$g_{\sigma \Lambda\Lambda}=8.46$} & $g_{\rho \Lambda\Lambda}=3.25$ & $g_{\omega \Lambda\Lambda}=5.28$ & $g_{\phi \Lambda\Lambda}=9.05$ \\
& & & & & $f_{\rho \Lambda\Lambda}=-7.81$ & $f_{\omega \Lambda\Lambda}=-12.70$ & $f_{\phi \Lambda\Lambda}=16.36$ \\
\hline
\multirow{2}{*}{$\Sigma$} & \multirow{2}{*}{$M_{\Lambda}=1193.15$} & \multirow{2}{*}{$g_{\pi \Lambda\Lambda}=13.29$} & \multirow{2}{*}{$g_{\eta \Lambda\Lambda}=5.70$} & \multirow{2}{*}{$g_{\sigma \Lambda\Lambda}=8.46$} & $g_{\rho \Lambda\Lambda}=6.50$ & $g_{\omega \Lambda\Lambda}=10.57$ & $g_{\phi \Lambda\Lambda}=4.60$ \\
& & & & & $f_{\rho \Lambda\Lambda}=13.27$ & $f_{\omega \Lambda\Lambda}=16.29$ & $f_{\phi \Lambda\Lambda}=-10.44$ \\
\hline\hline
\end{tabular}
\end{center}
\end{table}

Given the Lagrangians above, we can derive the effective potentials for the two-body interactions of the tritonlike systems. Expanding the $T$ matrices with external momenta to the leading order, one obtain the effective potentials for the two-body interactions. Then the effective potentials in coordinate space can be derived by Fourier transformation 
\begin{equation}
V(\vec{r})=\frac{1}{(2\pi)^3}\int d^3\vec{q}\, e^{i\vec{q}\cdot\vec{r}}\, T(\vec{q})F^{2}(\vec{q})~,\label{FT}
\end{equation}
where $F(\vec{q})$ is the monopole form factor attached to each scattering vertex. The form is
\begin{equation}
F(q)=\frac{\lambda^2-m_{\alpha}^2}{\lambda^2-q^2}=\frac{\lambda^2-m_{\alpha}^2}{{\lambda}^2+\vec{q}^2},
\end{equation}
with $m_\alpha$ the mass of exchange bosons. The $\lambda$ is cutoff parameter which cannot be well determined from fundamental theories due to the non-perturbative effect. It is a rough way to reflect the non-point-like hadronic structures and suppress the contribution from UV energies. 

In order to make clear of the binding properties of a three-body system, we should define some quantities for discussion. One is the total three-body energy eigenvalue relative to the three-free fermions threshold. Another one is the reduced three-body energy eigenvalue relative to the break-up state of the three-free fermions threshold. We need to discuss the minimum of the BO potential $V_{BO}(0)$ to represent its strength.
To emphasize the size of the system is large enough to keep the hadronic picture, we should define the root-mean-square radius of any two fermions in the system by using $r_{rms}$. At last, we introduce the probabilities for $S$-wave and $D$-wave components in any two fermions in the tri-fermion system.  

\subsection{Numerical results for the \texorpdfstring{$NNN$}{$NNN$}}\label{subsec32}

The OBE model is very successful in describing the binding properties of deuteron. One may believe that it  also describe the dynamics of triton well. Now we apply the BOP method and OBE model to the three-nucleon system. Since the experiment accumulate sufficient data on the system, the application on them can illustrate the feasibility of our formalism. There are two bound states of the three-nucleon system have been observed in experiment. One is triton, the other one is helium-3 nucleus. The two have the same binding energy and structure without considering the isospin symmetry breaking effect. For simplicity, we investigate the bound state of the $NNN$ system under the isospin symmetry. We list the isospin wave functions of the three-nucleon system in Appendix B. 

Applying the Lagrangians in Eq. (\ref{LNN}) for nucleon, one gets the $T$ matrix for the nucleon-nucleon scattering via boson exchanges. After the Fourier transformations as Eq. (\ref{FT}), one obtain the effective potentials of the two-body force for the nucleon-nucleon system in coordinate space. The effective potentials contain the contributions from long-range $\pi/\eta$ exchange, medium-range $\sigma$ exchange and short-range $\rho/\omega$ exchange. In general, the effective potentials derived from nucleon-nucleon scattering consist of the central term, the spin-spin force term, the spin-orbit force term and the tensor force term. Since the tensor force term leads to the S-D wave mixing, the contribution from D wave should be taken into account during the computations of the bound state for the $NNN$ system.

\begin{figure}[ht]
  \begin{center}
  \rotatebox{0}{\includegraphics*[width=0.45\textwidth]{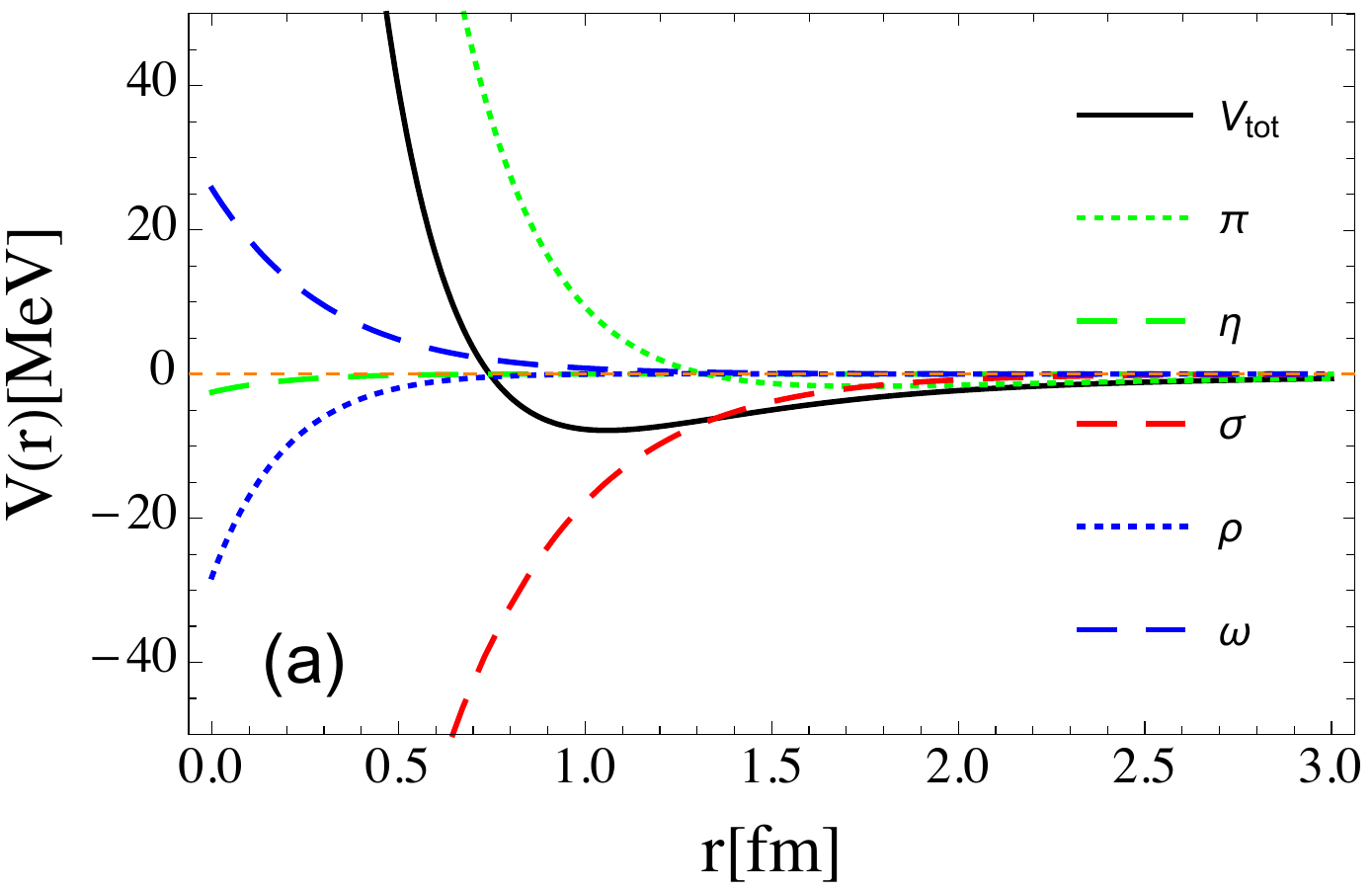}}
    \rotatebox{0}{\includegraphics*[width=0.45\textwidth]{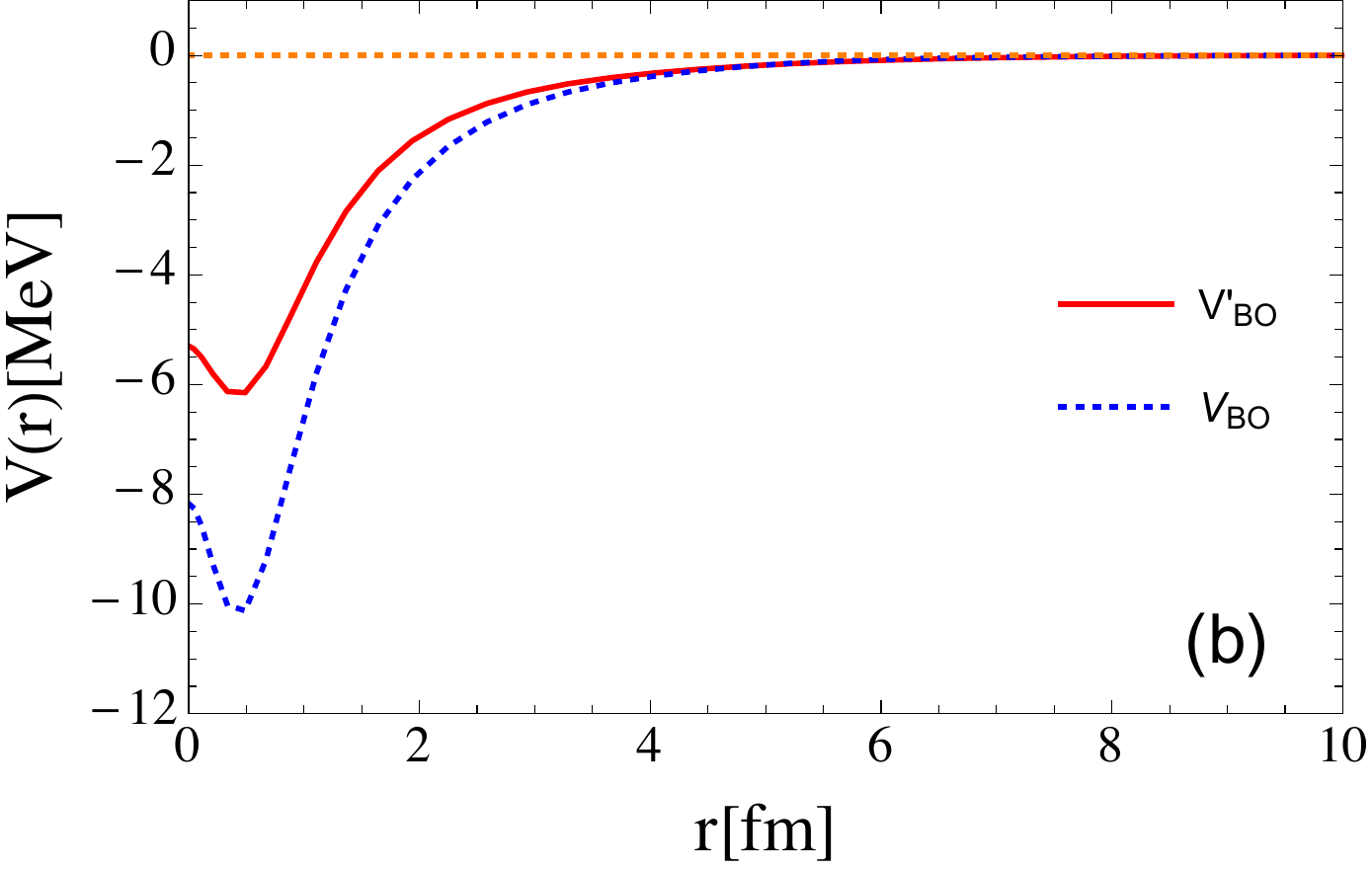}}
     \caption{(a) The effective potentials of the isospin-singlet force for the nucleon-nucleon system with the comparison of the contributions of each exchanged boson. (b) The BO potentials for the $NNN$ in the isospin state $|0,\frac{1}{2},\pm\frac{1}{2}\rangle$. The dotted and solid lines are the BO potentials before and after distortion correction, respectively. }
    \label{Npotential}
  \end{center}
\end{figure}

In our calculations, there is only one free parameter $\lambda$ in the monopole form factor introduced to reflect the inner structure of the interacting hadrons. It is still quite hard to determine the value of $\lambda$ from the fundamental theory. However, we can fix it by using the binding energy of deuteron which is $2.23$ MeV from experimental data. When the two-body isospin $I_2=0$ and the parameter $\lambda=811.80$ MeV, we reproduce the binding energy of deuteron $-E_2=2.23$ MeV.      
In this case, we plot the effective potentials of the isospin-singlet force for the nucleon-nucleon system in Fig.~\ref{Npotential}(a). 
In order to make a rough estimation of the specific roles of the exchanged boson in the effective potential, we also plot the effective potentials of each of them in the figure. As shown in Fig.~\ref{Npotential}(a), the $\pi$ exchange provides repulsive force in the short range but shallow attractive force in the medium range. The $\eta$ and $\sigma$ exchange create shallow and deep attraction, respectively. The $\rho$ exchange provides attractive force while the $\omega$ exchange provides repulsive force. Thus the total effective potential is repulsive in the short range while attractive in the medium range.   

Based on the formalism we have built in Sec.~\ref{sec2}, we calculate the $NNN$ system for all of cases of isospin configurations with the parameter $\lambda=811.80$ MeV. We only find a bound state for the case $|0,\frac{1}{2},\pm\frac{1}{2}\rangle$, where the total isospin $I_3=\frac{1}{2}$ and the isospin of the two-body subsystem $I_2=0$. In order to illustrate the contribution of one of the nucleons on the dynamics of the other two, we plot the BO potentials for the state $|0,\frac{1}{2},\pm\frac{1}{2}\rangle$ in Fig.~\ref{Npotential}(b). The blue-dotted and red-solid curves are the BO potentials before and after distortion correction, respectively. Both curves have a deep around $0.4$ fm as shown in the figure. The correction on the BO potential make it become weak. When the distance is larger than a certain value, the BO potential equal to zero. It corresponds to the break-up state for the $NNN$ system, which consists of a deuteron and a free nucleon.

\begin{table}[htbp]
\caption{Bound state solutions of the $NNN$ system with  isospin $I_3=1/2$. 
$E_2$ is the energy eigenvalue of its subsystem. 
 $E_3$ is the reduced three-body energy eigenvalue relative to the break-up state of the $NNN$ system.  
$E_T$ is the total three-body energy eigenvalue relative to the $NNN$ threshold. 
 $V_{BO}(0)$ is the minimum of the BO potential. 
$r_{rms}$ represents the root-mean-square radius of any two $N$ in the $NNN$ system. 
The $S$-wave and $D$-wave represent the probabilities for $S$-wave and
 $D$-wave components in any two $N$ in the $NNN$ system.}\label{E3EbN}
\begin{center}
\begin{tabular}{ | c | c | c | c  | c  | c | c  | c  | c  }
\hline\hline 
  $\Lambda$(MeV)  &  $E_2$(MeV)  & $E_3$(MeV) & $E_T$(MeV)  & $V_{BO}(0)$(MeV)  & S wave(\%) & D wave(\%) & $r_{rms}$(fm)   \\
\hline
       780.00 &  -0.18 & -1.37 & -1.55 & -3.27 & 97.51 & 2.49 & 4.55 \\
\hline       
       800.00 &  -1.42 & -3.62 & -5.05 & -4.97 & 96.78 & 3.22 & 4.09 \\
\hline
       811.80 &  -2.23 & -5.25 & -7.49 & -6.15 & 96.32 & 3.68 & 3.81\\
\hline       
       820.00 &  -2.82 & -6.48 & -9.30 & -7.00 & 96.00 & 4.00 & 3.62  \\
\hline       
       860.00 &  -5.70 & -12.76 & -18.46 & -11.17 & 94.69 & 5.31 & 2.91 \\
\hline       
       880.00 &  -7.04 & -15.71 & -22.75 & -12.98 & 94.19 & 5.81 & 2.68 \\
\hline
       900.00 &  -8.26 & -18.40 & -26.66 & -14.56 & 93.77 & 6.23 & 2.52 \\
\hline
       920.00 &  -9.37 & -20.80 & -30.17 & -15.94 & 93.41 & 6.59 & 2.41 \\
\hline\hline
\end{tabular}
\end{center}
\end{table}

Through the BOP method with the careful treatment on the S-D wave mixing, one can get the dependence of the binding properties on the parameter $\lambda$ as shown in Table~\ref{E3EbN}. There is a three-body bound state with total three-body binding energy in the range of $1.55-30.17$ MeV, when the parameter $\lambda$ varies from 780 MeV to 920 MeV. The corresponding binding energy of its two-body subsystem changes from $0.18$ MeV to $9.37$ MeV. Within the range of the $\lambda$, the reduced three-body binding energy increases from $1.37$ MeV to $20.80$ MeV. The root-mean-square radius of the system decreases from 4.55 fm to 2.41 fm when the parameter $\lambda$ grows. 
The three-body bound state of the $NNN$ system is a mixture of S and D wave due to the tensor force in the effective potentials. The proportion of the S wave state is more than 93\%. If we shut down the D wave, we cannot find the binding solution, which means that the S-D wave mixing is very crucial in the formation of the three-body bound state. 
If we fix the parameter $\lambda=811.80$ MeV by reproducing the binding energy of deuteron, the reduced three-body binding energy and total three-body binding energy are 5.25 MeV and 7.49 MeV, respectively. As we know, the empirical binding energies of the triton and helium-3 nucleus are 8.48 MeV and 7.80 MeV, respectively. 
Our numerical result 7.49 MeV is comparable with the empirical binding energies. 
Since we neglect the isospin breaking, there is no numerical difference between the binding energies of the triton and helium-3 nucleus in our computation.  

\begin{figure}[ht]
  \begin{center}
     \rotatebox{0}{\includegraphics*[width=0.45\textwidth]{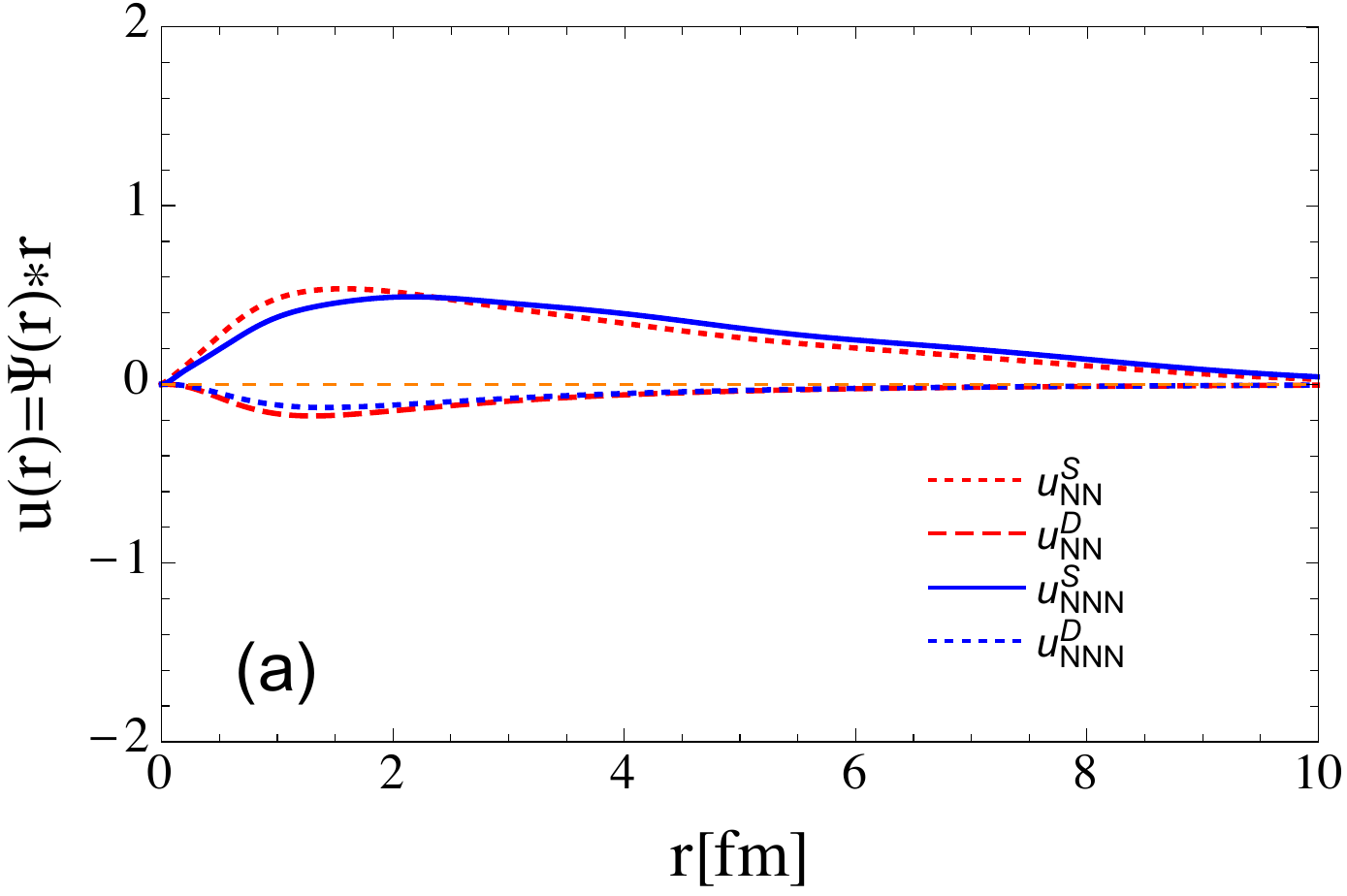}}
      \rotatebox{0}{\includegraphics*[width=0.45\textwidth]{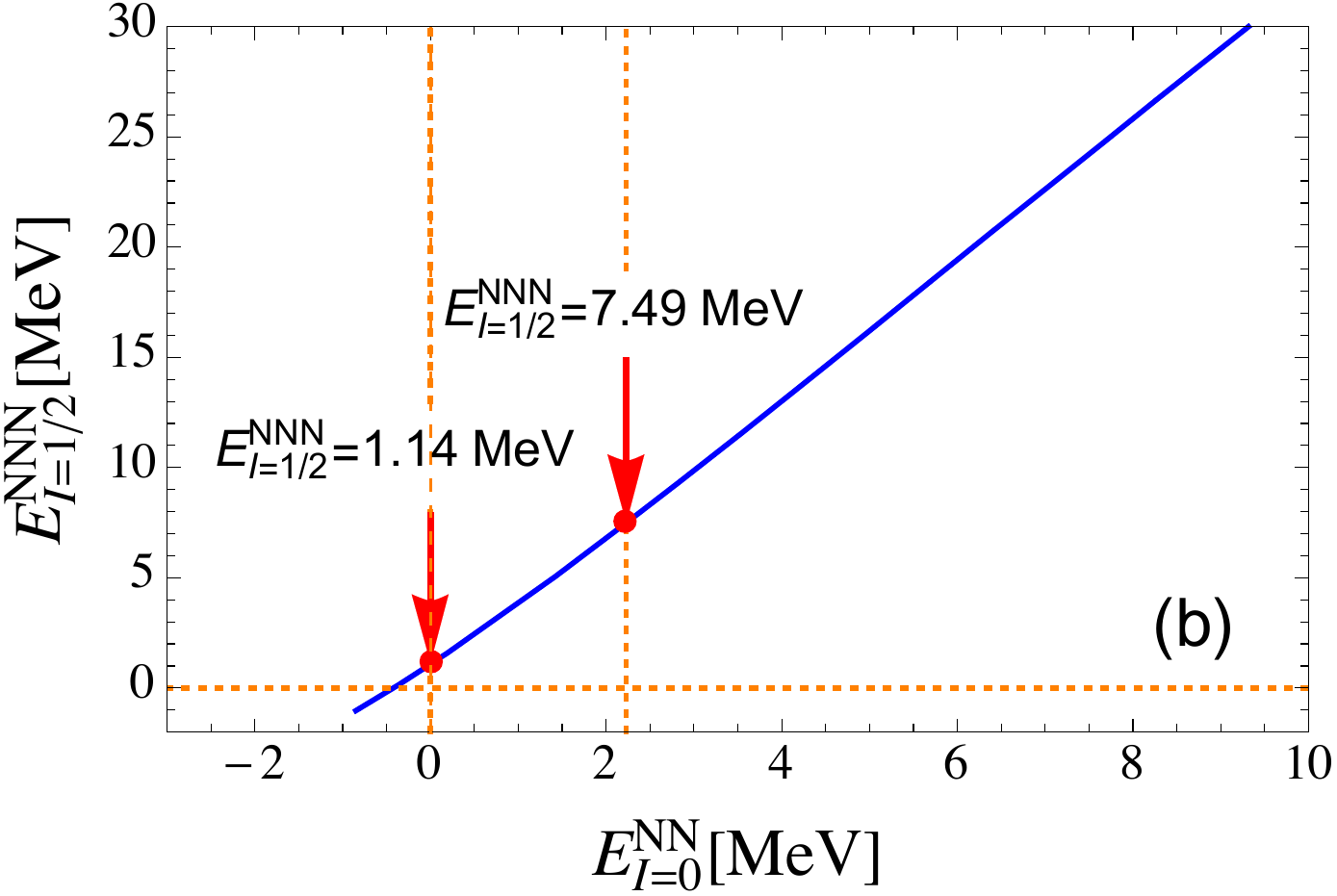}}
    \caption{(a) Plot of wave functions for the $NNN$ system in the isospin state $|0,\frac{1}{2},\pm\frac{1}{2}\rangle$. The blue lines denote the wave functions for any two nucleon in the $NNN$ system. The red lines represent the wave functions for its subsystem deuteron. (b) Dependence of the total three-body binding energy on the two-body binding energy of its subsystem $NN$. The left red point indicates the Borromean state of the system. The right one is our numerical result corresponding to the triton or helium-3 nucleus.}
    \label{Nwave}
  \end{center}
\end{figure}

To verify the binding solution we get is a bound state, we plot the wave functions for any two constituents in the $NNN$ in Fig.~\ref{Nwave}(a). For comparison, we also plot the wave functions for its subsystem $NN$. As shown in the figure, the bound state is dominated by the S-wave state with the proportion 96.32 $\%$ if the parameter $\Lambda=811.80$ MeV. 
The numerical results show that the total three-body binding energy of the $NNN$ system grows as the two-body binding energy of its subsystem $NN$ increases. To show it explicitly, we plot the dependence in Fig.~\ref{Nwave}(b), where $E_{I=1/2}^{NNN}$ and $E_{I=0}^{NN}$ denote the total three-body binding energy and two-body binding energy, respectively. One may wonder whether there is a critical value of $E_{I=0}^{NN}$, below which the system $NNN$ has no three-body bound state. In fact, it turns out no such critical value for the system with isospin configuration $|0,\frac{1}{2},\pm\frac{1}{2}\rangle$. When the two-body binding energy approaches 0 MeV, there is still a tiny value 1.14 MeV of the three-body binding energy for the $NNN$ system. No matter how small the two-body binding energy of the subsystem is, the whole system always has a shallow bound state. It is a reminiscent of a Borromean state, in which a three-body system may forms a three-body bound state despite none of its subsystem has a bound state. There are two red points in Fig.~\ref{Nwave}(b), where the left one indicates the Borromean state of the system. The right one is the numerical result of triton or helium-3 nucleus. It is a little below the experiment data. Since in our calculations we use the BOP method to construct the interpolating wave functions for diagonalizing the Hamiltonian of the three-body system, it always gives an upper limit of the energy for a system.

\subsection{Numerical results for the \texorpdfstring{$\Lambda\Lambda\Lambda$}{$\Lambda\Lambda\Lambda$}}\label{subsec33}

\begin{figure}[ht]
  \begin{center}
     \rotatebox{0}{\includegraphics*[width=0.45\textwidth]{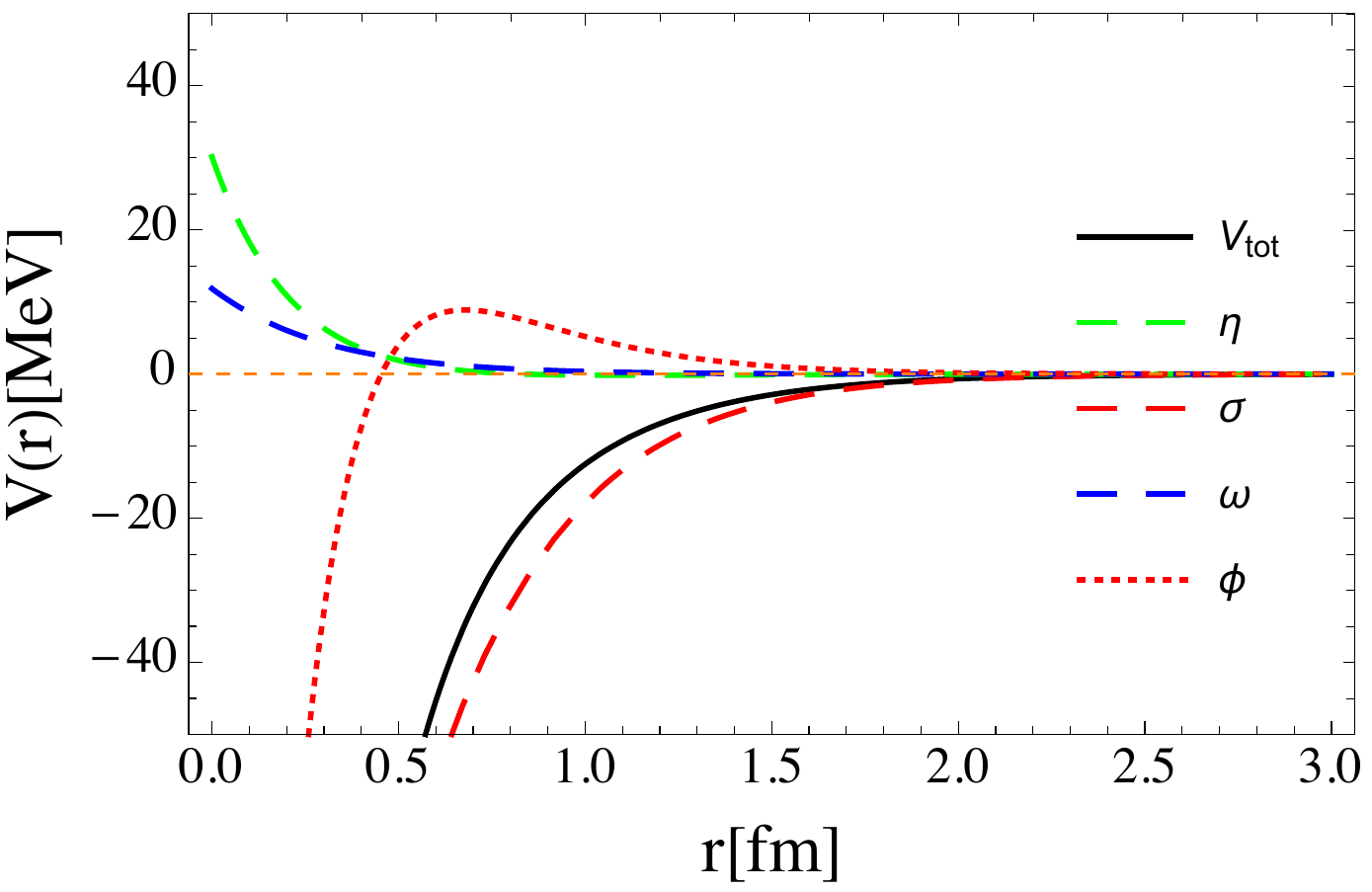}}
    \caption{The effective potentials of the two-body force for the $\Lambda\Lambda\Lambda$ system with the comparison of the effective potentials from each exchanged boson.}
    \label{Lambda}
  \end{center}
\end{figure}

There is only one isospin configuration $|0,0,0\rangle$ for the system $\Lambda\Lambda\Lambda$, as $\Lambda$ is an isospin singlet. Now we only have a free parameter $\lambda$ which is undetermined in our calculations. 
In general, the parameter $\lambda$ is within the range $800-1500$ MeV when investigate the binding properties of deuteron. In our calculation for the $NNN$ system, It is chosen at 811.80 MeV to reproduce the binding energy of deuteron within the OBE mechanism. 
One may expect that a heavier system has a smaller size which leads to a larger parameter $\lambda$. 
Thus we change the parameter $\lambda$ in the range of $800-2000$ MeV to search for the binding solutions of this system. It is crucial to investigate the two-body interaction before searching for the bound state of the system. 
The two-body force of the $\Lambda\Lambda\Lambda$ arises from the $\eta$, $\sigma$, $\omega$ and $\phi$ exchange. To highlight the contribution of each exchanged boson, we plot the total effective potential and the effective potentials of each exchanged boson in Fig.~\ref{Lambda} when we fix the $\lambda=811.80$ MeV.   
Both the $\eta$ and $\omega$ exchange provide repulsive force, while the $\sigma$ exchange is attractive. The $\phi$ exchange is attractive in the short range but repulsive in the medium range. The total effective potential is attractive. However, we fail to find any bound solution for the system $\Lambda\Lambda\Lambda$ and its two-body subsystem $\Lambda\Lambda$ within the range of $\lambda$, since the attraction between any two constituents is not strong enough to bind them.  

\subsection{Numerical results for the \texorpdfstring{$\Xi\Xi\Xi$}{$\Xi\Xi\Xi$}}\label{subsec34}

There are three isospin states for the $\Xi\Xi\Xi$ system, i.e. $|0,\frac{1}{2}\rangle$, $|1,\frac{1}{2}\rangle$ and $|1,\frac{3}{2}\rangle$. 
The general identity principle constricts the cases of the two-body force for the $\Xi\Xi\Xi$ system to be  $(1,0)$ and $(0,1)$. 
The two-body force $(1,0)$ governs the isospin states $|1,\frac{1}{2}\rangle$ and $|1,\frac{3}{2}\rangle$. The two-body force $(0,1)$ governs the state $|0,\frac{1}{2}\rangle$. 
Based on the OBE mechanism, the two-body force of the $\Xi\Xi\Xi$ is generated from the $\pi$, $\eta$, $\sigma$, $\rho$, $\omega$ and $\phi$ exchanges. 
We should search for the bound states of its two-body subsystem first, since one need the two-body wave functions to construct the interpolating wave functions for the whole system within the BOP method.   
The Fourier transformation as Eq. (\ref{FT}) yields the effective potentials of the two-body force in coordinate space, which is $\lambda$ dependent. 
With the variety of the parameter $\lambda$, we find binding solutions for both cases $(1,0)$ and $(0,1)$ of the subsystem $\Xi\Xi$ as shown in Table~\ref{E3Ebxi01} and Table~\ref{E3Ebxi10}. If the parameter $\lambda$ is chosen at 896.54 MeV, one finds a binding solution for the case $(0,1)$ with a binding energy of 2.23 MeV. The case $(1.0)$ also has a bound state with the same energy of 2.23 MeV, when the parameter $\lambda$ is fixed at 937.70 MeV.

\begin{figure}[ht]
  \begin{center}
  \rotatebox{0}{\includegraphics*[width=0.45\textwidth]{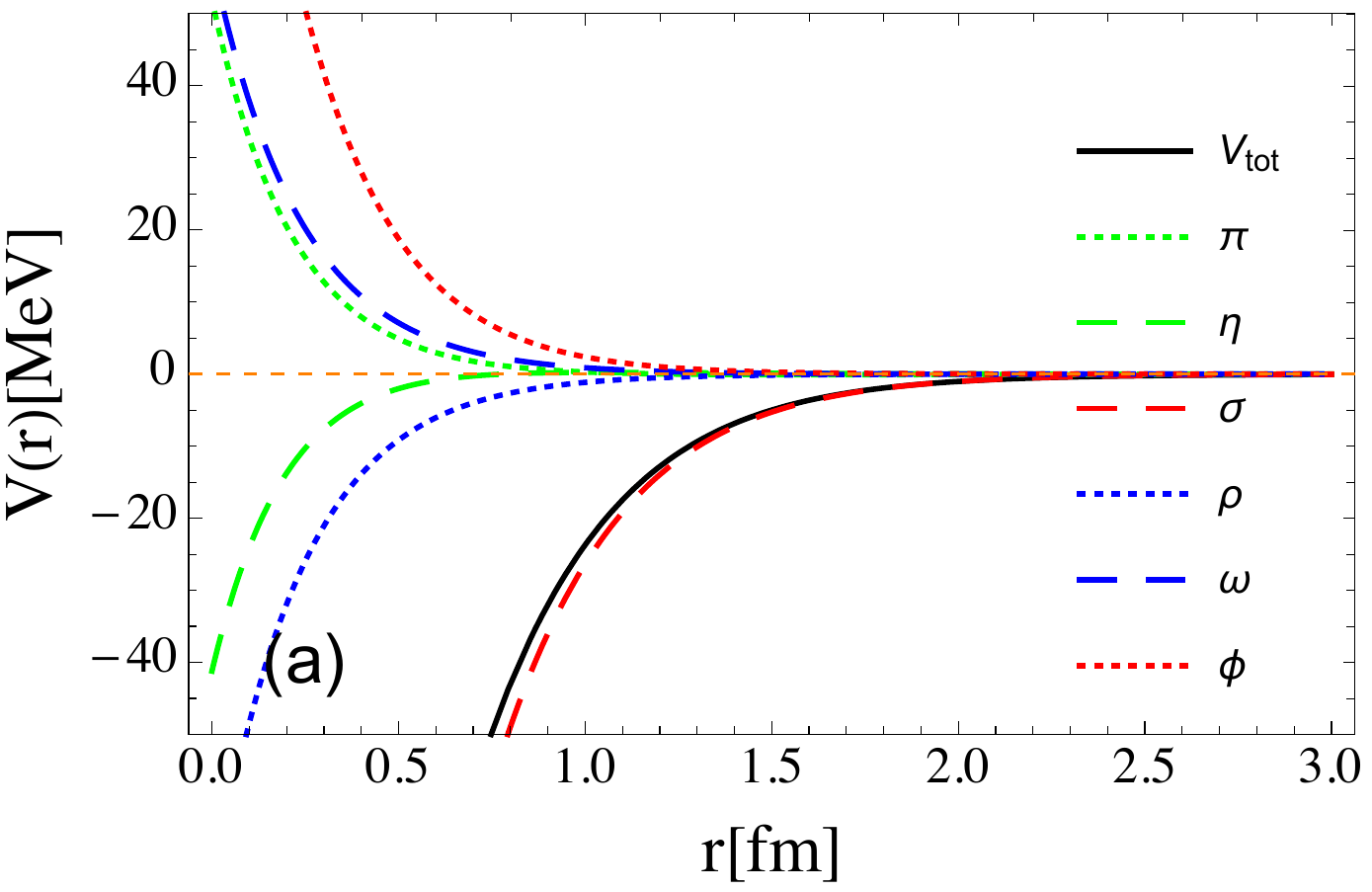}}
    \rotatebox{0}{\includegraphics*[width=0.45\textwidth]{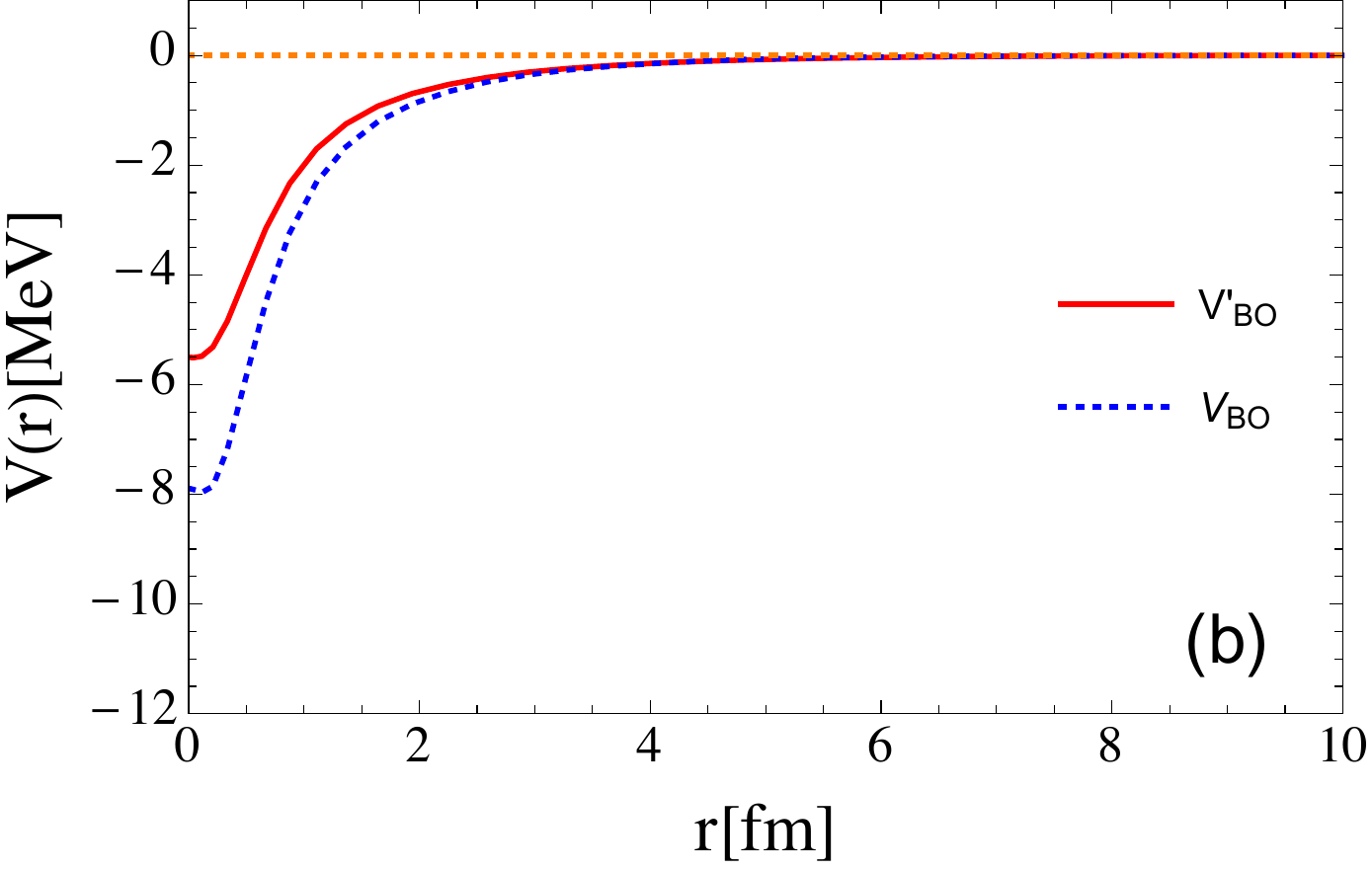}}
      \rotatebox{0}{\includegraphics*[width=0.45\textwidth]{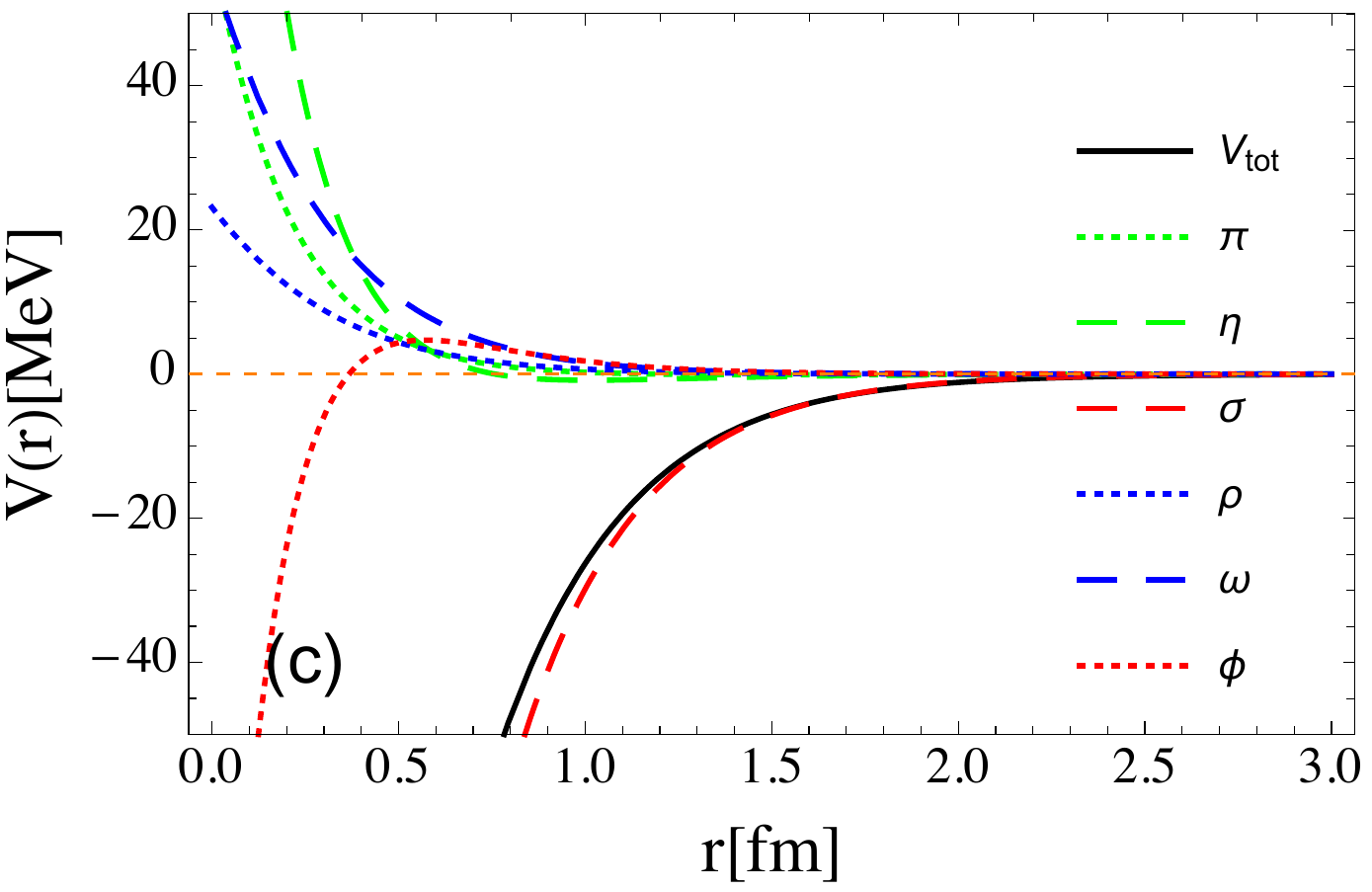}}
    \rotatebox{0}{\includegraphics*[width=0.45\textwidth]{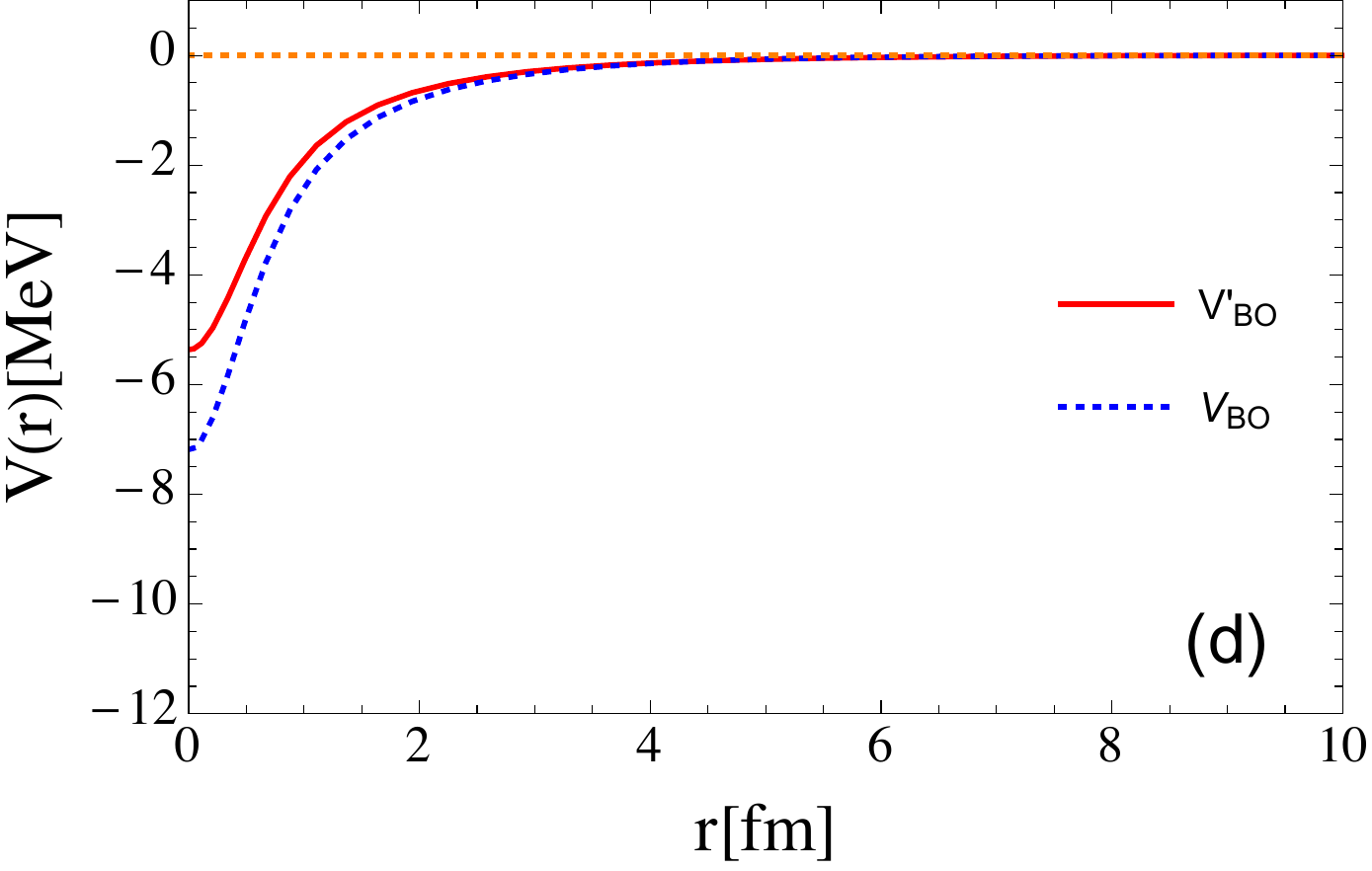}}
     \caption{(a) The effective potentials of the two-body force $(0,1)$ for the system $\Xi\Xi\Xi$ with the comparison of the contributions of each exchanged boson. (b) The BO potentials for the system $\Xi\Xi\Xi$ provided by the two-body force $(0,1)$. (c) The effective potentials of the two-body force $(1,0)$ for the system $\Xi\Xi\Xi$ with the comparison of the contributions of each exchanged boson. (d) The BO potentials for the system $\Xi\Xi\Xi$ provided by the two-body force $(1,0)$. The dotted and solid lines in (b) and (d) are the BO potentials before and after distortion correction, respectively.}
    \label{xipotential}
  \end{center}
\end{figure}

In order to show the properties of the two-body interactions, we plot the effective potentials for both cases when their subsystem $\Xi\Xi$ have the same binding energy of 2.23 MeV as shown in Fig.~\ref{xipotential}. The Fig.~\ref{xipotential}(a) corresponds to the case $(0,1)$, where the $\pi$, $\omega$ and $\phi$ provide the repulsive force, while the $\eta$, $\rho$ and $\sigma$ give the attractive force. Since the repulsion of $\pi$, $\omega$ and $\phi$ exchanges almost cancel the attraction of $\eta$ and $\rho$, the total effective potential is dominated by the contribution of $\sigma$ exchange. We plot the effective potentials for the case $(1,0)$ in Fig.~\ref{xipotential}(c), where the repulsion is mainly from $\pi$, $\eta$, $\rho$ and $\omega$ exchanges. The $\phi$ exchange provide a deep attraction in the short range but a slight repulsion in the medium range. The $\pi$, $\eta$, $\rho$, $\omega$ and $\phi$ almost cancel out and do not contribute the total effective potential. Thus the $\sigma$ exchange is mainly contribute the total effective potential. Based on the formalism discussed in Sec.~\ref{sec2}, we get the BO potentials for the $\Xi\Xi\Xi$ system as shown in Fig.~\ref{xipotential}(b) and Fig.~\ref{xipotential}(d), where the figure (b) and (d) correspond to the cases of $(0,1)$ and $(1,0)$, respectively. Similar with the discussions on the three-nucleon system, the corrections on the interpolating wave functions weaken the BO potentials to some extent.

\begin{table}[htbp]
\caption{Bound state solutions of the $\Xi\Xi\Xi$ system with  isospin $I_3=1/2$. 
$E_2$ is the energy eigenvalue of its subsystem. 
 $E_3$ is the reduced three-body energy eigenvalue relative to the break-up state of the$\Xi\Xi\Xi$ system.  
$E_T$ is the total three-body energy eigenvalue relative to the $\Xi\Xi\Xi$ threshold. 
 $V_{BO}(0)$ is the minimum of the BO potential. 
$r_{rms}$ represents the root-mean-square radius of any two $\Xi$ in the $\Xi\Xi\Xi$ system. 
The $S$-wave and $D$-wave represent the probabilities for $S$-wave and
 $D$-wave components in any two $\Xi$ in the $\Xi\Xi\Xi$ system.}\label{E3Ebxi01}
\begin{center}
\begin{tabular}{ | c | c | c | c  | c  | c | c  | c  | c  }
\hline\hline 
  $\Lambda$(MeV)  &  $E_2$(MeV)  & $E_3$(MeV) & $E_T$(MeV)  & $V_{BO}(0)$(MeV)  & S wave(\%) & D wave(\%) & $r_{rms}$(fm)   \\
\hline
       890.00 & -1.03 & -2.91 & -3.93 & -3.63 & 99.76 & 0.24 & 3.62 \\
\hline
       896.54 & -2.23 & -5.63 & -7.86 & -5.51 & 99.65 & 0.35 & 3.02  \\
\hline       
       900.00 & -3.00 & -7.46 & -10.46 & -6.68 & 99.59 & 0.41 & 2.72 \\
\hline
       910.00 & -5.71 & -14.11 & -19.81 & -10.51 & 99.40 & 0.60 & 2.07 \\
\hline       
       920.00 & -9.10 & -22.28 & -31.38 & -14.70 & 99.23 & 0.77 & 1.68  \\
\hline       
       930.00 & -13.07 & -31.60 & -44.68 & -19.11 & 99.07 & 0.93 & 1.45 \\
\hline       
       940.00 & -17.54 & -41.81 & -59.35 & -23.66 & 98.93 & 1.07 & 1.29 \\
\hline
       960.00 & -27.65 & -64.14 & -91.79 & -33.03 & 98.68 & 1.32 & 1.09 \\
\hline\hline
\end{tabular}
\end{center}
\end{table}

When the parameter $\lambda$ lies between 890.00 MeV and 960.00 MeV, there is bound solution for the state $|0,\frac{1}{2}\rangle$ with the total binding energies between 3.93 MeV and 91.79 MeV as shown in Table~\ref{E3Ebxi01}. The corresponding two-body binding energy is about 1.03-27.65 MeV. The reduced three-body binding energy increases from 2.91MeV to 64.14 MeV as the parameter $\lambda$ grows, while the root-mean-square radius of the system decreases from 3.62 fm to 1.09 fm. The three-body bound state of the state $|0,\frac{1}{2}\rangle$ is a mixture of S and D wave due to the tensor force in the effective potentials. The proportion of the S wave state is more than 98\%. For a better comparison with the numerical results of the $NNN$ system, we choose the parameter $\lambda$ at 896.54 MeV which yields a two-body binding energy of 2.23 MeV for the subsystem $\Xi\Xi$. Then the total three-body binding energy and the reduced three-body binding energy are 7.86 and 5.63 MeV, respectively.

\begin{table}[htbp]
\caption{Bound state solutions of the $\Xi\Xi\Xi$ system with  isospin $I_3=3/2$. 
$E_2$ is the energy eigenvalue of its subsystem. 
 $E_3$ is the reduced three-body energy eigenvalue relative to the break-up state of the$\Xi\Xi\Xi$ system.  
$E_T$ is the total three-body energy eigenvalue relative to the $\Xi\Xi\Xi$ threshold. 
 $V_{BO}(0)$ is the minimum of the BO potential. 
$r_{rms}$ represents the root-mean-square radius of any two $\Xi$ in the $\Xi\Xi\Xi$ system. 
The $S$-wave and $D$-wave represent the probabilities for $S$-wave and
 $D$-wave components in any two $\Xi$ in the $\Xi\Xi\Xi$ system.}\label{E3Ebxi10}
\begin{center}
\begin{tabular}{ | c | c | c | c  | c  | c | c  | c  | c  }
\hline\hline 
  $\Lambda$(MeV)  &  $E_2$(MeV)  & $E_3$(MeV) & $E_T$(MeV)  & $V_{BO}(0)$(MeV)  & S wave(\%) & D wave(\%) & $r_{rms}$(fm)   \\
\hline
       910.00 &  -0.35 & -1.76 & -2.10 & -2.60 & 100.00 & 0 & 3.94 \\
\hline       
       920.00 &  -0.96 & -3.04 & -4.00 & -3.48 & 100.00 & 0 & 3.57 \\
\hline
       930.00 &  -1.65 & -4.63 & -6.28 & -4.50 & 100.00 & 0 & 3.20 \\
\hline       
       937.70 & -2.23 & -6.04 & -8.27 & -5.36 & 100.00 & 0 & 2.93   \\
\hline       
       940.00 &  -2.42 & -6.48 & -8.90 & -5.62 & 100.00 & 0 & 2.86  \\
\hline       
       950.00 &  -3.24 & -8.54 & -11.78 & -6.79 & 100.00 & 0 & 2.57 \\
\hline
       960.00 &  -4.11 & -10.75 & -14.86 & -7.98 & 100.00 & 0 & 2.33 \\
\hline      
       970.00 &  -5.03 & -13.05 & -18.08 & -9.16 & 100.00 & 0 & 2.13 \\
\hline\hline
\end{tabular}
\end{center}
\end{table}

When the parameter $\lambda$ increases between 910.00 MeV and 970.00 MeV, we find a degenerate bound solution for the states $|1,\frac{1}{2}\rangle$ and $|1,\frac{3}{2}(\frac{1}{2})\rangle$. The corresponding two-body binding energy is about 1.11-6.19 MeV, and their three-body binding energies vary from 2.10 MeV to 18.08 MeV. The reduced three-body binding energy grows from 1.76 MeV to 13.05 MeV with the parameter $\lambda$ in the range of 910.00-970.00 MeV. The root-mean-square radius of the system decreases from 3.94 fm to 2.13 fm. 
The three-body bound states of the states $|1,\frac{1}{2}\rangle$ and $|1,\frac{3}{2}\rangle$ only have S-wave state due to their effective potentials has no tensor force, which is different from the case $|0,\frac{1}{2}\rangle$.  
From Table II we can see that the total three-body binding energy of the states $|1,\frac{1}{2},\rangle$ and $|1,\frac{3}{2}\rangle$ states is about 8.27 MeV when the two-body binding energy is 2.23 MeV with the corresponding parameter $\lambda$ is fixed at 937.70 MeV.

\begin{figure}[ht]
  \begin{center}
     \rotatebox{0}{\includegraphics*[width=0.45\textwidth]{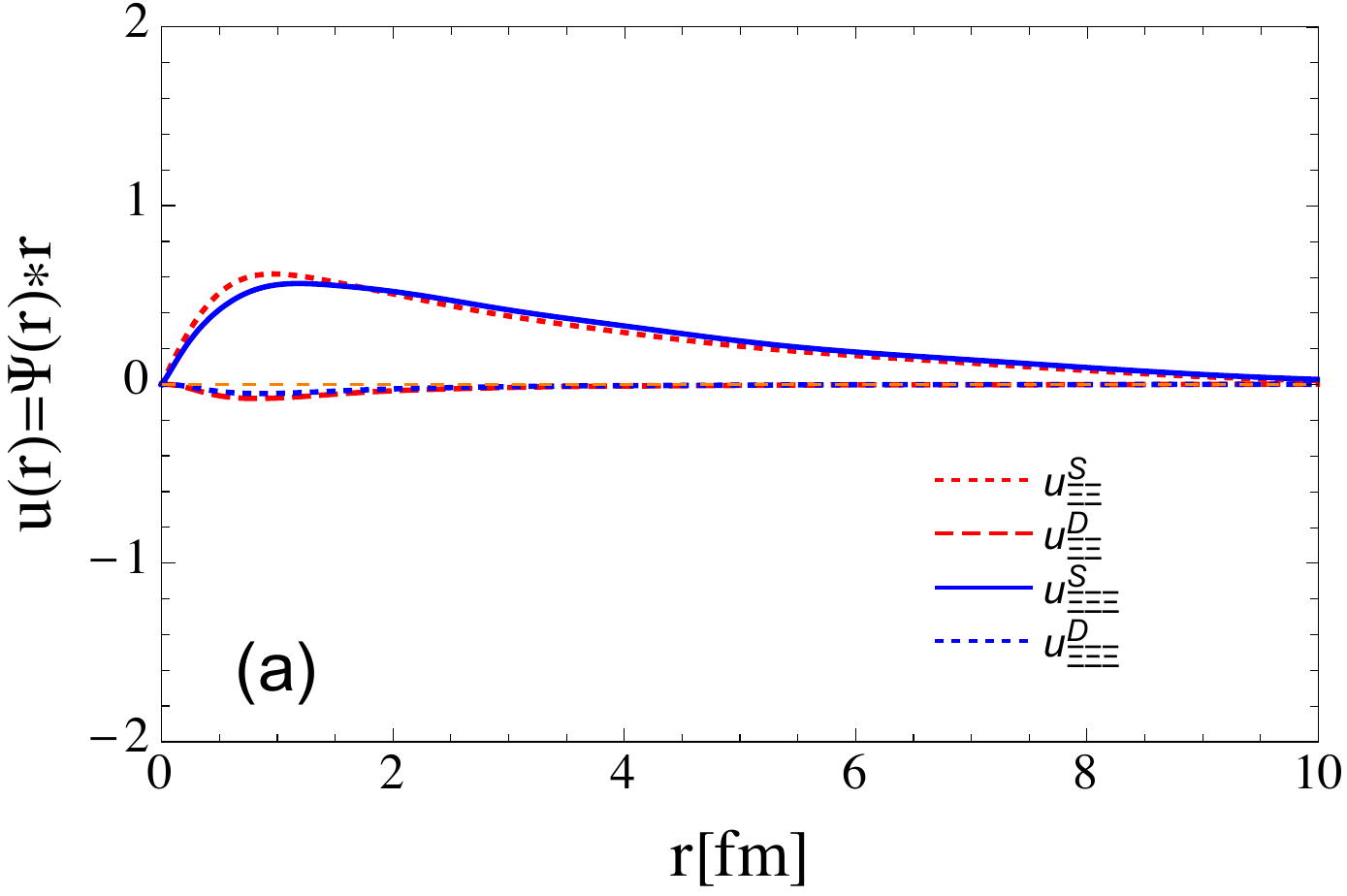}}
      \rotatebox{0}{\includegraphics*[width=0.45\textwidth]{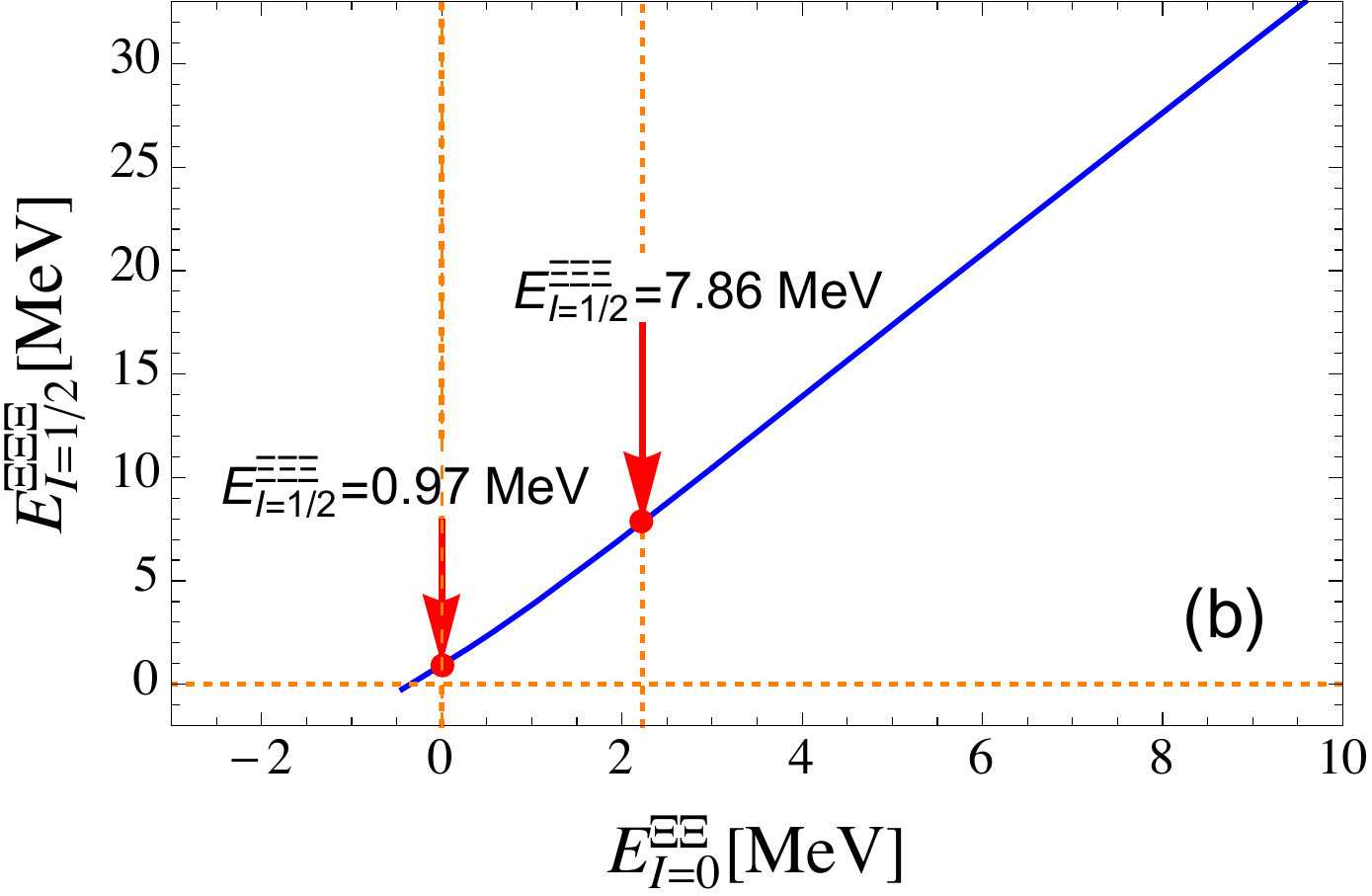}}
       \rotatebox{0}{\includegraphics*[width=0.45\textwidth]{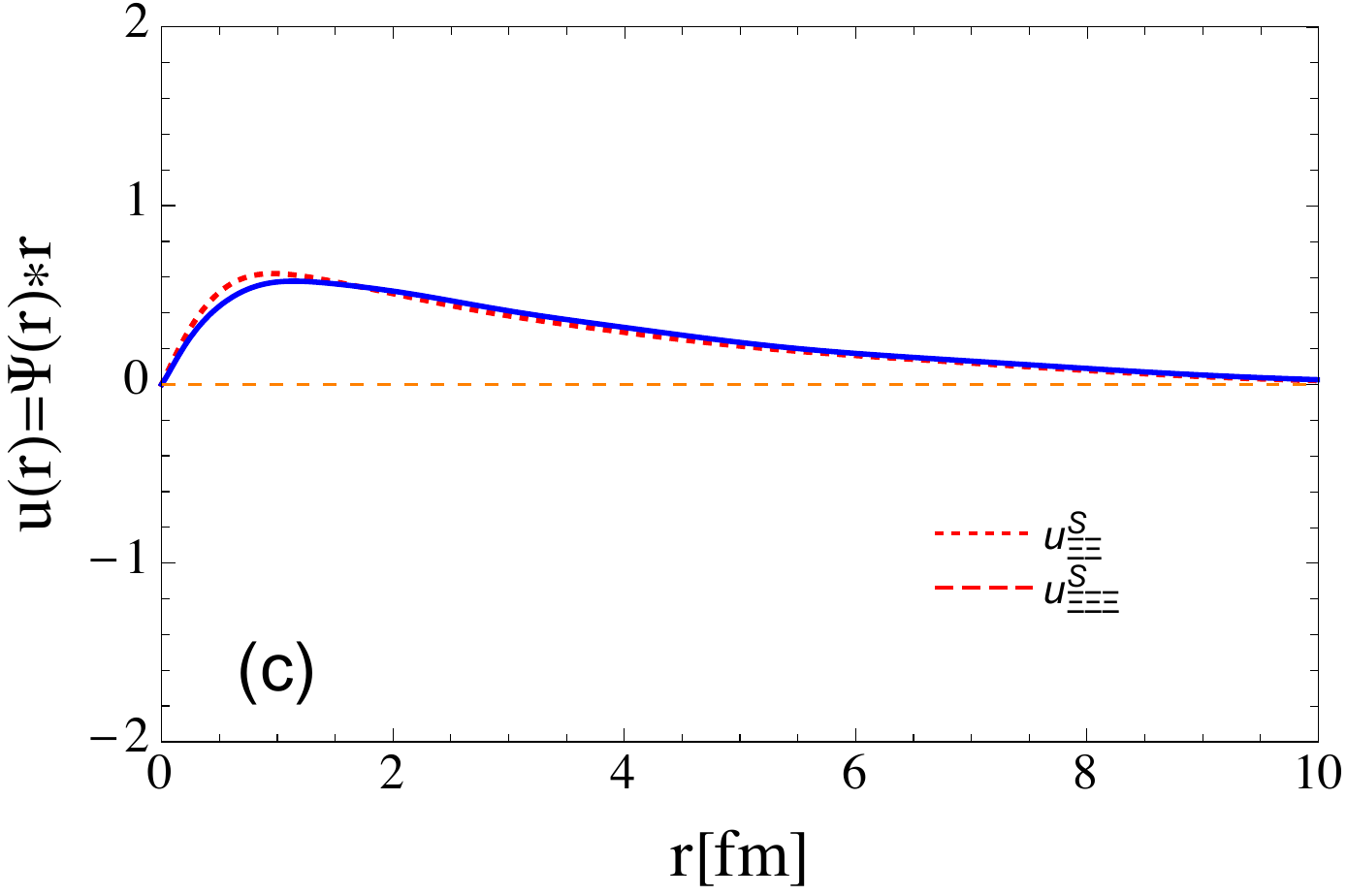}}
      \rotatebox{0}{\includegraphics*[width=0.45\textwidth]{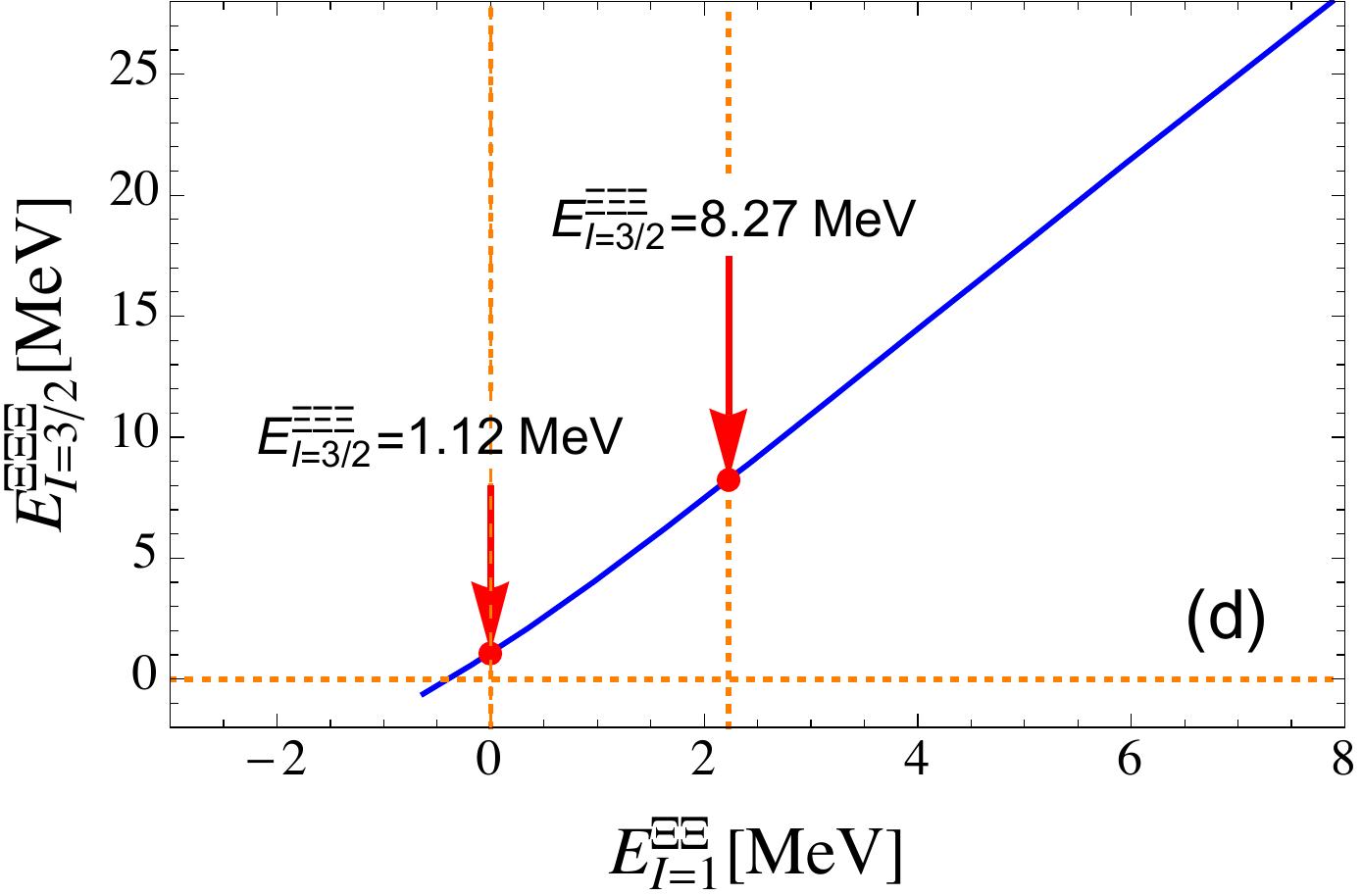}}
    \caption{(a) Plot of wave functions for the $\Xi\Xi\Xi$ system in the isospin state $|0,\frac{1}{2}\rangle$. The blue lines denote the wave functions for any two $\Xi$ in the $\Xi\Xi\Xi$ system. The red lines represent the wave functions for its subsystem $\Xi\Xi$. (b) Dependence of the total three-body binding energy on the two-body binding energy of its subsystem $\Xi\Xi$ for the state $|0,\frac{1}{2}\rangle$. The left red point indicates the Borromean state of the system. The right one is our numerical result when the two-body binding energy is chosen at 2.23 MeV. (c) Plot of wave functions for the $\Xi\Xi\Xi$ system in the isospin state $|1,\frac{3}{2}(\frac{1}{2})\rangle$. (d) Dependence of the total three-body binding energy on the two-body binding energy of its subsystem $\Xi\Xi$ for the state $|1,\frac{3}{2}(\frac{1}{2})\rangle$.}
    \label{xiwave}
  \end{center}
\end{figure}

We plot the wave functions for the $\Xi\Xi\Xi$ in Fig.~\ref{xiwave}(a) and Fig.~\ref{xiwave}(c) to verify the solutions we get are bound states. The wave functions in Fig.~\ref{xiwave}(a) correspond to the state $|0,\frac{1}{2}\rangle$, while the wave functions in Fig.~\ref{xiwave}(c) correspond to the states $|1,\frac{1}{2}\rangle$ and $|1,\frac{3}{2}\rangle$. 
We plot the dependence of the three-body binding energy of the $\Xi\Xi\Xi$ system on its corresponding two-body binding energy in Fig.~\ref{xiwave}(b) and Fig.~\ref{xiwave}(d), where Fig.~\ref{xiwave}(b) and Fig.~\ref{xiwave}(d) correspond to state $|0,\frac{1}{2}\rangle$ and $|1,\frac{3}{2}(\frac{1}{2})\rangle$, respectively. Similar with the three-nucleon system, the $\Xi\Xi\Xi$ system for both cases also have a Borromean state. There are two red points in both figures, where the left one indicate the Borromean state of the system. For the state $|0,\frac{1}{2}\rangle$ as shown in Fig.~\ref{xiwave}(b), there is still a shallow three-body bound state with a tiny binding energy of 1.06 MeV when the two-body binding energy approaches to 0 MeV.  
Similarly, the isospin states $|1,\frac{1}{2}\rangle$ and $|1,\frac{3}{2}\rangle$ have a three-body bound state with a tiny binding energy of 1.12 MeV when the two-body binding energy approaches to 0 MeV.   
The right red points in both figures represent our numerical results when the two-body binding energy is chosen at 2.23 MeV. Since $\Xi$ is a little heavier than nucleon, the three-body bound state for the $\Xi\Xi\Xi$ system is more likely to have a deeper bound state.

\subsection{Numerical results for the \texorpdfstring{$\Sigma\Sigma\Sigma$}{$\Sigma\Sigma\Sigma$}}\label{subsec35}

The $\Sigma\Sigma\Sigma$ system has seven isospin states $|0,1\rangle$, $|1,(0,1,2)\rangle$ and $|2,(1,2,3)\rangle$. Since the constraint from the general identity principle, there are three possible two-body force for the $\Sigma\Sigma\Sigma$ system, i.e. $(0,0)$, $(1,1)$ and $(2,0)$. The state $|0,1\rangle$ is governed by the two-body force $(0,0)$. The states $|1,(0,1,2)\rangle$ are governed by the two-body force $(1,1)$. The states $|2,(1,2,3)\rangle$ is dominated by the two-body force $(2,0)$.  
The $\pi$, $\eta$, $\sigma$, $\rho$, $\omega$ and $\phi$ exchanges all contribute to the total effective potentials. Based on the formalism we construct in Sec.~\ref{sec2}, we need to search for the two-body bound states of its subsystem first. The effective potentials for the subsystem $\Sigma\Sigma$ can be derived from the Lagrangians in Eq. (\ref{LSS}). After the Fourier transformation in Eq. (\ref{FT}), we get the effective potentials of the two-body force in coordinate space, which rely on the parameter $\lambda$. When the parameter $\lambda$ is more than certain value, we find bound solutions for all of three cases as shown in Table~\ref{E3Ebsi01}-\ref{E3Ebsi21}. For the case $(0,0)$, we find a bound state with a binding energy of 2.23 MeV when the parameter $\lambda$ is fixed at 741.66 MeV; For the case $(1,1)$, there is a bound state with a binding energy of 2.23 MeV when the parameter $\lambda$ is chosen at 915.71 MeV. It turns out be a bound state with the same binding energy for the case $(2,0)$ when we fix the parameter $\lambda$ at 1104.42 MeV.

\begin{figure}[ht]
  \begin{center}
  \rotatebox{0}{\includegraphics*[width=0.32\textwidth]{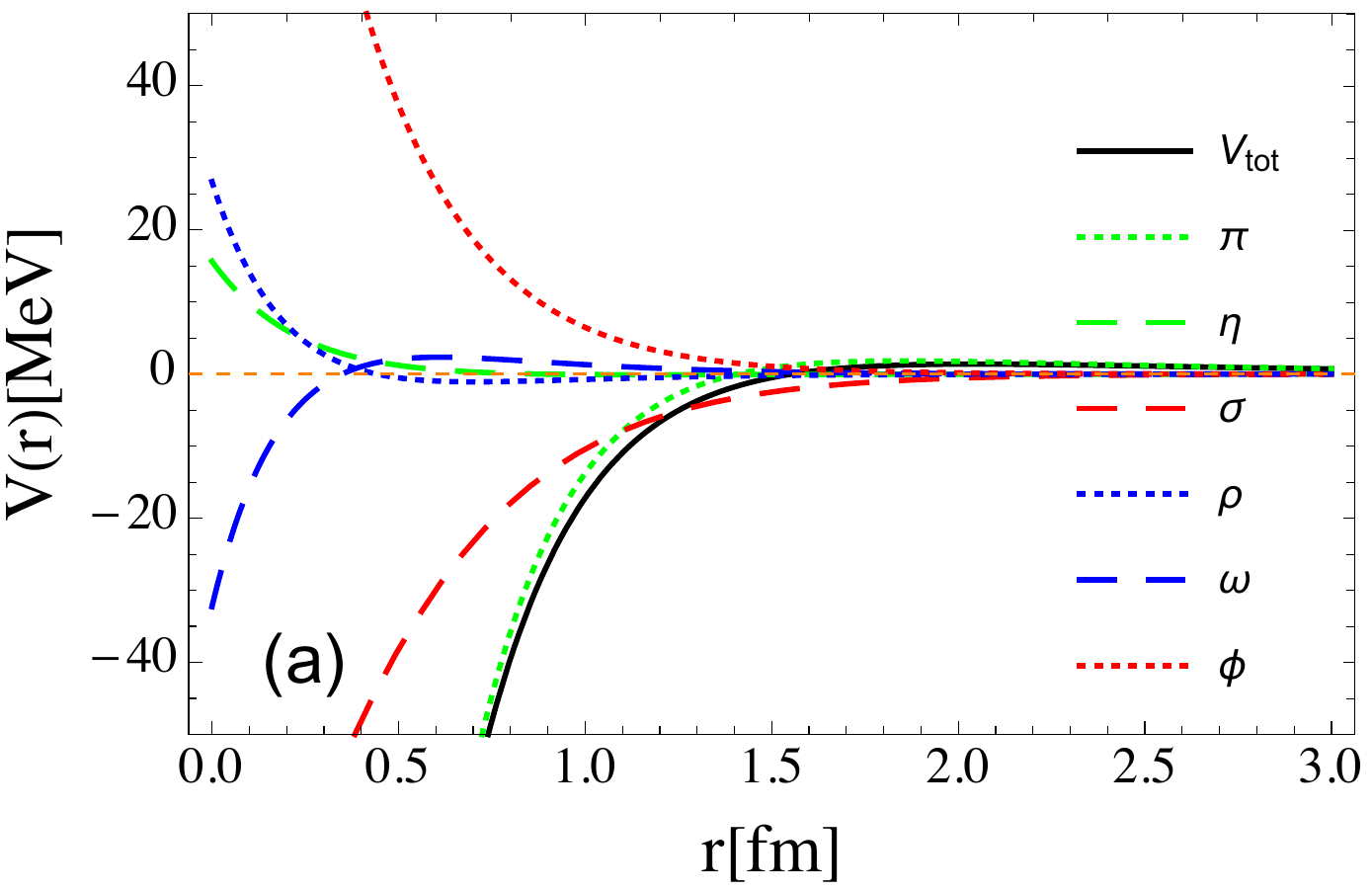}}
  \rotatebox{0}{\includegraphics*[width=0.32\textwidth]{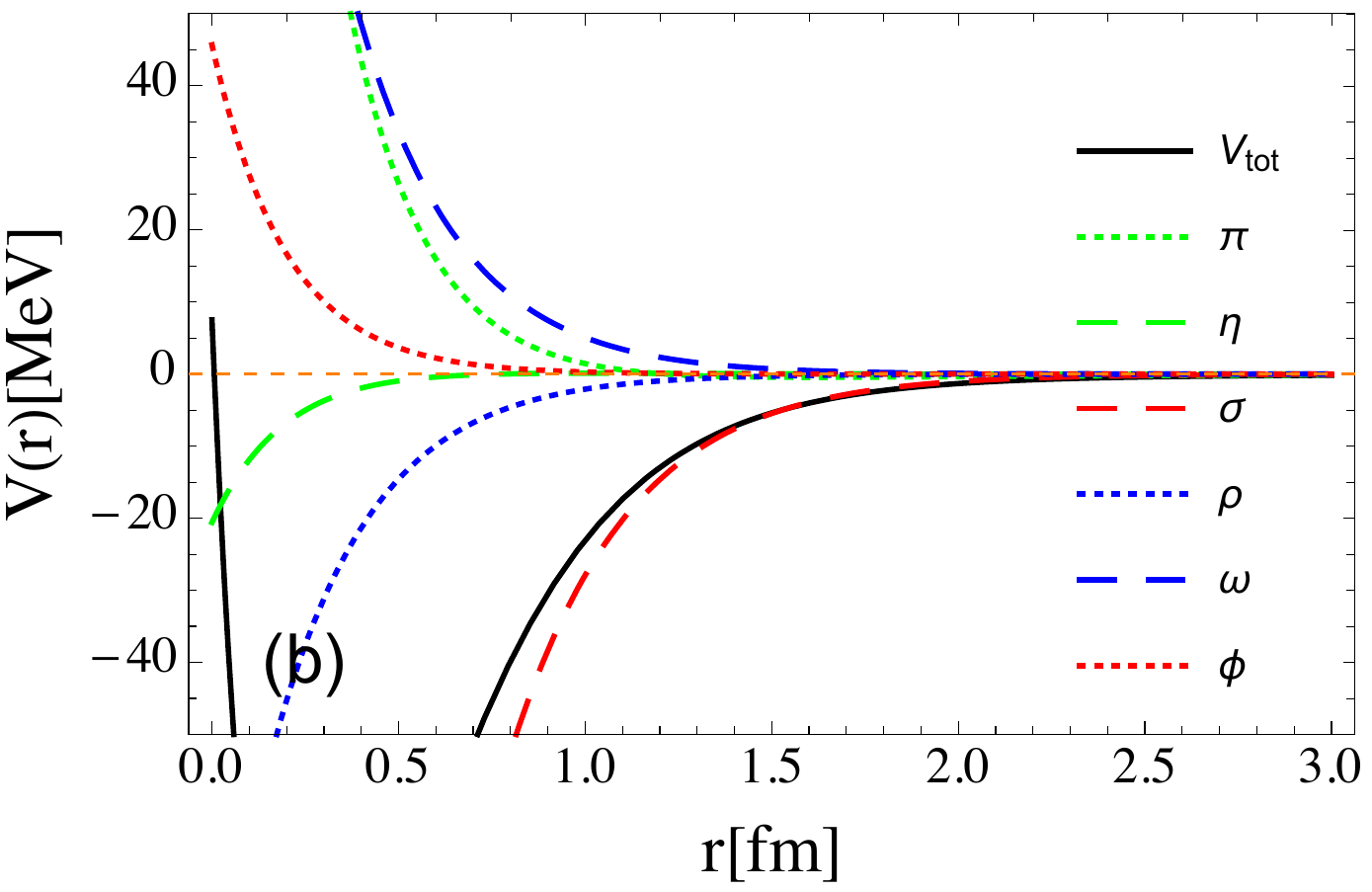}}
  \rotatebox{0}{\includegraphics*[width=0.32\textwidth]{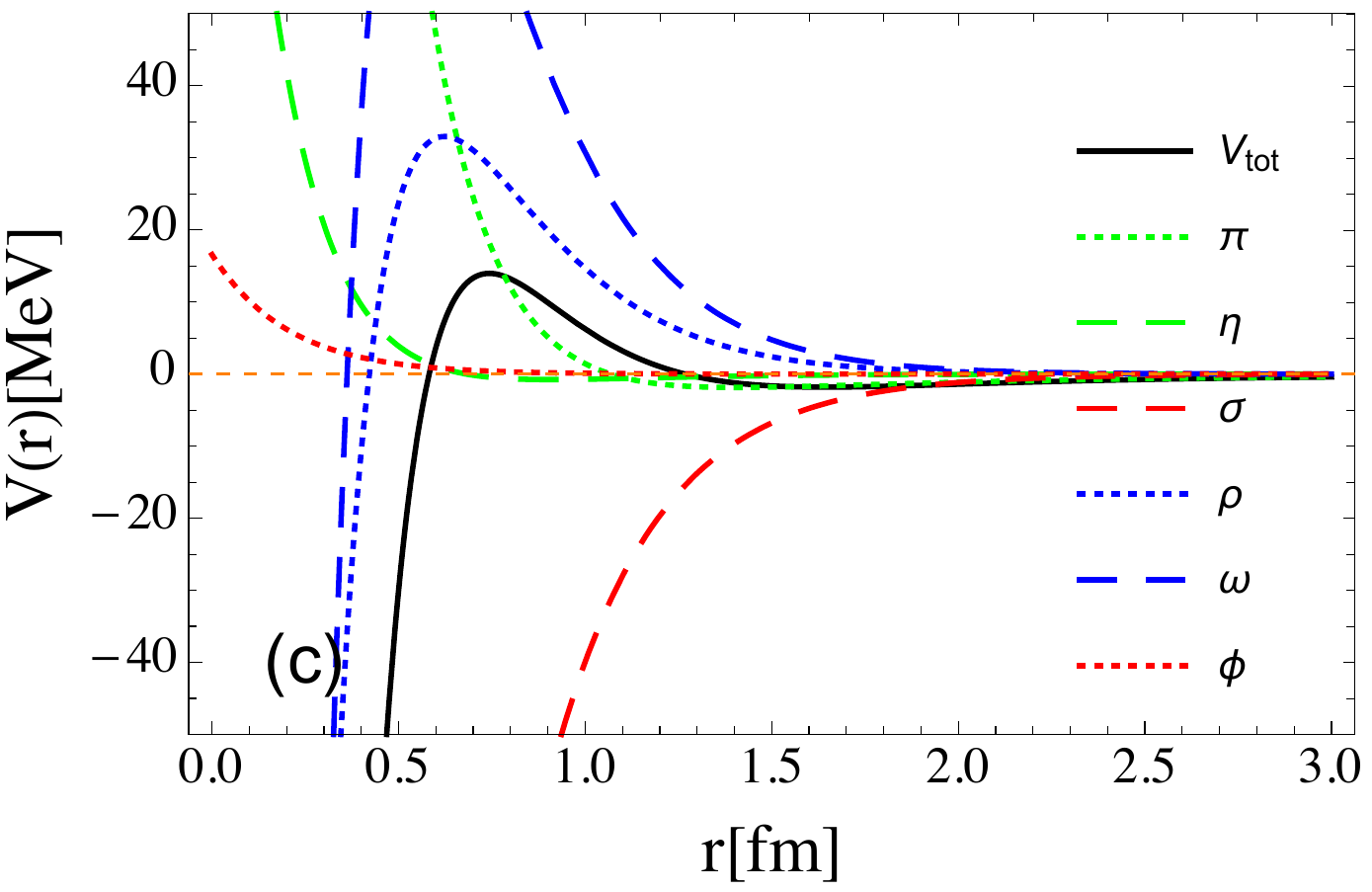}}
  \rotatebox{0}{\includegraphics*[width=0.32\textwidth]{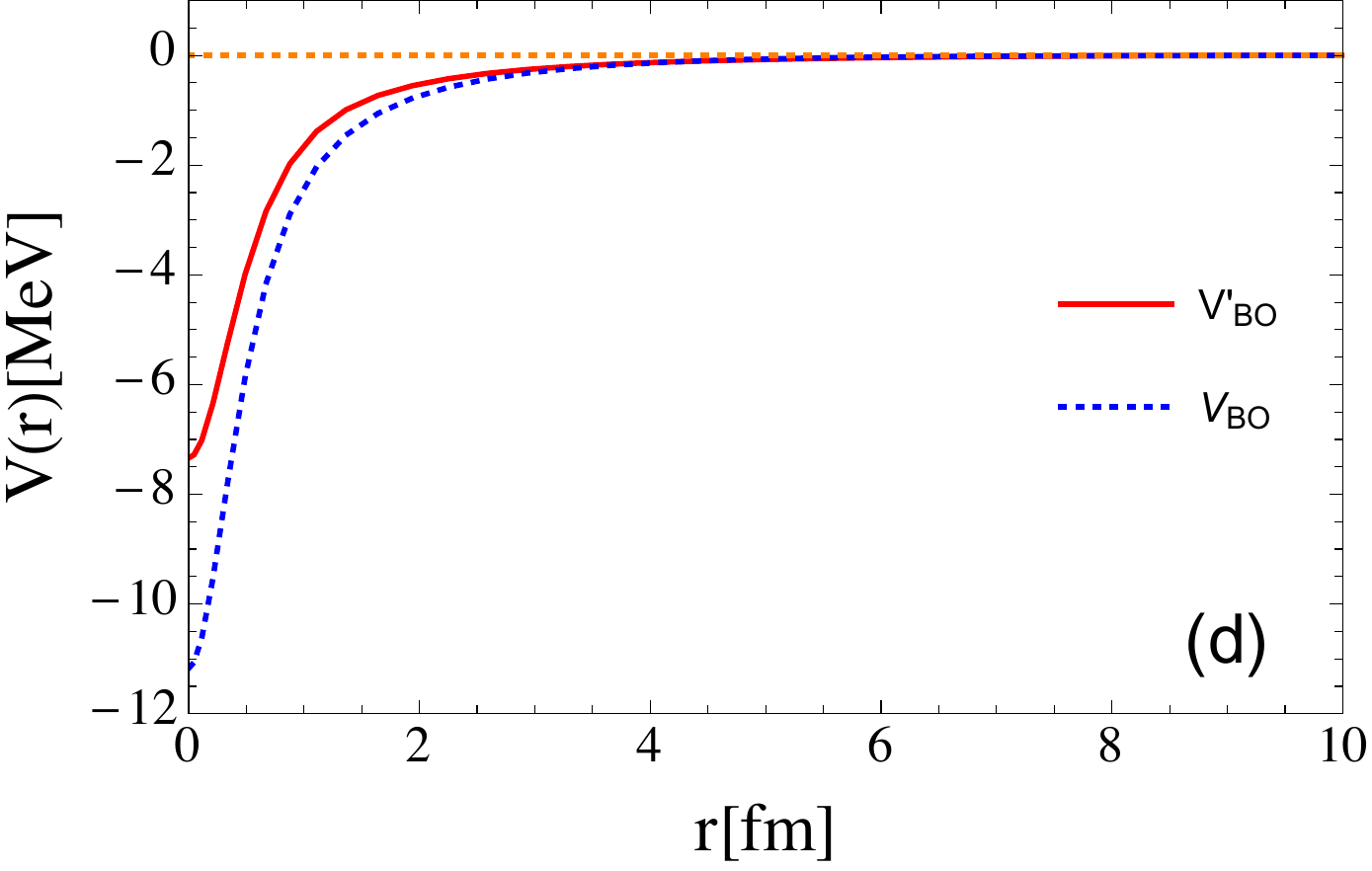}}
  \rotatebox{0}{\includegraphics*[width=0.32\textwidth]{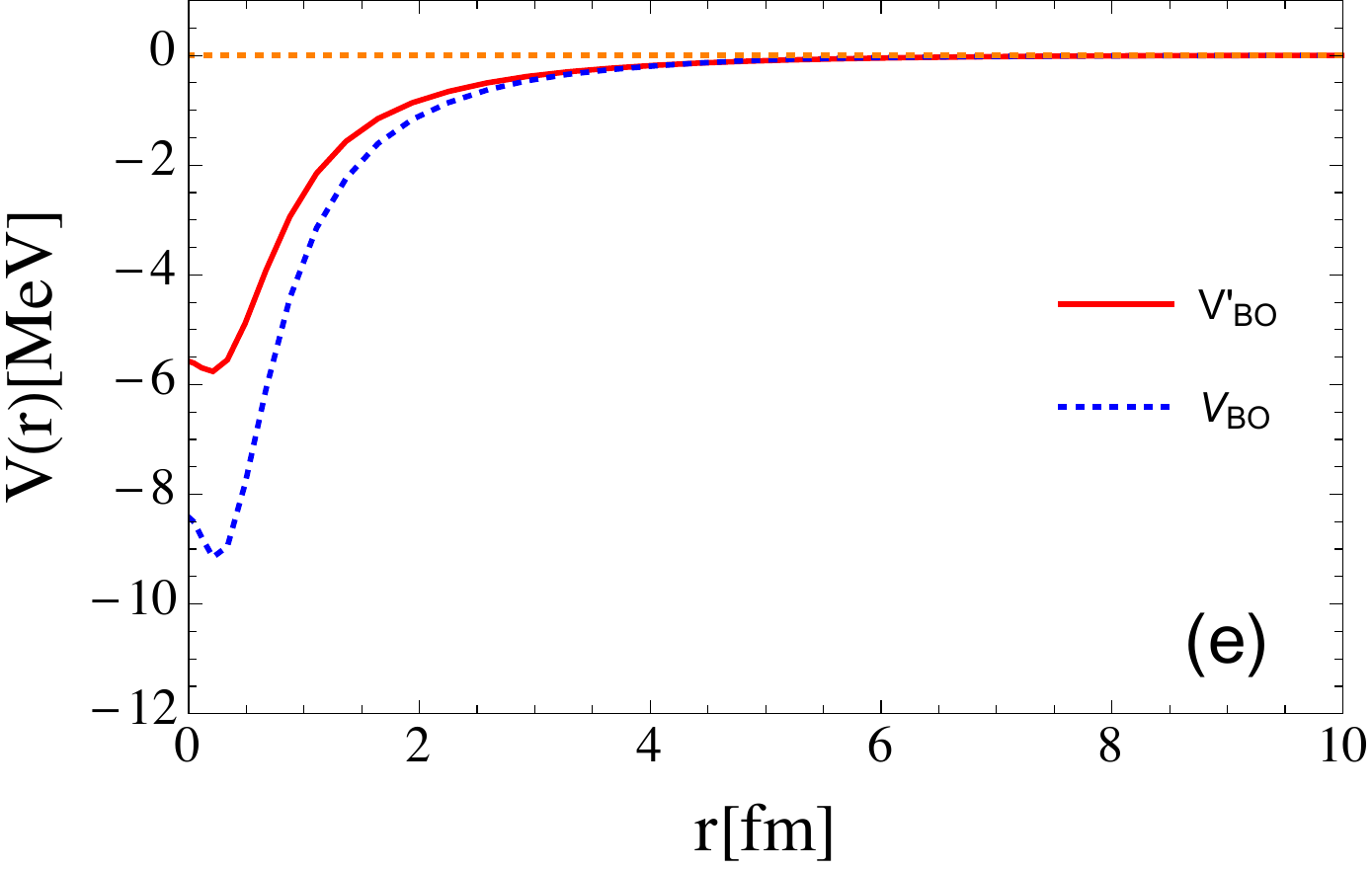}}
  \rotatebox{0}{\includegraphics*[width=0.32\textwidth]{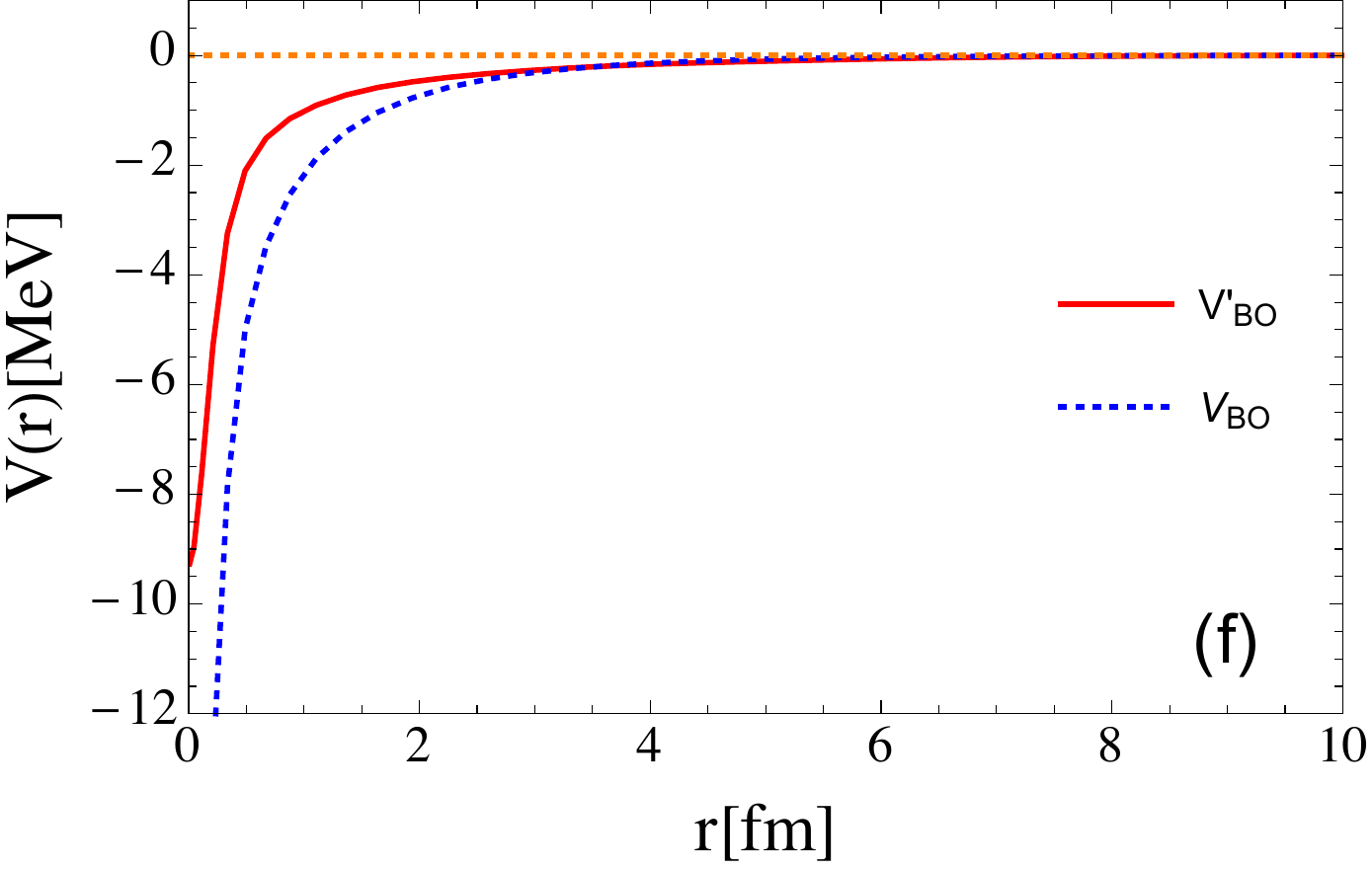}}
     \caption{(a), (b) and (c) are plots of the effective potentials of the two-body force $(0,0)$, $(1,1)$ and $(2,0)$ for the system $\Sigma\Sigma\Sigma$, respectively. (d), (e) and (f) are plots of the BO potentials for the system $\Sigma\Sigma\Sigma$ provided by the two-body force $(0,1)$, $(1,1)$ and $(2,0)$, respectively. The dotted and solid lines in (d), (e) and (f) are the BO potentials before and after distortion corrections, respectively.}
    \label{sipotential}
  \end{center}
\end{figure}

To make the individual role of the exchanged bosons clear, we plot the effective potentials for all of the three cases when their subsystem $\Sigma\Sigma$ have a same binding energy of 2.23 MeV. We show them in Fig.~\ref{sipotential}, where the figure (a), (b) and (c) correspond to the case $(0,0)$, $(1,1)$ and $(2,0)$, respectively. In Fig.~\ref{sipotential}(a), the $\eta$, $\rho$ and $\phi$ exchanges are repulsive, while the $\pi$ and $\sigma$ provide attraction. The $\omega$ exchange gives a shallow repulsion in the medium range but attraction in the short range. In Fig.~\ref{sipotential}(b), the $\pi$, $\omega$ and $\phi$ exchanges provide repulsion, while the $\eta$, $\rho$ and $\sigma$ exchanges are attractive. In Fig.~\ref{sipotential}(c), the $\eta$ and $\phi$ exchanges are repulsive, while the $\sigma$ exchange is attractive. The $\rho$ and $\omega$ provide deep attractions in the short range but strong repulsion in the medium range. The $\pi$ exchange is repulsive in the short range but slightly attractive in the medium range. We also plot the corresponding BO potentials for the three cases in Fig.~\ref{sipotential}(d)-(f). Similarly, the BO potentials become shallow after distortion corrections.  
 
\begin{table}[htbp]
\caption{Bound state solutions of the $\Sigma\Sigma\Sigma$ system with isospin $(I_2,I_3)=(0,1)$. 
$E_2$ is the energy eigenvalue of its subsystem. 
 $E_3$ is the reduced three-body energy eigenvalue relative to the break-up state of the$\Sigma\Sigma\Sigma$ system.  
$E_T$ is the total three-body energy eigenvalue relative to the $\Sigma\Sigma\Sigma$ threshold. 
 $V_{BO}(0)$ is the minimum of the BO potential. 
$r_{rms}$ represents the root-mean-square radius of any two $\Sigma$ in the $\Sigma\Sigma\Sigma$ system. 
The $S$-wave and $D$-wave represent the probabilities for $S$-wave and
 $D$-wave components in any two $\Sigma$ in the $\Sigma\Sigma\Sigma$ system.}\label{E3Ebsi01}
\begin{center}
\begin{tabular}{ | c | c | c | c  | c  | c | c  | c  | c  }
\hline\hline 
  $\Lambda$(MeV)  &  $E_2$(MeV)  & $E_3$(MeV) & $E_T$(MeV)  & $V_{BO}(0)$(MeV)  & S wave(\%) & D wave(\%) & $r_{rms}$(fm)   \\
\hline
       735.00 &  -0.49 & -1.89 & -2.38 & -4.02 & 100.00 & 0 & 3.91 \\
\hline
       740.00 &  -1.75 & -4.83 & -6.57 & -6.40 & 100.00 & 0 & 3.21  \\
\hline       
       741.66 &  -2.23 & -6.08 & -8.31 & -7.33 & 100.00 & 0 & 2.98 \\
\hline
       745.00 &  -3.32 & -8.97 & -12.29 & -9.30 & 100.00 & 0 & 2.57 \\
\hline       
       750.00 &  -5.21 & -14.07 & -19.28 & -12.42 & 100.00 & 0 & 2.09  \\
\hline       
       755.00 &  -7.37 & -19.86 & -27.24 & -15.62 & 100.00 & 0 & 1.78 \\
\hline       
       760.00 &  -9.78 & -26.18 & -35.96 & -18.85 & 100.00 & 0 & 1.56 \\
\hline
       765.00 &  -12.42 & -32.91 & -45.33 & -22.11 & 100.00 & 0 & 1.41  \\
\hline\hline
\end{tabular}
\end{center}
\end{table}

A three-body bound state appears for the state $|0,1\rangle$ with the total binding energy about 2.38-45.33 MeV, when the parameter is around 735.00-765.00 MeV as shown in Table~\ref{E3Ebsi01}. The corresponding two-body binding energy of its two-body subsystem is around 0.49-12.42 MeV. The root-mean-square radius of the system decreases from 3.91 fm to 1.41 fm. Since the potentials in this case do not contain tensor force, the three-body binding solution is dominated by the S-wave state. For comparison with the numerical results of the $NNN$ and $\Xi\Xi\Xi$ systems, we fix the two-body binding energy at 2.23 MeV which yields a total three-body binding energy of 8.31 MeV with the parameter $\lambda=741.66$ MeV.

\begin{table}[htbp]
\caption{Bound state solutions of the $\Sigma\Sigma\Sigma$ system with  isospin $(I_2,I_3)=(1,1)$. 
$E_2$ is the energy eigenvalue of its subsystem. 
 $E_3$ is the reduced three-body energy eigenvalue relative to the break-up state of the$\Sigma\Sigma\Sigma$ system.  
$E_T$ is the total three-body energy eigenvalue relative to the $\Sigma\Sigma\Sigma$ threshold. 
 $V_{BO}(0)$ is the minimum of the BO potential. 
$r_{rms}$ represents the root-mean-square radius of any two $\Sigma$ in the $\Sigma\Sigma\Sigma$ system. 
The $S$-wave and $D$-wave represent the probabilities for $S$-wave and
 $D$-wave components in any two $\Sigma$ in the $\Sigma\Sigma\Sigma$ system.}\label{E3Ebsi11}
\begin{center}
\begin{tabular}{ | c | c | c | c  | c  | c | c  | c  | c  }
\hline\hline 
  $\Lambda$(MeV)  &  $E_2$(MeV)  & $E_3$(MeV) & $E_T$(MeV)  & $V_{BO}(0)$(MeV)  & S wave(\%) & D wave(\%) & $r_{rms}$(fm)   \\
\hline
       900.00 &  -1.11 & -2.87 & -3.98 & -3.99 & 99.11 & 0.89 & 3.78  \\
\hline
       910.00 &  -1.80 & -4.35 & -6.15 & -5.08 & 98.94 & 1.06 & 3.44   \\
\hline       
       915.71 &  -2.24 & -5.31 & -7.55 & -5.76 & 98.84 & 1.16 & 3.25 \\
\hline
       920.00 &  -2.58 & -6.08 & -8.66 & -6.29 & 98.77 & 1.23 & 3.11  \\
\hline       
       930.00 &  -3.42 & -8.02 & -11.43 & -7.57 & 98.61 & 1.39 & 2.81  \\
\hline       
       940.00 &  -4.31 & -10.10 & -14.41 & -8.87 & 98.45 & 1.55 & 2.56  \\
\hline       
       950.00 &  -5.24 & -12.29 & -17.53 & -10.18 & 98.31 & 1.69 & 2.36 \\
\hline
       960.00 &  -6.19 & -14.53 & -20.72 & -11.47 & 98.18 & 1.82 & 2.19  \\
\hline\hline
\end{tabular}
\end{center}
\end{table}

After solving the three-body equation in Eq. (\ref{SDE}), we find that all of the states $|1,0\rangle$, $|1,1\rangle$ and $|1,2\rangle$ have a three-body bound state as shown in Table~\ref{E3Ebsi11}. They are degenerate due to the negligence of three-body force in our formalism. For simplicity, we write the states $|1,0\rangle$, $|1,1\rangle$ and $|1,2\rangle$ together as $|1,(0,1,2)\rangle$. 
The total binding energy of the $\Sigma\Sigma\Sigma$ corresponding to the $|1,(0,1,2)\rangle$ state is about 3.98-20.72 MeV when the parameter $\lambda$ is around 900.00-960.00 MeV. The corresponding two-body binding energy of its two-body subsystem is around 1.11-6.19 MeV. The root-mean-square radius of the system in this isospin state decreases from 3.78 fm to 2.19 fm. Since the existence of the tensor force in the effective potentials, the S-D wave mixing should be taken into account. Within the variety of the parameter $\lambda$, one can see that the bound state is dominated by the S wave state, which has a proportion more than 98 \%. If we choose the parameter $\lambda$ at 915.71 MeV, then the total three-body binding energy is fixed at MeV with the corresponding two-body binding energy of 2.23 MeV.

\begin{table}[htbp]
\caption{Bound state solutions of the $\Sigma\Sigma\Sigma$ system with isospin $(I_2,I_3)=(2,1)$. 
$E_2$ is the energy eigenvalue of its subsystem. $E_3$ is the reduced three-body energy eigenvalue relative to the break-up state of the$\Sigma\Sigma\Sigma$ system.  
$E_T$ is the total three-body energy eigenvalue relative to the $\Sigma\Sigma\Sigma$ threshold. 
 $V_{BO}(0)$ is the minimum of the BO potential. 
$r_{rms}$ represents the root-mean-square radius of any two $\Sigma$ in the $\Sigma\Sigma\Sigma$ system. 
The $S$-wave and $D$-wave represent the probabilities for $S$-wave and $D$-wave components in any two $\Sigma$ in the $\Sigma\Sigma\Sigma$ system.}\label{E3Ebsi21}
\begin{center}
\begin{tabular}{ | c | c | c | c  | c  | c | c  | c  | c  }
\hline\hline 
  $\Lambda$(MeV)  &  $E_2$(MeV)  & $E_3$(MeV) & $E_T$(MeV)  & $V_{BO}(0)$(MeV)  & S wave(\%) & D wave(\%) & $r_{rms}$(fm)   \\
\hline
       1100.00 &  -0.85 & -1.23 & -2.09 & -5.30 & 100.00 & 0 & 4.43  \\
\hline
       1104.42 &  -2.23 & -3.63 & -5.86 & -9.27 & 100.00 & 0 & 3.84   \\
\hline       
       1105.00 &  -2.45 & -4.07 & -6.52 & -9.96 & 100.00 & 0 & 3.74 \\
\hline
       1110.00 &  -4.86 & -9.60 & -14.46 & -17.51 & 100.00 & 0 & 2.80  \\
\hline       
       1115.00 &  -8.20 & -18.91 & -27.11 & -27.42 & 100.00 & 0 & 1.98 \\
\hline       
       1120.00 &  -12.53 & -32.12 & -44.65 & -38.78 & 100.00 & 0 & 1.47 \\
\hline\hline
\end{tabular}
\end{center}
\end{table}

We also find that all of the states $|2,1\rangle$, $|2,2\rangle$ and $|2,3\rangle$ have a degenerate three-body bound state as shown in Table~\ref{E3Ebsi21}. Similarly, we write the states $|2,1\rangle$, $|2,2\rangle$ and $|2,3\rangle$ together as $|2,(1,2,3)\rangle$ for simplicity. When the parameter $\lambda$ is about 1100.00-1120.00 MeV, the three-body binding energy is around 2.09-44.65 MeV. The corresponding two-body binding energy of the two-body subsystem is around 0.85-12.53 MeV. The root-mean-square radius of the system decreases from 4.43 fm to 1.47 fm. Similar with the case of the $|0,1\rangle$ state, the binding solution only has the S wave state. The total three-body binding energy is fixed at 5.86 MeV with the same two-body binding energy of 2.23 MeV, when the parameter $\lambda$ is 1104.42 MeV.

\begin{figure}[ht]
  \begin{center}
  \rotatebox{0}{\includegraphics*[width=0.32\textwidth]{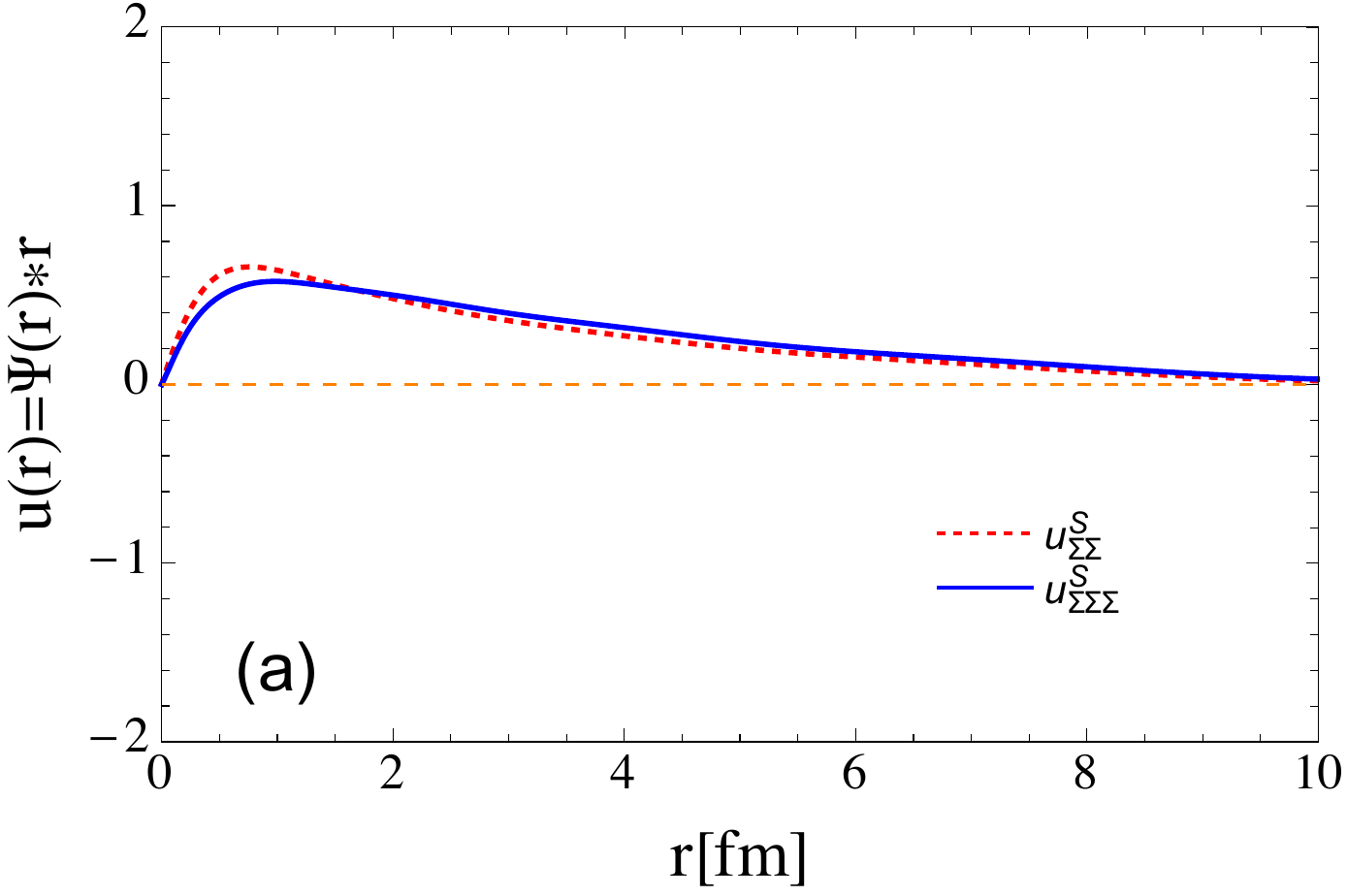}}
  \rotatebox{0}{\includegraphics*[width=0.32\textwidth]{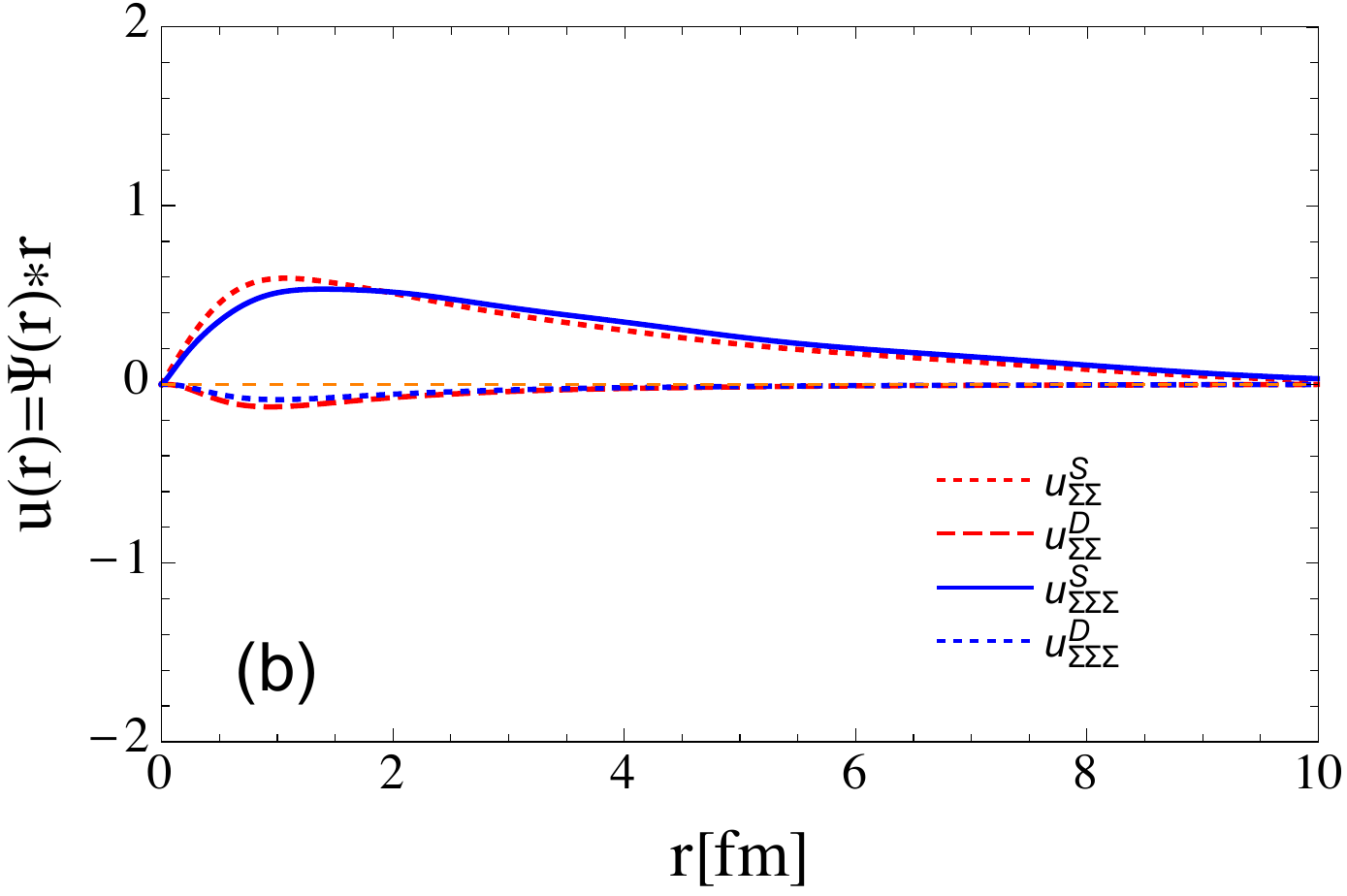}}
  \rotatebox{0}{\includegraphics*[width=0.32\textwidth]{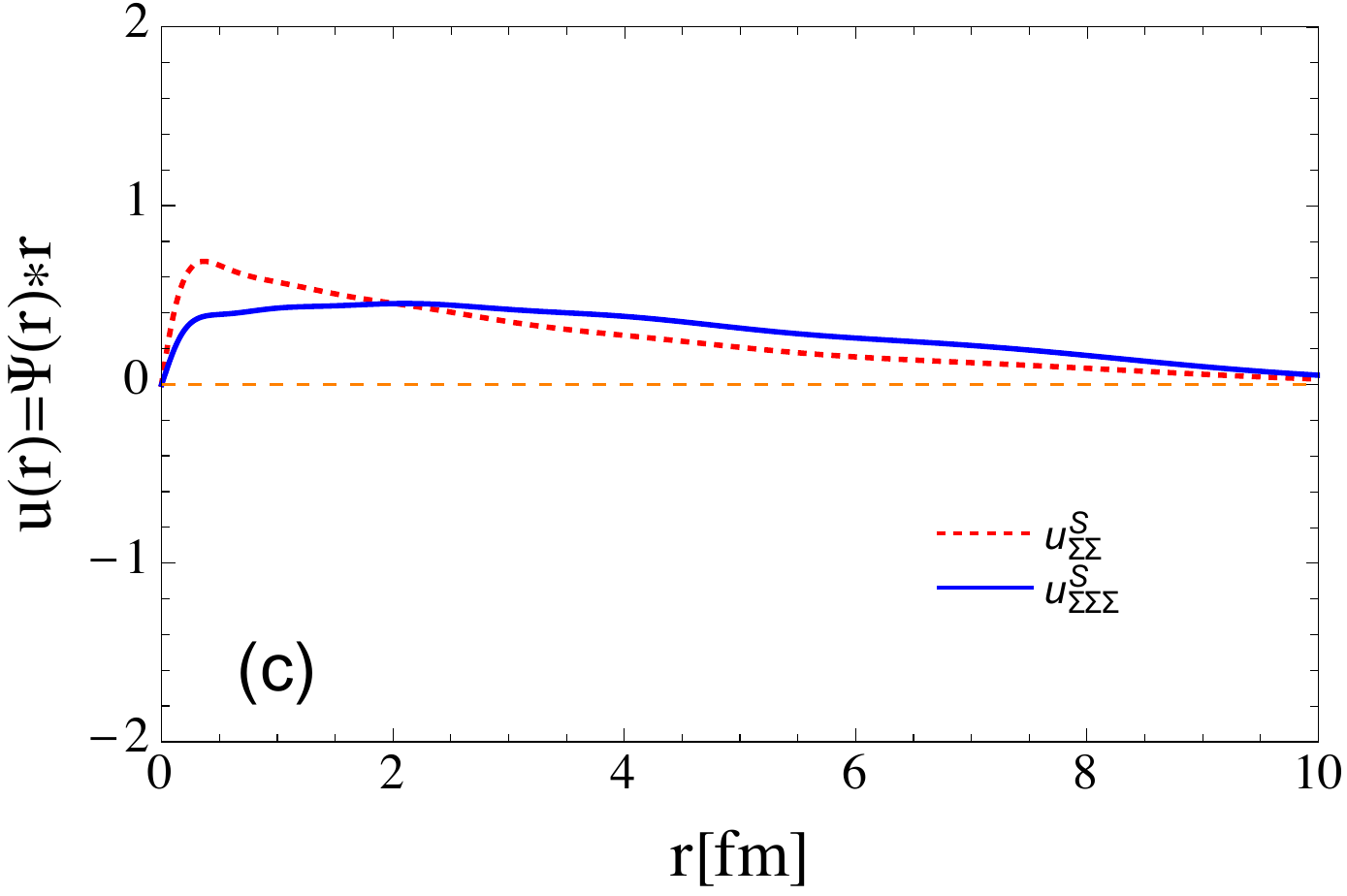}}
  \rotatebox{0}{\includegraphics*[width=0.32\textwidth]{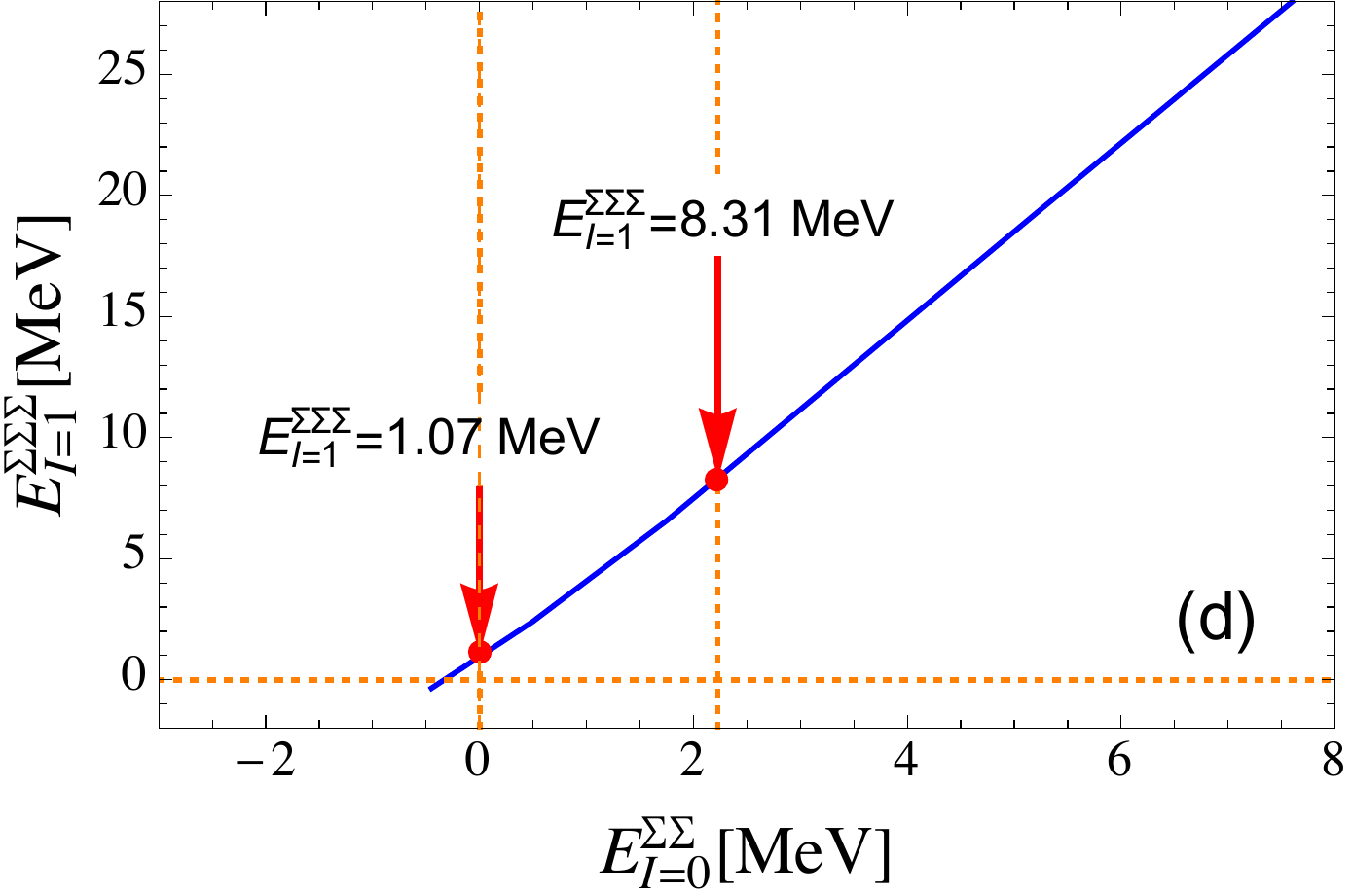}}
  \rotatebox{0}{\includegraphics*[width=0.32\textwidth]{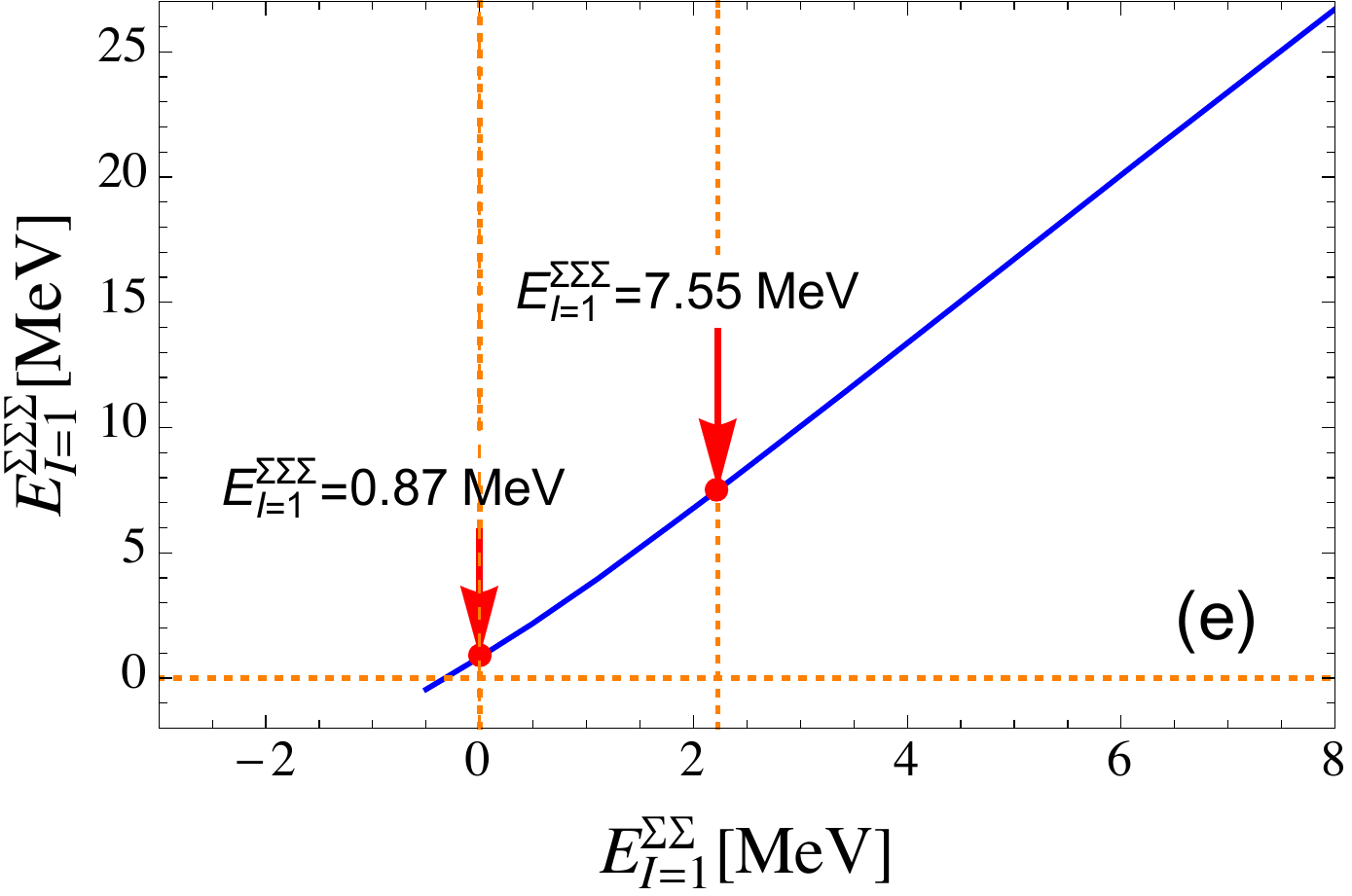}}
  \rotatebox{0}{\includegraphics*[width=0.32\textwidth]{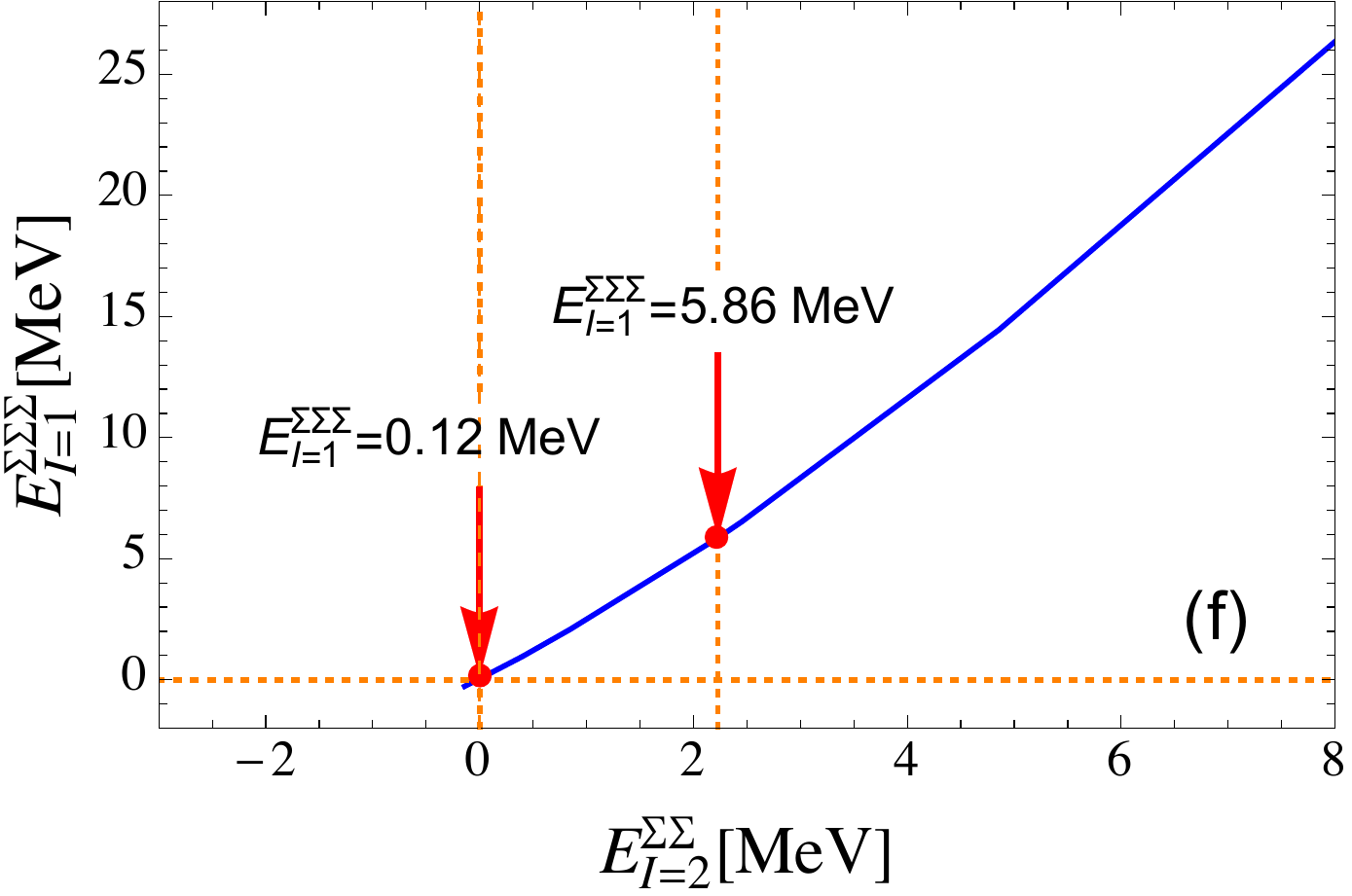}}
     \caption{(a), (b) and (c) are plots of wave functions for the $\Sigma\Sigma\Sigma$ system in the isospin states $|0,1\rangle$, $|1,(0,1,2)\rangle$ and $|2,(1,2,3)\rangle$, respectively. The blue lines denote the wave functions for any two $\Sigma$ in the $\Sigma\Sigma\Sigma$ system. The red lines represent the wave functions for its subsystem $\Sigma\Sigma$. (d), (e) and (f) are plots of the dependence of the total three-body binding energy on the two-body binding energy of the subsystem $\Sigma\Sigma$ for the states $|0,1\rangle$, $|1,(0,1,2)\rangle$ and $|2,(1,2,3)\rangle$, respectively. The left red point indicates the Borromean state of the system. The right one is our numerical result when the two-body binding energy is chosen at 2.23 MeV.}
    \label{siwave}
  \end{center}
\end{figure}

The wave functions for the $\Sigma\Sigma\Sigma$ system are shown in Fig.~\ref{siwave}(a)-(c), where the (a), (b) and (c) are corresponding to the isospin states $|0,1\rangle$, $|1,(0,1,2)\rangle$ and $|2,(1,2,3)\rangle$, respectively. 
The numerical results show that the three-body binding energy increases as the two-body binding energy increases. One may wonder whether there is a Borromean state for the $\Sigma\Sigma\Sigma$ system. After lots of calculations, it turns out that the Borromean states for all of the isospin states are existent. We label two red points in all of the figures, where the left one indicates the Borromean state of the system. In Fig.~\ref{siwave}(d) we can see that the Borromean state of the $|0,1\rangle$ state has a small three-body binding energy of 1.07 MeV when the two-body binding energy approaches to 0 MeV. The isospin state $|1,(0,1,2)\rangle$ also has a small three-body binding energy of 1.06 MeV when we decrease the two-body binding energy to 0 MeV. Similarly, it appears a Borromean state for the $|2,(1,2,3)\rangle$ state when the two-body binding energy approaches to 0 MeV. 
The right red points in the three figures represent our numerical results when the two-body binding energy is chosen at 2.23 MeV. Since our formalism is a version of variational principle, the numerical results of the three-body binding energies should be lower than the strict values.

\section{Summary}\label{sec4}

The contradiction between the reported "XYZ" states and the traditional quark model arouses interest in searching for the possible deuteronlike molecules. The molecules scenario become more and more popular, and has been identified to be a good candidate to interpret some of the "XYZ" states. The molecular structures of triton and helium-3 nucleus inspire us to study the tritonlike systems. 
First we have constructed the formalism of the BOP method for a general system consists of the three identical fermions with half spin. 
Then based on the BOP method as well as the OBE model, we have performed an extensive investigation on the tritonlike systems composed of three identical hadrons of baryon octet, i.e. the $\Lambda\Lambda\Lambda$, $\Xi\Xi\Xi$ and $\Sigma\Sigma\Sigma$. In our formalism, there is only one free parameter $\lambda$ introduced in the monopole form factor to suppress the contribution from UV energies. The parameter $\lambda$ is hard to be pinned down from fundamental theories. Thus we choose the parameter $\lambda$ in a reasonable range to show various binding solutions of the tritonlike systems. 
Before calculations on the tri-hyperon systems, we apply our formalism on the $NNN$ system to verify the feasibility of the BOP method. 
We find a binding solution in the isospin state $|0,\frac{1}{2}\rangle$ with a total binding energy of 7.49 MeV, which is comparable with empirical binding energies of the triton and helium-3 nucleus. 
In our calculations, there is no numerical difference between the binding energies of the triton and helium-3 nucleus, as we do not consider the isospin breaking effect.

Then we return to the calculations on the tri-hyperon system. 
For the $\Lambda\Lambda\Lambda$ system, we fail to find any bound solution with the variety of the parameter $\lambda$. 
After the treatments of the S-D wave mixing within the OBE mechanism, we find that all of the isospin eigenstates have bound state solutions for the $\Xi\Xi\Xi$ system. 
For better comparisons with the three-nucleon system, we fix the two-body binding energy of its subsystem at 2.23 MeV which is the same with deuteron. 
Then the total binding energy of the $\Xi\Xi\Xi$ system in the state $|0,\frac{1}{2}\rangle$ is 7.86 MeV, while it is 8.27 MeV for the state $|1,\frac{1}{2}\rangle$ and $|1,\frac{3}{2}\rangle$.  
A three-body bound state appears for the $\Sigma\Sigma\Sigma$ system in all of the isospin eigenstates as the parameter $\lambda$ grows.  
When the two-body binding energy is chosen at 2.23 MeV, the total binding energy of the $\Sigma\Sigma\Sigma$ system in the state $|0,1\rangle$ is 8.31 MeV. For the isospin states $|1,(0,1,2)\rangle$ and $|2,(1,2,3)\rangle$, the total binding energies are 7.55 MeV and 5.86 MeV, respectively. 
Since total binding energy of the tri-hyperon system depend on the two-body binding energy of their corresponding subsystems, we plot the dependence lines for each systems. All of the $NNN$, $\Xi\Xi\Xi$ and $\Sigma\Sigma\Sigma$ have a Borromean state when the two-body binding energies of their corresponding two-body subsystems approach to 0 MeV. It means that no matter how small their two-body binding energies are, as long as their corresponding two-body subsystems can form bound states, the three-body bound states are most likely to exist.  

In our formalism, we use the BO potential to describe the contribution of one of the particles on the dynamics of the other two. 
Through the BOP method, one can divide a three-body system into three two-body systems with the BO potentials created by the omitted particles. The three different simplification schemes lead to the three different configurations. After simplest combinations of the three configurations, we construct the interpolating wave functions for diagonalizing the Hamiltonian of the three-body system. 
In fact, the strict solutions should be the a more complicated combinations of all possible configurations, which require further investigations.   
One should notice that the BOP method we have used is a version of the variational principle. It always gives an upper limit of the energy for a system. Therefore, the solutions we solved in the last section are approximate solutions. The binding energies we obtained may be less than the strict values of the three-body systems. 

In short summary, we have perform an extensive investigation on the tritonlike system with three identical hadrons of baryon octet. i.e. the $NNN$, $\Lambda\Lambda\Lambda$, $\Xi\Xi\Xi$ and $\Sigma\Sigma\Sigma$. Except the $\Lambda\Lambda\Lambda$, all of the tritonlike systems have Borromean states. It indicates that the $NNN$, $\Xi\Xi\Xi$ and $\Sigma\Sigma\Sigma$ systems could be bound states as long as their corresponding two-body subsystems have binding solutions. 
The efforts in this work might be helpful to understand the three-body systems in future. 

\section*{Acknowledgements}

This work is supported in part by the Fundamental Research Funds for the Central Universities under Grant No. 2020RC005 and No. 2020JBM073. 

\section*{Appendix A: The matrix element of the \texorpdfstring{$\mathcal{H}_{\slashed{a}\slashed{a}}$}{$\mathcal{H}_{\slashed{a}\slashed{a}}$}}

The matrix element of the $\mathcal{H}_{\slashed{a}\slashed{a}}$ has the form  
\begin{eqnarray}
\mathcal{H}_{\slashed{a}\slashed{a}}^{ij} &=& \langle \tilde{\psi}_{\slashed{a}}^i | \mathcal{H} | \tilde{\psi}_{\slashed{a}}^j \rangle = \Big{\langle} \frac{1}{N_i}\big[ (\psi_{\slashed{a}}^i+\psi_{\slashed{b}}^i+\psi_{\slashed{c}}^i)-\sum_{i} x_{im}\psi_{\slashed{a}}^m \big] \Big{|} \mathcal{H} \Big{|} \frac{1}{N_j}\big[ (\psi_{\slashed{a}}^j+\psi_{\slashed{b}}^j+\psi_{\slashed{c}}^j)-\sum_{i} x_{jn}\psi_{\slashed{a}}^n \big] \Big{\rangle} \nonumber\\
&=& 3\frac{1}{N_i N_j} \langle \psi_{\slashed{a}}^i | \mathcal{H} | \psi_{\slashed{a}}^j \rangle + 6\frac{1}{N_i N_j} \langle \psi_{\slashed{b}}^i | \mathcal{H} | \psi_{\slashed{a}}^j \rangle -x_{im} \frac{1}{N_i N_j} \langle \psi_{\slashed{a}}^m | \mathcal{H} | \psi_{\slashed{a}}^j \rangle -x_{jn}\frac{1}{N_i N_j} \langle \psi_{\slashed{a}}^i | \mathcal{H} | \psi_{\slashed{a}}^n \rangle \nonumber\\
&-& 2 x_{im} \frac{1}{N_i N_j} \langle \psi_{\slashed{b}}^m | \mathcal{H} | \psi_{\slashed{a}}^j \rangle -2 x_{jn} \frac{1}{N_i N_j} \langle \psi_{\slashed{b}}^i | \mathcal{H} | \psi_{\slashed{a}}^n \rangle + x_{im}x_{jn} \frac{1}{N_i N_j} \langle \psi_{\slashed{a}}^m | \mathcal{H} | \psi_{\slashed{a}}^n \rangle,
\end{eqnarray}
where, the interchange symmetry in the $\mho\mho\mho$ system is used in the last step. 
Similarly, we have 
\begin{eqnarray}
\mathcal{H}_{\slashed{b}\slashed{a}}^{ij} &=& \langle \tilde{\psi}_{\slashed{b}}^i | \mathcal{H} |\tilde{\psi}_{\slashed{a}}^j \rangle = \Big{\langle} \frac{1}{N_i}\big[ (\psi_{\slashed{a}}^i+\psi_{\slashed{b}}^i+\psi_{\slashed{c}}^i)-\sum_{i} x_{im}\psi_{\slashed{b}}^m \big] \Big{|} \mathcal{H} \Big{|} \frac{1}{N_j}\big[ (\psi_{\slashed{a}}^j+\psi_{\slashed{b}}^j+\psi_{\slashed{c}}^j)-\sum_{i} x_{jn}\psi_{\slashed{a}}^n \big] \Big{\rangle} \nonumber\\
&=& 3\frac{1}{N_i N_j} \langle \psi_{\slashed{a}}^i | \mathcal{H} | \psi_{\slashed{a}}^j \rangle + 6\frac{1}{N_i N_j} \langle \psi_{\slashed{b}}^i | \mathcal{H} | \psi_{\slashed{a}}^j \rangle -x_{im} \frac{1}{N_i N_j} \langle \psi_{\slashed{a}}^m | \mathcal{H} | \psi_{\slashed{a}}^j \rangle -x_{jn}\frac{1}{N_i N_j} \langle \psi_{\slashed{a}}^i | \mathcal{H} | \psi_{\slashed{a}}^n \rangle \nonumber\\
&-& 2 x_{im} \frac{1}{N_i N_j} \langle \psi_{\slashed{b}}^m | \mathcal{H} | \psi_{\slashed{a}}^j \rangle -2 x_{jn} \frac{1}{N_i N_j} \langle \psi_{\slashed{b}}^i | \mathcal{H} | \psi_{\slashed{a}}^n \rangle + x_{im}x_{jn} \frac{1}{N_i N_j} \langle \psi_{\slashed{b}}^m | \mathcal{H} | \psi_{\slashed{a}}^n \rangle. 
\end{eqnarray}
Since the interchange symmetry in the $\mho\mho\mho$ system, only two matrices in Eq. (\ref{SDE}) are independent, which have the form
\begin{eqnarray*}
\langle \psi_{\slashed{a}}^i | \mathcal{H} | \psi_{\slashed{a}}^j \rangle &=& |N|^2  \int d\vec{r}_{bc}  \Big{\{}    
  (2 \langle \psi_{ab} | \psi_{ab} \rangle+ \langle \psi_{ab} | \psi_{ac} \rangle)[ \phi_{bc}^i (T'_{\ast}+V_{BO}^{bc}) \phi_{bc}^j ] + (\langle \psi_{ab} | \psi_{ac} \rangle)[ \phi_{bc}^i V_{1}^{bc} \phi_{bc}^j ]  \Big{\}} \\
 &=&  \int d\vec{r}_{bc}  \Big{\{} \phi_{bc}^i (T'_{\ast}+V_{BO}^{bc}) \phi_{bc}^j + (1-\frac{2}{2+\langle \psi_{ab} | \psi_{ac} \rangle}) \phi_{bc}^i V^{bc} \phi_{bc}^j,
\end{eqnarray*}
\begin{eqnarray*}
\langle \psi_{\slashed{b}}^i | \mathcal{H} | \psi_{\slashed{a}}^j \rangle &=& \frac{ |N|^2}{2}  \int d\vec{r}_{bc}  \Big{\{} \langle \psi_{ab}\phi_{ac}^i | T'_{\ast}+V_{BO}^{bc} | \phi_{bc}^j (\psi_{ab}+\psi_{ac})\rangle + \langle (\psi_{ab}+\psi_{bc})\phi_{ac}^i | T'_{\ast}+V_{BO}^{bc} | \phi_{bc}^j \psi_{ab}\rangle \\
&+& \langle \psi_{bc}\phi_{ac}^i | T'_{\ast}+V_{BO}^{bc} | \phi_{bc}^j \psi_{ac}\rangle + \langle \psi_{bc}\phi_{ac}^i | V^{bc} | \phi_{bc}^j \psi_{ab} \rangle + \langle (\psi_{ab}+\psi_{bc})\phi_{ac}^i | V^{bc} | \phi_{bc}^j \psi_{ac} \rangle \Big{\}} \\
&=& \frac{1}{4} \frac{1}{1+\frac{1}{2}\langle \psi_{ab}| \psi_{ac}\rangle} \Big{\{} 2 \langle \psi_{ab} \phi_{ac}^i | T'_{\ast}+V_{BO}^{bc} | \psi_{bc}^j \psi_{ab}\rangle + \langle \psi_{ab} \phi_{ac}^i | T'_{\ast}+V_{BO}^{bc} | \phi_{bc}^j \psi_{ac}\rangle \\
&+& \langle \psi_{bc} \phi_{ac}^i | T'_{\ast}+V_{BO}^{bc} | \psi_{bc}^j \psi_{ab} + \langle \psi_{bc} \phi_{ac}^i | T'_{\ast}+V_{BO}^{bc} | \phi_{bc}^j \psi_{ac} \rangle \\
&+& \langle \psi_{bc} \phi_{ac}^i | V^{bc} | \phi_{bc}^j \psi_{ab} \rangle + \langle \psi_{ab} \phi_{ac}^i | V^{bc} | \phi_{bc}^j \psi_{ac} \rangle + \langle \psi_{bc} \phi_{ac}^i | V^{bc} | \phi_{bc}^j \psi_{ac} \rangle \Big{\}},
\end{eqnarray*}
where we have introduced the abbreviations $\phi_{ab}^i$, $\phi_{bc}^i$, $\phi_{ac}^i$, $\psi_{ab}$,
$\psi_{bc}$, $\psi_{ac}$ for $\phi(\vec{r}_{ab})^i$, $\phi(\vec{r}_{bc})^i$, $\phi(\vec{r}_{ac})^i$,
$\psi(\vec{r}_{ab})$, $\psi(\vec{r}_{bc})$, $\psi(\vec{r}_{ac})$, respectively. 
The expression for the $\mathcal{H}_{\slashed{c}\slashed{a}}$ can be derived by the
replacement $c\rightarrow b, ~b\rightarrow c$ on the expression for the
$\mathcal{H}_{\slashed{b}\slashed{a}}$. Similarly, the expression for the
$\mathcal{H}_{\slashed{c}\slashed{b}}$ is derived by the replacement $c\rightarrow b, ~b\rightarrow a,
~a\rightarrow c$ on the expression for the $\mathcal{H}_{\slashed{b}\slashed{a}}$. 
The interchange invariance of the $\mho\mho\mho$ system is help to simplify the calculation,
i.e. $\mathcal{H}_{\slashed{c}\slashed{a}}=\mathcal{H}_{\slashed{c}\slashed{b}}=\mathcal{H}_{\slashed{b}\slashed{a}}$
and $\mathcal{H}_{\slashed{a}\slashed{a}}=\mathcal{H}_{\slashed{b}\slashed{b}}=\mathcal{H}_{\slashed{c}\slashed{c}}$.

\section*{Appendix B: The isospin wave functions of the relevant systems}

We define the isospin wave function of a tritonlike system as $|I_2,I_3,I_{3z}\rangle$, where the $I_2$ is the isospin of its two-body subsystem, the $I_3$ and $I_{3z}$ denote the total isospin of the three-body system and its $z$ direction, respectively. 
The isospin wave functions of the $NNN$ system read
\begin{eqnarray*}
\left|1, \frac{3}{2}, \frac{3}{2}\right\rangle  &=& (pp)p, \\
\left|1, \frac{3}{2}, \frac{1}{2}\right\rangle &=& \frac{1}{\sqrt{3}} [(np)p+(pn)p+(pp)n], \\
\left|1, \frac{3}{2}, -\frac{1}{2}\right\rangle &=& \frac{1}{\sqrt{3}} [(nn)p+(np)n+(pn)n],\\
\left|1, \frac{3}{2}, -\frac{3}{2}\right\rangle  &=& (nn)n, \\
\left|1, \frac{1}{2}, \frac{1}{2}\right\rangle &=& \frac{1}{\sqrt{6}} [2(pp)n-(pn)p-(np)p],\\
\left|1, \frac{1}{2}, -\frac{1}{2}\right\rangle &=& \frac{1}{\sqrt{6}} [(pn)n+(np)n-2(nn)p],\\
\left|0, \frac{1}{2}, \frac{1}{2}\right\rangle &=& \frac{1}{\sqrt{2}}[| (pn)p \rangle-| (np)p \rangle], \\
\left|0, \frac{1}{2}, -\frac{1}{2}\right\rangle &=& \frac{1}{\sqrt{2}}[| (pn)n \rangle-| (np)n \rangle]. 
\end{eqnarray*}
The isospin wave functions of the $\Xi\Xi\Xi$ system read
\begin{eqnarray*}
\left|1, \frac{3}{2}, \frac{3}{2}\right\rangle  &=& (\Xi^0 \Xi^0)\Xi^0, \\
\left|1, \frac{3}{2}, \frac{1}{2}\right\rangle &=& \frac{1}{\sqrt{3}} [(\Xi^-\Xi^0)\Xi^0+(\Xi^0\Xi^-)\Xi^0+(\Xi^0\Xi^0)\Xi^-], \\
\left|1, \frac{3}{2}, -\frac{1}{2}\right\rangle &=& \frac{1}{\sqrt{3}} [(\Xi^-\Xi^-)\Xi^0+(\Xi^-\Xi^0)\Xi^-+(\Xi^0\Xi^-)\Xi^-],\\
\left|1, \frac{3}{2}, -\frac{3}{2}\right\rangle  &=& (\Xi^-\Xi^-)\Xi^-, \\
\left|1, \frac{1}{2}, \frac{1}{2}\right\rangle &=& \frac{1}{\sqrt{6}} [2(\Xi^0\Xi^0)\Xi^--(\Xi^0\Xi^-)\Xi^0-(\Xi^-\Xi^0)\Xi^0],\\
\left|1, \frac{1}{2}, -\frac{1}{2}\right\rangle &=& \frac{1}{\sqrt{6}} [(\Xi^0\Xi^-)\Xi^-+(\Xi^-\Xi^0)\Xi^--2(\Xi^-\Xi^-)\Xi^0],\\
\left|0, \frac{1}{2}, \frac{1}{2}\right\rangle &=& \frac{1}{\sqrt{2}}[| (\Xi^0\Xi^-)\Xi^0 \rangle-| (\Xi^-\Xi^0)\Xi^0 \rangle], \\
\left|0, \frac{1}{2}, -\frac{1}{2}\right\rangle &=& \frac{1}{\sqrt{2}}[| (\Xi^0\Xi^-)\Xi^- \rangle-| (\Xi^-\Xi^0)\Xi^- \rangle]. 
\end{eqnarray*}
The isospin wave functions of the $\Sigma\Sigma\Sigma$ system read
\begin{eqnarray*}
\left|2, 3, 3\right\rangle  &=& (\Sigma^+\Sigma^+)\Sigma^+,\\
\left|2, 3, 2\right\rangle  &=& \frac{1}{\sqrt{3}} [(\Sigma^+\Sigma^+)\Sigma^0+(\Sigma^+\Sigma^0)\Sigma^++(\Sigma^0\Sigma^+)\Sigma^+],\\
\left|2, 3, 1\right\rangle  &=& \frac{1}{\sqrt{15}} [(\Sigma^+\Sigma^+)\Sigma^-+2(\Sigma^+\Sigma^0)\Sigma^0+2(\Sigma^0\Sigma^+)\Sigma^0+(\Sigma^+\Sigma^-)\Sigma^++2(\Sigma^0\Sigma^0)\Sigma^++(\Sigma^-\Sigma^+)\Sigma^+],\\
\left|2, 3, 0\right\rangle  &=& \frac{1}{\sqrt{10}} [(\Sigma^+\Sigma^0)\Sigma^-+(\Sigma^0\Sigma^+)\Sigma^-+(\Sigma^+\Sigma^-)\Sigma^0+2(\Sigma^0\Sigma^0)\Sigma^0+(\Sigma^-\Sigma^+)\Sigma^0+(\Sigma^0\Sigma^-)\Sigma^++(\Sigma^-\Sigma^0)\Sigma^+], \\
\left|2, 3, -1\right\rangle  &=& \frac{1}{\sqrt{15}} [(\Sigma^+\Sigma^-)\Sigma^-+2(\Sigma^0\Sigma^0)\Sigma^-+(\Sigma^-\Sigma^+)\Sigma^-+2(\Sigma^0\Sigma^-)\Sigma^0+2(\Sigma^-\Sigma^0)\Sigma^0+(\Sigma^-\Sigma^-)\Sigma^+],\\
\left|2, 3, -2\right\rangle  &=& \frac{1}{\sqrt{3}} [(\Sigma^0\Sigma^-)\Sigma^-+(\Sigma^-\Sigma^0)\Sigma^-+(\Sigma^-\Sigma^-)\Sigma^0],\\
\left|2, 3, -3\right\rangle  &=& (\Sigma^-\Sigma^-)\Sigma^-,
\end{eqnarray*}
\begin{eqnarray*}
\left|2, 2, 2\right\rangle  &=& \frac{1}{\sqrt{6}} [(\Sigma^+\Sigma^+)\Sigma^0-(\Sigma^+\Sigma^0)\Sigma^+-(\Sigma^0\Sigma^+)\Sigma^+],\\
\left|2, 2, 1\right\rangle  &=& \frac{1}{2\sqrt{3}} [2(\Sigma^+\Sigma^+)\Sigma^-+(\Sigma^+\Sigma^0)\Sigma^0+(\Sigma^0\Sigma^+)\Sigma^0-(\Sigma^+\Sigma^-)\Sigma^+-(\Sigma^0\Sigma^0)\Sigma^+-(\Sigma^-\Sigma^+)\Sigma^+], \\
\left|2, 2, 0\right\rangle  &=& \frac{1}{2} [(\Sigma^+\Sigma^0)\Sigma^-+(\Sigma^0\Sigma^+)\Sigma^--(\Sigma^0\Sigma^-)\Sigma^+-(\Sigma^-\Sigma^0)\Sigma^+],\\
\left|2, 2, -1\right\rangle  &=&  \frac{1}{2\sqrt{3}} [(\Sigma^+\Sigma^-)\Sigma^-+2(\Sigma^0\Sigma^0)\Sigma^-+(\Sigma^-\Sigma^+)\Sigma^--(\Sigma^0\Sigma^-)\Sigma^0-(\Sigma^-\Sigma^0)\Sigma^0-2(\Sigma^-\Sigma^-)\Sigma^+],\\
\left|2, 2, -2\right\rangle  &=& \frac{1}{\sqrt{6}} [(\Sigma^0\Sigma^-)\Sigma^-+(\Sigma^-\Sigma^0)\Sigma^--2(\Sigma^-\Sigma^-)\Sigma^0],
\end{eqnarray*}
\begin{eqnarray*}
\left|2, 1, 1\right\rangle  &=& \frac{\sqrt{15}}{30} [6(\Sigma^+\Sigma^+)\Sigma^--3(\Sigma^+\Sigma^0)\Sigma^0-3(\Sigma^0\Sigma^+)\Sigma^0+(\Sigma^+\Sigma^-)\Sigma^++2(\Sigma^0\Sigma^0)\Sigma^++(\Sigma^-\Sigma^+)\Sigma^+],\\
\left|2, 1, 0\right\rangle  &=& \frac{1}{2\sqrt{15}} [3(\Sigma^+\Sigma^0)\Sigma^-+3(\Sigma^0\Sigma^+)\Sigma^--2(\Sigma^+\Sigma^-)\Sigma^0-4(\Sigma^0\Sigma^0)\Sigma^0-2(\Sigma^-\Sigma^+)\Sigma^0+3(\Sigma^0\Sigma^-)\Sigma^+3(\Sigma^-\Sigma^0)\Sigma^+],\\
\left|2, 1, -1\right\rangle  &=& \frac{1}{2\sqrt{15}} [(\Sigma^+\Sigma^-)\Sigma^-+2(\Sigma^0\Sigma^0)\Sigma^-+(\Sigma^-\Sigma^+)\Sigma^--3(\Sigma^0\Sigma^-)\Sigma^0-3(\Sigma^-\Sigma^0)\Sigma^0+6(\Sigma^-\Sigma^-)\Sigma^+],
\end{eqnarray*}
\begin{eqnarray*}
\left|1, 2, 2\right\rangle  &=& \frac{1}{\sqrt{2}} [(\Sigma^+\Sigma^0)\Sigma^+-(\Sigma^0\Sigma^+)\Sigma^+],\\
\left|1, 2, 1\right\rangle  &=& \frac{1}{2} [(\Sigma^+\Sigma^0)\Sigma^0-(\Sigma^0\Sigma^+)\Sigma^0+(\Sigma^+\Sigma^-)\Sigma^+-(\Sigma^-\Sigma^+)\Sigma^+],\\
\left|1, 2, 0\right\rangle  &=& \frac{1}{2\sqrt{3}} [(\Sigma^+\Sigma^0)\Sigma^--(\Sigma^0\Sigma^+)\Sigma^-+2(\Sigma^+\Sigma^-)\Sigma^0-2(\Sigma^-\Sigma^+)\Sigma^0+(\Sigma^0\Sigma^-)\Sigma^+-(\Sigma^-\Sigma^0)\Sigma^+],\\
\left|1, 2, -1\right\rangle  &=& \frac{1}{2} [(\Sigma^+\Sigma^-)\Sigma^-+(\Sigma^0\Sigma^-)\Sigma^0-(\Sigma^-\Sigma^+)\Sigma^--(\Sigma^-\Sigma^0)\Sigma^0],\\
\left|1, 2, -2\right\rangle  &=& \frac{1}{\sqrt{2}} [(\Sigma^0\Sigma^-)\Sigma^--(\Sigma^-\Sigma^0)\Sigma^-],
\end{eqnarray*}
\begin{eqnarray*}
\left|1, 1, 1\right\rangle  &=& \frac{1}{2} [(\Sigma^+\Sigma^0)\Sigma^0-(\Sigma^0\Sigma^+)\Sigma^0-(\Sigma^+\Sigma^-)\Sigma^++(\Sigma^-\Sigma^+)\Sigma^+],\\
\left|1, 1, 0\right\rangle  &=& \frac{1}{2} [(\Sigma^+\Sigma^0)\Sigma^-+(\Sigma^-\Sigma^0)\Sigma^+-(\Sigma^0\Sigma^+)\Sigma^--(\Sigma^0\Sigma^-)\Sigma^+],\\
\left|1, 1, -1\right\rangle  &=& \frac{1}{2} [(\Sigma^+\Sigma^-)\Sigma^--(\Sigma^-\Sigma^+)\Sigma^--(\Sigma^0\Sigma^-)\Sigma^0+(\Sigma^-\Sigma^0)\Sigma^0],\\
\left|1, 0, 0\right\rangle  &=& \frac{1}{\sqrt{6}} [(\Sigma^+\Sigma^0)\Sigma^--(\Sigma^0\Sigma^+)\Sigma^--(\Sigma^+\Sigma^-)\Sigma^0+(\Sigma^-\Sigma^+)\Sigma^0+(\Sigma^0\Sigma^-)\Sigma^+-(\Sigma^-\Sigma^0)\Sigma^+]. 
\end{eqnarray*}

\end{document}